\newif\ifpublicversion
\publicversiontrue 
\ifpublicversion
\documentclass[sigconf,nonacm,balance=false]{acmart}
\else
\documentclass[sigconf,anonymous,review,balance=false]{acmart}
\fi
\usepackage{popets}
\renewcommand\footnotetextcopyrightpermission[1]{} % removes footnote with conference information in first column
\setcopyright{none}
\acmConference{}
\usepackage{amsfonts}
\usepackage{amsmath}
\usepackage{arydshln} %dashed line in tabular
\usepackage{booktabs}
\usepackage[hypcap = false]{caption}
\usepackage{color}
\usepackage{cuted}
\usepackage{enumerate}
\usepackage{enumitem}
\usepackage{float}
\usepackage{graphicx}
\usepackage{makecell}
\usepackage{mathtools}
\usepackage{multirow}
\usepackage{multicol}
\usepackage{nicematrix}
\usepackage{orcidlink}
\usepackage{pgf}
\usepackage{pgfplots}
\pgfplotsset{compat=1.18}
\usepackage{pgfplotstable}
\usepgfplotslibrary{groupplots}

\usepackage{pifont}
\usepackage{ragged2e}
\usepackage{stmaryrd}
\usepackage{stfloats}
\usepackage{subcaption}
\usepackage{siunitx}
\usepackage{tabu}
\usepackage{tabularx}
\usepackage{threeparttable}
\usepackage{tikz}
\usepackage[normalem]{ulem}
\usepackage{url}
\usepackage{verbatim}
\usepackage{wasysym}
\usepackage{wrapfig}
\usepackage{xcolor}
\usepackage{xspace}
\usepackage[english]{babel}
\usepackage [autostyle, english = american]{csquotes}
\MakeOuterQuote{"}
\usepackage[framemethod=tikz]{mdframed} 
\usepackage{algpseudocode}
\usepackage{nccmath}
\usetikzlibrary[calc,shapes,arrows.meta,arrows,positioning]
\usepackage{standalone}

\definecolor{lightblue}{RGB}{100,200,255}
\definecolor{lightred}{RGB}{255,100,100}
\definecolor{lightgreen}{RGB}{100,255,100}

\definecolor{UniBlau}{cmyk}{1,0.7,0,0}
\definecolor{UniGruen}{cmyk}{0.6,0,1,0}
\definecolor{UniOrange}{cmyk}{0,0.3,1,0}
\definecolor{UniRot}{cmyk}{0.4,1,0,0}
\definecolor{darkred}{rgb}{.6,0,0}
\definecolor{darkgreen}{rgb}{0,.4,0}
\definecolor{darkblue}{rgb}{0,0,.6}

\definecolor{absolutezero}{rgb}{0.0,0.28,0.73}
\definecolor{acidgreen}{rgb}{0.69,0.75,0.1}
\definecolor{amazon}{rgb}{0.23,0.48,0.34}
\definecolor{ambersaeece}{rgb}{1.0,0.49,0.0}
\definecolor{antiquebronze}{rgb}{0.4,0.36,0.12}
\definecolor{aoenglish}{rgb}{0.0,0.5,0.0}
\definecolor{brightmaroon}{rgb}{0.76,0.13,0.28}
\definecolor{cadmiumred}{rgb}{0.89,0.0,0.13}
\definecolor{britishracinggreen}{rgb}{0.0,0.26,0.15}

\definecolor{lightgray}{gray}{0.9}

\definecolor{alabaster}{rgb}{0.93,0.92,0.88}
\definecolor{antiquewhite}{rgb}{0.98,0.92,0.84}
\definecolor{beige}{rgb}{0.96,0.96,0.86}
\definecolor{burlywood}{rgb}{0.87,0.72,0.53}
\definecolor{camel}{rgb}{0.76,0.6,0.42}
\definecolor{darkkhaki}{rgb}{0.74,0.72,0.42}

\newif\ifsubmission
\submissionfalse % editing (draft) version
\newif\iffullversion
\fullversiontrue % full version (ePrint)

\newif\ifanonymous
\newif\ifworkingdraft
\workingdraftfalse
\ifsubmission
	\newcommand{\TODO}[1]{}
	\newcommand{\ASSIGN}[1]{}
	\newcommand{\TODOM}[1]{}
	
	\newcommand{\del}[1]{}
	
	\newcommand{\TOASK}[1]{}
	\newcommand{\ALL}[1]{}
	\newcommand{\tsc}[1]{}
    \newcommand{\mycomm}[3]{}
    \newcommand{\mynote}[1]{}
\else
	\newcommand{\TODO}[1]{\textcolor{red}{TODO: #1}}
	\newcommand{\ASSIGN}[1]{\textcolor{darkred}{(Assigned to #1)}}
	\newcommand{\TODOM}[1]{\marginpar{\textcolor{red}{TODO: #1}}}
	
	\newcommand{\del}[1]{\textcolor{purple}{\sout{#1}}}
	
	\newcommand{\TOASK}[1]{\marginpar{\textcolor{green}{?: #1}}}
	\newcommand{\ALL}[1]{\textcolor{violet}{ALL: #1}}
	\newcommand{\tsc}[1]{\textcolor{darkgreen}{#1}}
    \newcommand{\mycomm}[3]{{\footnotesize{{\color{#2} \textbf{[#1: #3]}}}}}
    \newcommand{\mynote}[1]{\mycomm{note}{cadmiumred}{#1}} 
\fi

\newcommand{\myemail}[1]{\href{mailto:#1}{ \textcolor{darkblue}{#1}}}

\newcommand{\csec}{\kappa}

\newcommand{\abort}{\ensuremath{\mathtt{abort}}}

\newcommand{\continue}{\ensuremath{\mathtt{continue}}}

\newcommand{\select}{\ensuremath{\mathsf{Select}}}
\newcommand{\Input}{\ensuremath{\mathsf{Input}}}
\newcommand{\Output}{\ensuremath{\mathsf{Output}}}

\newcommand{\Party}[1]{\ensuremath{{P}_{#1}}}
\newcommand{\Partyverify}[1]{\ensuremath{{P}_{#1}^{V}}}
\newcommand{\Partyset}{\ensuremath{\mathcal{P}}}
\newcommand{\Partysubset}[1]{\ensuremath{\Partyset_{#1}}}
\newcommand{\Partysubsetverify}[1]{\ensuremath{\Partyset_{#1}^{V}}}

\newcommand{\bitb}{\ensuremath{\mathsf{b}}}

\newcommand{\Hash}{\ensuremath{\mathsf{H}}}

\newcommand{\Z}[1]{\ensuremath{\mathbb{Z}}_{2^{#1}}}

\setlist[description]{style=unboxed,leftmargin=0cm}

\setlist[enumerate]{itemsep=0mm}

\newcommand{\makeparafit}{\looseness=-1}

\newcommand{\tabref}[1]{Table~\protect\ref{tab:#1}}
\newcommand{\secref}[1]{\S\protect\ref{sec:#1}}

\newcommand{\figref}[1]{Figure~\ref{fig:#1}}
\newcommand{\figlab}[1]{\label{fig:#1}}

\newenvironment{boxfig*}[2]{% {#1}{#2} = {Caption}{label}
	\begin{figure*}[h!]		
		\fontsize{5}{5}\selectfont
		\newcommand{\FigCaption}{#1}
		\newcommand{\FigLabel}{#2}
		\vspace{-.05cm}
		\begin{center}
			\begin{small}			 
				\begin{adjustbox}{max width=\textwidth}
					\begin{tabular}{@{}|@{~~}l@{~~}|@{}}
						\hline
						\rule[-1ex]{0pt}{1ex}\begin{minipage}[b]{.95\linewidth}
							\vspace{1ex}	
						}{%
						\end{minipage}\\
						\hline
					\end{tabular}	
				\end{adjustbox}		
			\end{small}
			\vspace{-0.1cm}
			\caption{\FigCaption}
			\figlab{\FigLabel}
		\end{center}
		\vspace{-.38cm}
	\end{figure*}
}

\newenvironment{myboxfig*}[2]{% {#1}{#2} = {Caption}{label}
	\begin{figure*}[!htb]		
		\fontsize{5}{5}\selectfont
		\newcommand{\FigCaption}{#1}
		\newcommand{\FigLabel}{#2}
		\vspace{-.10cm}
		\begin{center}
			\caption{\FigCaption}
			\begin{small}			 
				\begin{adjustbox}{max width=\textwidth}
					\begin{tabular}{@{}|@{~~}l@{~~}|@{}}
						\hline
						\rule[-1ex]{0pt}{1ex}\begin{minipage}[b]{.95\linewidth}
							\vspace{1ex}	
						}{%
						\end{minipage}\\
						\hline
					\end{tabular}	
				\end{adjustbox}		
			\end{small}
			\vspace{-0.25cm}
			\figlab{\FigLabel}
		\end{center}
		\vspace{-.38cm}
	\end{figure*}
}

\RequirePackage{mdframed}

\newenvironment{acmarttitlebox}[5]
{\mdfsetup{
		style=#2,
		innertopmargin=1.1\baselineskip,
		skipabove={\dimexpr0.2\baselineskip+\topskip\relax},
		skipbelow={-1.2em},needspace=3\baselineskip,
		singleextra={\node[#3,right=10pt,overlay] at (P-|O){~{\sffamily\bfseries #1 }};},%
		firstextra={\node[#3,right=10pt,overlay] at (P-|O) {~{\sffamily\bfseries #1 }};},
		frametitleaboveskip=9em,
		innerrightmargin=5pt
	}
	\newcommand{\TitleCaption}{#4}
	\newcommand{\TitleLabel}{#5}
	\begin{mdframed}[font=\small]
		\setlist[itemize]{leftmargin=13pt}\setlist[enumerate]{leftmargin=13pt}\raggedright% 
	}
	{\end{mdframed}
	{\captionof{figure}{\small \TitleCaption}\label{\TitleLabel}}
}

\mdfdefinestyle{acmartcommonbox}{%
    align=center, 
    middlelinewidth=1.1pt,
    userdefinedwidth=\linewidth,
    innerrightmargin=0pt,
    innerleftmargin=2pt,
    innertopmargin=5pt,
    splittopskip=7pt,
    splitbottomskip=3pt
}

\newenvironment{titlebox}[5]
{\mdfsetup{
		style=#2,
		innertopmargin=1.1\baselineskip,
		skipabove={\dimexpr0.2\baselineskip+\topskip\relax},
		skipbelow={1em},needspace=3\baselineskip,
		singleextra={\node[#3,right=10pt,overlay] at (P-|O){~{\sffamily\bfseries #1 }};},%
		firstextra={\node[#3,right=10pt,overlay] at (P-|O) {~{\sffamily\bfseries #1 }};},
		frametitleaboveskip=9em,
		innerrightmargin=5pt
	}
	\newcommand{\TitleCaption}{#4}
	\newcommand{\TitleLabel}{#5}
	\begin{mdframed}[font=\small]
		\setlist[itemize]{leftmargin=13pt}\setlist[enumerate]{leftmargin=13pt}\raggedright% 
	}
	{\end{mdframed}
	\vspace{-2em} %CCS/PETS
	{\captionof{figure}{\small \TitleCaption}\label{\TitleLabel}}
	\smallskip
}

\tikzstyle{normal} = [thick, fill=white, text=black, draw, rounded corners, rectangle, minimum height=.7cm, inner sep=3pt]
\tikzstyle{gray} = [thick, fill=gray!90, text=white, rounded corners, rectangle, minimum height=.7cm, inner sep=3pt]

\mdfdefinestyle{commonbox}{%
	align=center, middlelinewidth=1.1pt,userdefinedwidth=\linewidth
	innerrightmargin=0pt,innerleftmargin=2pt,innertopmargin=5pt,
	splittopskip=7pt,splitbottomskip=3pt%,
}

\mdfdefinestyle{roundbox}{style=commonbox,roundcorner=5pt,userdefinedwidth=\linewidth}

\newenvironment{systembox}[3]
{\vspace{\baselineskip}\begin{titlebox}{Functionality \normalfont #1}{roundbox}{normal}{#2}{#3}}
	{\end{titlebox}}

\newenvironment{protocolbox}[3]
{\begin{titlebox}{Protocol \normalfont #1}{commonbox}{normal}{#2}{#3}}
	{\end{titlebox}}

\newenvironment{simulatorbox}[3]
{\begin{titlebox}{Simulator \normalfont #1}{commonbox}{normal}{#2}{#3}}
	{\end{titlebox}}
\newenvironment{splittitlebox}[5]
{\mdfsetup{
		style=#2,
		innertopmargin=1.1\baselineskip,
		skipabove={\dimexpr0.2\baselineskip+\topskip\relax},
		skipbelow={1em},needspace=3\baselineskip,
		singleextra={\node[#3,right=10pt,overlay] at (P-|O){~{\sffamily\bfseries #1 }};},%
		firstextra={\node[#3,right=10pt,overlay] at (P-|O) {~{\sffamily\bfseries #1 }};},
		frametitleaboveskip=9em,
		innerrightmargin=5pt
	}
	\newcommand{\TitleCaption}{#4}
	\newcommand{\TitleLabel}{#5}
	\begin{mdframed}[font=\small]
		\setlist[itemize]{leftmargin=13pt}\setlist[enumerate]{leftmargin=13pt}\raggedright% 
	}
	{\end{mdframed}
	\vspace{-1em} %CCS/PETS
	{\captionof{figure}{\small \TitleCaption}\label{\TitleLabel}}
	\smallskip
}

\mdfdefinestyle{commonsplitbox}{%
	align=center, middlelinewidth=1.1pt,userdefinedwidth=\linewidth
	innerrightmargin=0pt,innerleftmargin=2pt,innertopmargin=5pt,
	splittopskip=7pt,splitbottomskip=3pt%,
}

\newenvironment{systembox*}[3]
{\begin{strip}
		\vspace{\baselineskip}\begin{titlebox}{Functionality \normalfont #1}{roundbox}{normal}{#2}{#3}}
		{\end{titlebox}
\end{strip}}

\newenvironment{gsystembox*}[3]
{\begin{strip}
		\vspace{\baselineskip}\begin{titlebox}{Global Functionality \normalfont #1}{roundbox}{normal}{#2}{#3}}
		{\end{titlebox}
\end{strip}}

\newenvironment{protocolbox*}[3]
{\begin{strip}
		\begin{titlebox}{Protocol \normalfont #1}{commonbox}{normal}{#2}{#3}}
		{\end{titlebox}
\end{strip}}

\newenvironment{algobox*}[3]
{\begin{strip}
		\begin{titlebox}{Algorithm \normalfont #1}{commonbox}{normal}{#2}{#3}}
		{\end{titlebox}
\end{strip}}

\newenvironment{reductionbox*}[3]
{\begin{strip}
		\begin{titlebox}{Reduction \normalfont #1}{commonbox}{normal}{#2}{#3}}
		{\end{titlebox}
\end{strip}}

\newenvironment{gamebox*}[3]
{\begin{strip}
		\begin{titlebox}{Game \normalfont #1}{commonbox}{gray}{#2}{#3}}
		{\end{titlebox}
\end{strip}}

\newenvironment{simulatorbox*}[3]
{\begin{strip}
		\begin{titlebox}{Simulator \normalfont #1}{commonbox}{normal}{#2}{#3}}
		{\end{titlebox}
\end{strip}}

\newenvironment{protocolsplitbox*}[3]
{\begin{strip}
		\begin{splittitlebox}{Protocol \normalfont #1}{commonbox}{normal}{#2}{#3}}
		{\end{splittitlebox}
\end{strip}}

\mdfdefinestyle{mycommonbox}{%
	align=center, middlelinewidth=1.1pt,userdefinedwidth=\linewidth
	innerrightmargin=0pt,innerleftmargin=2pt,innertopmargin=3pt,
	splittopskip=7pt,splitbottomskip=3pt
}
\mdfdefinestyle{myroundbox}{style=mycommonbox,roundcorner=5pt,userdefinedwidth=\linewidth}

\newenvironment{titlebox*}[5]
{\mdfsetup{
		style=#2,
		innertopmargin=0.3\baselineskip,
		skipabove={0.4em},
		skipbelow={1em},needspace=3\baselineskip,
		frametitleaboveskip=5em,
		innerrightmargin=5pt
	}
	\newcommand{\TitleCaption}{#4}
	\newcommand{\TitleLabel}{#5}
	\begin{mdframed}[font=\small]
		\setlist[itemize]{leftmargin=13pt}\setlist[enumerate]{leftmargin=13pt}\raggedright% 
	}
	{\end{mdframed}
	\vspace{-2em}
	{\captionof{figure}{\normalfont \TitleCaption}\label{\TitleLabel}}
	\medskip
}

\newenvironment{mysystembox*}[3]
{\begin{strip}
		\vspace{\baselineskip}\begin{titlebox*}{Functionality \normalfont #1}{myroundbox}{normal}{#2}{#3}}
		{\end{titlebox*}
\end{strip}}

\newenvironment{mygsystembox*}[3]
{\begin{strip}
		\vspace{\baselineskip}\begin{titlebox*}{Global Functionality \normalfont #1}{myroundbox}{normal}{#2}{#3}}
		{\end{titlebox*}
\end{strip}}

\newenvironment{myprotocolbox*}[3]
{\begin{strip}
		\begin{titlebox*}{Protocol \normalfont #1}{mycommonbox}{normal}{#2}{#3}}
		{\end{titlebox*}
\end{strip}}

\newenvironment{myalgobox*}[3]
{\begin{strip}
		\begin{titlebox*}{Algorithm \normalfont #1}{mycommonbox}{normal}{#2}{#3}}
		{\end{titlebox*}
\end{strip}}

\newenvironment{myreductionbox*}[3]
{\begin{strip}
		\begin{titlebox*}{Reduction \normalfont #1}{mycommonbox}{normal}{#2}{#3}}
		{\end{titlebox*}
\end{strip}}

\newenvironment{mygamebox*}[3]
{\begin{strip}
		\begin{titlebox*}{Game \normalfont #1}{mycommonbox}{gray}{#2}{#3}}
		{\end{titlebox*}
\end{strip}}

\newenvironment{mysimulatorbox*}[3]
{\begin{strip}
		\begin{titlebox*}{Simulator \normalfont #1}{mycommonbox}{normal}{#2}{#3}}
		{\end{titlebox*}
\end{strip}}

\newcommand{\algoHeadacmart}[1]{\underline{\textbf{#1}}}

\makeatletter
\algnewcommand{\ExtendedState}[1]{\State
	\parbox[t]{\dimexpr\linewidth-\ALG@thistlm}{\hangindent=\algorithmicindent\strut\hangafter=3#1\strut}}
\makeatother

\algnewcommand\algorithmicinput{\textbf{Input:}}
\algnewcommand\Input{\item[\algorithmicinput]}

\algrenewcommand{\algorithmiccomment}[1]{{\color{gray}// #1}}

\newcommand{\xmath}[1]{\ensuremath{#1}\xspace}

\let\emptyset\varnothing

\newcommand{\Func}[1][\relax]{\xmath{\mathcal{F}_{\textsc{#1}}}}
\DeclarePairedDelimiterX{\dbrackets}[1]{\lbrack}{\rbrack}{
	\nhphantom{\lbrack}{#1} \delimsize\lbrack \mathopen{} #1 \mathclose{} \delimsize\rbrack \nhphantom{\rbrack}{#1}
}
\DeclarePairedDelimiterX{\dbraces}[1]{\lbrace}{\rbrace}{
	\nhphantom{\lbrace}{#1} \delimsize\lbrace \mathopen{} #1 \mathclose{} \delimsize\rbrace \nhphantom{\rbrace}{#1}
}
\DeclarePairedDelimiterX{\dparens}[1]{\lparen}{\rparen}{
	\nhphantom{\lparen}{#1} \delimsize\lparen \mathopen{} #1 \mathclose{} \delimsize\rparen \nhphantom{\rparen}{#1}
}

\newcommand{\nhphantom}[2]{\sbox0{$\left#1\vphantom{#2}\right.$}\hspace{-0.58\wd0}}

\newcommand{\threepcname}{\textsf{Trio}}
\newcommand{\fourpcname}{\textsf{Quad}}

\newcommand{\lv}[2]{\ensuremath{\lambda_{#1}^{#2}}}
\newcommand{\lw}[2]{\ensuremath{\lambda_{#1}^{#2*}}}

\newcommand{\mv}[2]{\ensuremath{\mathsf{m}_{#1\ifx&#2&\else,#2\fi}}}
\newcommand{\mw}[2]{\ensuremath{\mathsf{m}_{#1}^{#2*}}}
\newcommand{\mverify}[2]{\ensuremath{\overline{\mathsf{m}}_{#1}^{#2}}}

\newcommand{\error}[1]{\ensuremath{\mathsf{e}_{#1}}}

\newcommand{\msg}[2]{\ensuremath{\mathsf{M}_{#1\ifx&#2&\else,#2\fi}}}
\newcommand{\msgset}[2]{\ensuremath{\mathsf{M}_{\{#1\}\ifx&#2&\else,#2\fi}}}
\newcommand{\msgerror}[3]{\ensuremath{\mathsf{M}_{\{#1\}\ifx&#2&\else,#2\fi}^{\error{#3}}}}

\newcommand{\cval}[2]{\ensuremath{\mathsf{V}_{#1\ifx&#2&\else,#2\fi}}}
\newcommand{\valset}[2]{\ensuremath{\mathsf{V}_{\{#1\}\ifx&#2&\else,#2\fi}}}
\newcommand{\valerror}[3]{\ensuremath{\mathsf{V}_{\{#1\}\ifx&#2&\else,#2\fi}^{\error{#3}}}}

\newcommand{\ptrunc}[1]{\ensuremath{\left(#1\right)}^{pt}}
\newcommand{\ftrunc}[1]{\ensuremath{\left(#1\right)}^t}

\newcommand{\sqr}[1]{\ensuremath{\left[#1\right]}}
\newcommand{\shr}[1]{\ensuremath{\llbracket#1\rrbracket}}
\newcommand{\sh}[1]{\ensuremath{\llbracket#1\rrbracket}}

\newcommand{\key}[1]{\ensuremath{\mathsf{k}_{#1}}}
\newcommand{\counter}[1]{\ensuremath{\mathsf{c}_{#1}}}
\newcommand{\buffer}[1]{\ensuremath{\mathsf{B}_{#1}}}
\newcommand{\buffertop}{\ensuremath{\mathsf{B}^{t}}}
\newcommand{\randval}[1]{\ensuremath{\mathsf{r}_{#1}}}

\newcommand{\bool}{\textsf{bool}}
\newcommand{\true}{\textsf{true}}
\newcommand{\false}{\textsf{false}}

\newcommand{\vl}[1]{\ensuremath{\mathsf{#1}}}

\newcommand{\Mult}{\ensuremath{\mathsf{Mult}}}
\newcommand{\MultH}{\ensuremath{\mathsf{Mult\_H}}}
\newcommand{\AtoB}{\ensuremath{\mathsf{A2B}}}

\newcommand{\BittoA}{\ensuremath{\mathsf{Bit2A}}}
\newcommand{\SRNG}{\ensuremath{\mathsf{SRNG}}}
\newcommand{\CompareView}{\ensuremath{\mathsf{CV}}}

\newcommand{\Share}{\ensuremath{\mathsf{Sh}}}
\newcommand{\Rec}{\ensuremath{\mathsf{Rec}}}
\newcommand{\FairRec}{\ensuremath{\mathsf{fairRec}}}

\newcommand{\piAtoB}{\ensuremath{\Pi_{\AtoB}}}

\newcommand{\piBittoA}{\ensuremath{\Pi_{\BittoA}}}
\newcommand{\piMult}{\ensuremath{\Pi_{\Mult}}}
\newcommand{\piSRNG}{\ensuremath{\Pi_{\SRNG}}}
\newcommand{\piCompareView}{\ensuremath{\Pi_{\CompareView}}}
\newcommand{\piMultH}{\ensuremath{\Pi_{\MultH}}}

\newcommand{\piShare}{\ensuremath{\Pi_{\Share}}}
\newcommand{\piRec}{\ensuremath{\Pi_{\Rec}}}
\newcommand{\piFairRec}{\ensuremath{\Pi_{\FairRec}}}

\newcommand{\function}{\mathcal{F}}

\newcommand{\Ideal}{\ensuremath{\textsc{ideal}}}
\newcommand{\Real}{\ensuremath{\textsc{real}}}
\newcommand{\funcn}[1]{\function_{\textsf{#1}}}

\newcommand{\circuit}{{\mathsf{ckt}}}
\newcommand{\PPT}{{\mathsf{PPT}}}
\newcommand{\func}[1]{{f(#1)}}

\newcommand{\Adv}{\ensuremath{\mathcal{A}}}
\newcommand{\Sim}{\ensuremath{\mathcal{S}}}

\newcommand{\FuncFour}[1]{\function^{\textsf{4pc}}_{\textsf{#1}}}

\newcommand{\FuncSRNG}{\ensuremath{\function_{\SRNG}}}
\newcommand{\Funcsetup}{\ensuremath{\function_{\mathsf{setup}}}}

\begin{document}
%----------------------------
\title[High-Throughput Secure Multiparty Computation]{High-Throughput Secure Multiparty Computation with an Honest Majority in Various Network Settings}

%%%%%%%%%%%%%%%% Authors' Info %%%%%%%%%%%%%%%%%

\author{Christopher Harth-Kitzerow\orcidlink{0000-0003-0792-7207}}
\affiliation{%
  \institution{Technical University of Munich, BMW Group}
  \country{\myemail{christopher.harth-kitzerow@tum.de}}
}

\author{Ajith Suresh\orcidlink{0000-0002-5164-7758}}
\affiliation{%
  \institution{Technology Innovation Institute, Abu~Dhabi}
  \country{\myemail{ajith.suresh@tii.ae}}
}

\author{Yongqin Wang\orcidlink{0009-0005-5076-7706}}
\affiliation{%
  \institution{University of Southern California}
  \country{\myemail{yongqin@usc.edu}}
}

\author{Hossein Yalame\orcidlink{0000-0001-6438-534X}}
\affiliation{%
  \institution{Robert Bosch GmbH, Germany}
  \country{\myemail{hossein.yalame@de.bosch.com}}
}

\author{Georg Carle\orcidlink{0000-0002-2347-1839}}
\affiliation{%
  \institution{Technical University of Munich}
  \country{\myemail{carle@net.in.tum.de}}
}

\author{Murali Annavaram\orcidlink{0000-0002-4633-6867}}
\affiliation{%
  \institution{University of Southern California}
  \country{\myemail{annavara@usc.edu}}
}


%\renewcommand{\shortauthors}{Harth-Kitzerow et al.}

%----------------------------
\begin{abstract}
In this work, we present novel protocols over rings for semi-honest secure three-party computation (3PC) and malicious four-party computation (4PC) with one corruption. While most existing works focus on improving total communication complexity, challenges such as network heterogeneity and computational complexity, which impact MPC performance in practice, remain underexplored. 

Our protocols address these issues by tolerating multiple arbitrarily weak network links between parties without any substantial decrease in performance. Additionally, they significantly reduce computational complexity by requiring up to half the number of basic instructions per gate compared to related work. These improvements lead to up to twice the throughput of state-of-the-art protocols in homogeneous network settings and up to eight times higher throughput in real-world heterogeneous settings. These advantages come at no additional cost: Our protocols maintain the best-known total communication complexity per multiplication, requiring 3 elements for 3PC and 5 elements for 4PC.

We implemented our protocols alongside several state-of-the-art protocols (Replicated 3PC, ASTRA, Fantastic Four, Tetrad) in a novel open-source C++ framework optimized for high throughput. Five out of six implemented 3PC and 4PC protocols achieve more than one billion 32-bit multiplications or over 32 billion AND gates per second using our implementation in a 25 Gbit/s LAN environment. This represents the highest throughput achieved in 3PC and 4PC so far, outperforming existing frameworks like MP-SPDZ, ABY3, MPyC, and MOTION by two to three orders of magnitude.
\end{abstract}

%----------------------------

%----------------------------
\maketitle
%----------------------------

%----------------------------
%============================
\section{Introduction}
\label{sec:Introduction}
%============================
Secure Multi-party Computation~(MPC) enables parties to execute functions on obliviously shared inputs without revealing them~\cite{CACM:Lindell21}. Consider multiple hospitals that want to study the adverse effects of a certain medication based on their patients' data. While joining these datasets could enable more statistically significant results, hospitals might be prohibited from sharing their private patient data with each other. MPC enables these hospitals to perform this study and only reveal the final output of the function evaluated. MPC deployments are gaining prominence, such as private inventory matching to match buyers and sellers of stocks privately~\cite{berkeley}.

MPC protocols fall into two categories: high-throughput~\cite{CCS:AFLNO16,CCSW:CCPS19,USENIX:Dalskov0K21,NDSS:ChaRacSur20} and low-latency~\cite{STOC:BeaverMR90,FOCS:Yao86}. Low-latency protocols often uses garbled circuits~\cite{FOCS:Yao86}, resulting in constant-round solutions. In contrast, high-throughput protocols are mainly based on secret-sharing~(SS) solutions, requiring communication rounds proportional to the multiplicative depth of the circuit. However, SS-based protocols generally involve less communication than garbled circuits, allowing multiple instances of SS-based protocols to be executed in parallel, thereby achieving high throughput. For instance, Araki et al.~\cite{CCS:AFLNO16} designed a high-throughput 3PC protocol capable of authenticating a login storm of 35,000 users, which requires computing a large number of parallel AES blocks. Also, several other MPC use cases such as privacy-preserving machine learning~\cite{SP:NC23} with large batch sizes can benefit from implementations that achieve high throughput.

Most SS-based MPC protocols are designed to operate either over a field, or over an arbitrary ring $\Z{\ell}$. Typical choices are the ring $\mathbb{Z}_{2}$ for Boolean circuits and $\mathbb{Z}_{2^{64}}$ for arithmetic circuits. While Boolean circuits can express comparison-based functions, arithmetic circuits can express arithmetic functions more compactly~\cite{CCS:KelOrsSch16}. Computation over $\mathbb{Z}_{2^{64}}$ is supported by 64-bit hardware natively and thus leads to efficient implementations~\cite{PPAI:SXL19}. Multiple approaches also allow share conversion between computation domains to evaluate mixed circuits~\cite{TCC:CraDamIsh05,CCS:MohRin18,USENIX:PSSY21,NDSS:ChaRacSur20}. 

MPC protocols can be secure against semi-honest and malicious adversaries~\cite{ISCI:ZhaoZZCGLT19}. A semi-honest adversary tries to break the privacy of the protocol. In contrast, a malicious adversary may try to break both the privacy and the correctness of the protocol by arbitrarily deviating from the protocol. Another important aspect of an MPC protocol is the maximum number of parties the adversary is allowed to corrupt. MPC protocols that can tolerate more than half of the participating parties are in the dishonest-majority class, while protocols that can tolerate less than half of the parties are in the honest-majority class. Typically, honest-majority protocols achieve significantly lower communication complexity than dishonest-majority ones.

Over the recent years, there has been a significant interest in fast and lightweight SS-based honest-majority MPC protocols for both the semi-honest \cite{FCW:KerLauRan16,CCSW:CCPS19,CCS:AFLNO16} and malicious adversaries \cite{USENIX:Dalskov0K21,NDSS:KotiPRS22,NDSS:PatSur20,USENIX:KPPS21,PETS:ByaliCPS20,NDSS:ChaRacSur20}. In the semi-honest setting, the 3PC~\cite{FCW:KerLauRan16, CCS:AFLNO16, CCS:MohRin18, CCSW:CCPS19} protocols that tolerate up to one corruption are particularly relevant due to their low bandwidth requirements. This setting allows the use of information-theoretic security techniques not applicable to two-party computation~\cite{BOOK:CDN2015}. Likewise, for malicious corruption, the 4PC~\cite{PETS:ByaliCPS20,USENIX:KPPS21,USENIX:Dalskov0K21,NDSS:ChaRacSur20} protocols are of particular interest as they allow exploiting the redundancy of secretly shared values to efficiently verify the correctness of exchanged messages, when compared with their 3PC counterparts. These properties of 3PC and 4PC protocols can also be utilized in the outsourced computation model~\cite{FC:DDNNT16}, where three or four fixed computation nodes perform a computation for any number of input parties.

All recently proposed 3PC~\cite{CCSW:CCPS19,CCS:AFLNO16,SP:BSSSY24} and 4PC~\cite{PETS:ByaliCPS20,USENIX:KPPS21,USENIX:Dalskov0K21,NDSS:ChaRacSur20} protocols share that they require at least three resp. six elements of global communication per multiplication gate. 
Only recently, a 4PC protocol achieved five elements of global communication per multiplication gate \cite{NDSS:KotiPRS22}. However, among other 4PC protocols \cite{PETS:ByaliCPS20,USENIX:KPPS21,NDSS:ChaRacSur20}, it is lacking an open-source implementation.
Additionally, many protocols offer a \emph{Preprocessing Phase} that is independent of actual inputs and can be evaluated prior to the \emph{Online Phase} which in turn handles input-dependent computation and communication.
Most existing honest-majority protocols work optimally in a homogeneous network setting \cite{STOC:BenGolWig88, FCW:KerLauRan16, CCS:AFLNO16, USENIX:Dalskov0K21}. While a few protocols also work well in heterogeneous network settings \cite{CCSW:CCPS19, NDSS:KotiPRS22}, they also do not provide an open-source implementation. 
In an orthogonal direction, a recent work proposed protocol to convert any honest-majority semi-honest protocol into a malicious one at no additional amortized communication costs~\cite{CCS:BGIN19}. However, the Distributed Zero-Knowledge Proofs~(DZKP) required for this conversion come with significant computational overhead. According to a recent benchmark~\cite{USENIX:Dalskov0K21}, their ring-based solution only achieves 22 multiplication gates per second. This is orders of magnitude lower than what state-of-the-art protocols achieve in practice.

%--------------------------------------
\begin{table}[htb!]
\small
\centering
\captionsetup{font=small}
 \caption{Operations and communication for multiplication}
 \label{tab:comp_score}
 \vspace{-3mm}
\begin{threeparttable}
 %\resizebox{0.42\textwidth}{!}{%
 \begin{tabular}{c|l|ll|lll}
 \hline
 %-------------------------
 \multirow{2}{*}{~} & \multirow{2}{*}{Protocol} & \multicolumn{2}{c|}{Operation} & \multicolumn{3}{c}{Communication\tnote{a}} \\ \cline{3-7}
 &  & Add & Mult & Pre.\tnote{b} & On \tnote{b} & Links \tnote{c} \\ \hline
 %-------------------------
 \multirow{3}{*}{3PC} 
 & Replicated~\cite{CCS:AFLNO16} & 12 & 6 (+3)\tnote{d} & 0 & 3 & 0B,0L   \\
 & ASTRA~\cite{CCSW:CCPS19} & 11 & 5\tnote{e} & 1 & 2 & 1B,2L   \\ \cline{2-7}
 & $\threepcname$~(\textbf{This work}) & 11 & 5 & 1 & 2 & 1B,2L    \\ \hline
%-------------------------
 \multirow{3}{*}{4PC}
 & Fantastic Four \cite{USENIX:Dalskov0K21} & 60 & 36 & 0 & 6 & 2-4B,2-4L   \\
 & Tetrad \cite{NDSS:KotiPRS22} & 52 & 30 & 2 & 3 & 2B,5L  \\ 
 & Trident \cite{NDSS:ChaRacSur20} & 51 & 30 & 3 & 3 & 3B,3L  \\ \cline{2-7}
 & $\fourpcname$~(\textbf{This work}) & 25 & 12 & 2 & 3 & 3-4B,5L  \\ \hline
 %-------------------------
 \end{tabular}
 %}
 \begin{tablenotes}%[para]
 \item[a] Total communication complexity for all parties combined.
 \item[b] \textit{Pre.} refers to Preprocessing, \textit{On.} refers to Online Phase.
 \item[c] B - \#low-bandwidth links tolerated, L - \#high-latency links tolerated. Ranges (e.g. 3-4) represent different protocol variations. 
 \item[d] Replicated 3PC additionally requires one division operation per party (+3) in the arithmetic domain.
 \item[e] Reduced from 6 using a straightforward optimization (c.f. \secref{3PC}).
 \end{tablenotes}
\end{threeparttable}
\end{table}
%--------------------------------------

The performance of MPC protocols is limited by both computational and communication complexity. \tabref{comp_score} shows the number of local additions and multiplications required to compute a single multiplication gate securely using existing works in 3PC and 4PC settings. As shown in~\tabref{comp_score}, even state-of-the-art 4PC protocols \cite{USENIX:Dalskov0K21,NDSS:ChaRacSur20,NDSS:KotiPRS22} require almost 100 local additions and multiplications for each multiplication gate on top of computing hashes and sampling shared random numbers. In~\secref{Bottlenecks}, we quantitatively show that introducing this large computational overhead can become a serious performance hurdle for certain workloads. 

\tabref{comp_score} also provides the total communication complexity and the number of network links that the protocol can tolerate in terms of bandwidth and latency~(see~\tabref{comp_full} for per-party analysis).  
Our quantitative analysis shows that MPC practitioners should expect significant variability in latency and bandwidth among network links, even in highly sophisticated cloud settings. Therefore, a versatile MPC protocol must address both computational complexity and network heterogeneity.
Since existing related works do not address these issues, we bridge this gap by proposing an efficient protocol for semi-honest 3PC, and then a malicious 4PC protocol building on the 3PC.

%============================
\subsection*{Our Contributions}
\label{sec:Contribution}
%============================
We approach the challenge of achieving high-throughput MPC in practical settings from two perspectives:
\begin{enumerate}[wide, labelwidth=!, labelindent=10pt, parsep=2pt, itemsep=1mm]
    %-------
    \item[--] \textbf{Protocol Design:} We develop new protocols that overcome existing bottlenecks in MPC workloads.
    %-------
    \item[--] \textbf{Framework Implementation:} We implement a C++ framework that accelerates any MPC protocol by efficiently utilizing hardware and networking resources.
    %-------
\end{enumerate}
This holistic, end-to-end approach allows us to maximize the potential of existing node setups and ensure that theoretical advancements in the communication and computational complexity of MPC protocols are reflected in practice.

Our novel protocols offer the following contributions:
\begin{enumerate}[wide, labelwidth=!, labelindent=10pt, parsep=2pt, itemsep=1mm]
    %-------
    \item We present a semi-honest 3PC protocol, $\threepcname$~(cf.~\secref{3PC}), and a maliciously secure 4PC protocol, $\fourpcname$~(cf.~\secref{4PC}). $\threepcname$ requires total communication of three elements per multiplication gate, and $\fourpcname$ requires five, aligning with the state-of-the-art.
    %-------
    \item By reducing the correlation between the shares of parties, we significantly reduce the computational complexity per gate compared to related work.  Our 4PC protocol, $\fourpcname$, achieves up to two times higher throughput for computationally intensive tasks like dot products. 
    %-------
    \item Our protocols exploit network heterogeneity by redistributing communication between parties, leveraging stronger network links and minimizing reliance on weaker ones without increasing communication complexity. Both our 3PC and 4PC protocols can tolerate all but two links having arbitrarily low bandwidth and all but one link having arbitrarily high latency, while still achieving fast runtimes. 
    %-------
\end{enumerate}
    
Our novel framework offers the following contributions : 
\begin{enumerate}[wide, labelwidth=!, labelindent=10pt, parsep=2pt, itemsep=1mm]
    %-------
    \item  Our C++ framework utilizes techniques such as Bitslicing, vectorization, message buffering, and load balancing. These implementation techniques are orthogonal to our primary protocol contributions and can accelerate existing protocols as well. By incorporating these optimizations, our framework achieves an unmatched throughput of more than 25 billion AND gates per second on a 25 Gbit/s network for each of the six currently implemented honest-majority protocols. This performance is two to three orders of magnitude higher than that of the open-source frameworks MP-SPDZ~\cite{CCS:Keller20}, ABY3~\cite{CCS:MohRin18}, MOTION~\cite{TOPS:MOTION}, and MpyC~\cite{TPMPC:Sch18} under the same setup.   
    %-------
    \item We have open-sourced our framework\footnote{Code Repository: 
    \ifpublicversion
       \href{https://github.com/chart21/hpmpc/}{\textcolor{darkblue}{https://github.com/chart21/hpmpc/}}
    \else
     \href{https://anonymous.4open.science/r/hpmpc-29C4/}{\textcolor{darkblue}{https://anonymous.4open.science/r/hpmpc-29C4/}}, Download: \href{https://drive.google.com/file/d/1JVf93159Yk9JsoTMWND91LI-SP-pBy2t/}{\textcolor{darkblue}{ZIP}}
    \fi
    }, which includes our novel protocols as well as several other state-of-the-art 3PC protocols~\cite{FCW:KerLauRan16,CCS:AFLNO16,CCSW:CCPS19}, 4PC protocols~\cite{USENIX:Dalskov0K21,NDSS:KotiPRS22}, and a trusted-third-party protocol. Some of these protocols have not been previously implemented in any open-source framework \cite{CCSW:CCPS19,NDSS:KotiPRS22}\footnote{In the meantime, ASTRA has also been implemented by~\cite{SP:BSSSY24}.}. For other protocols~\cite{CCS:AFLNO16,USENIX:Dalskov0K21}, we achieve up to three orders of magnitudes higher throughput compared to their current open-source implementations in MP-SPDZ~\cite{CCS:Keller20}. 
    %-------
\end{enumerate}
%============================
\section{Bottlenecks of MPC in Practice}
\label{sec:Bottlenecks}
%============================
In this work, we identify and tackle three challenges that limit the performance of MPC protocols in practice and are not related to the total communication complexity between parties. 

%-------------------------------
\begin{table*}[htb!]
\centering
\small
\captionsetup{font=small}
\caption{Throughput in Gates/s for Various Frameworks}
\label{tab:otherFrameworks}
\vspace{-3mm}
\resizebox{0.85\textwidth}{!}{%
\begin{tabular}{lccccc}
\toprule
\multirow{2}{*}{Framework} & \multirow{2}{*}{Measurement} & \multicolumn{2}{c}{$\mathbb{Z}_{2^{32}}$ Multiplication} & \multicolumn{2}{c}{ $\mathbb{Z}_{2}$ AND} \\ 
\cmidrule(lr){3-4} \cmidrule(lr){5-6}
& & 3PC & 4PC & 3PC & 4PC \\ 
\midrule
\multirow{2}{*}{MP-SPDZ \cite{CCS:Keller20}} 
& Internal & $(1.94 \pm 0.47) \times 10^7$ & $(7.07 \pm 0.22) \times 10^6$ & $(4.74 \pm 0.13) \times 10^6$ & $(1.84 \pm 0.03) \times 10^6$ \\ 
& External & $(7.46 \pm 0.50) \times 10^6$ & $(3.87 \pm 0.18) \times 10^6$ & $(2.61 \pm 0.06) \times 10^6$ & $(1.34 \pm 0.02) \times 10^6$ \\ 
\midrule
\multirow{2}{*}{ABY3 \cite{CCS:MohRin18}} 
& External (Vanilla) & $(1.49 \pm 0.04) \times 10^4$ & - & - & - \\ 
& External (Multithreading) & $(2.83 \pm 0.06) \times 10^5$ & - & - & - \\ 
\midrule
\multirow{2}{*}{MOTION \cite{TOPS:MOTION}} 
& Internal & $(2.22 \pm 0.10) \times 10^3$ & $(6.34 \pm 0.15) \times 10^4$ & $(8.27 \pm 0.17) \times 10^4$ & $(1.49 \pm 0.08) \times 10^3$ \\ 
& External & $(1.15 \pm 0.19) \times 10^3$ & $(4.06 \pm 0.07) \times 10^4$ & $(5.28 \pm 0.17) \times 10^4$ & $(8.22 \pm 0.03) \times 10^2$ \\ 
\midrule
\multirow{2}{*}{MPyC \cite{TPMPC:Sch18}} 
& Internal & $(4.41 \pm 0.50) \times 10^5$ & $(4.21 \pm 0.46) \times 10^5$ & $(5.82 \pm 0.01) \times 10^2$ & $(5.81 \pm 0.01) \times 10^2$ \\ 
& External & $(4.05 \pm 0.32) \times 10^5$ & $(3.83 \pm 0.27) \times 10^5$ & $(5.69 \pm 0.01) \times 10^2$ & $(5.68 \pm 0.01) \times 10^2$ \\ 
\bottomrule
\end{tabular}
}
\vspace{-2mm}
\end{table*}

%-------------------------------

%-------------------------------
\subsubsection*{\textbf{Network Heterogeneity is Ubiquitous}}
%-------------------------------
In real-world settings, network links between distributed parties are highly heterogeneous. To quantify this heterogeneity, we set up multiple AWS~C6in instances across different network settings: Same Data center~(LAN), Same City~(MAN), Same Continent~(WAN1), Different Continents (WAN2), and Mixed Constellations~(Mixed). As datacenter placements of a single cloud provider are often optimized in MAN settings, we also perform a cross-cloud MAN~(CMAN) measurement of the different providers AWS, GCP, Microsoft Azure, and Oracle Cloud. We measured both the bandwidth and latency for each link between the parties and identified the strongest and weakest links within each node setup in terms of latency and bandwidth. \tabref{node_location} presents the measured values. We observed significant variance in both the best and worst-case measurements within the same setting, with even more dramatic differences across different settings. 

%-------------------------------

\begin{table}[htb!]
\small
\centering
\captionsetup{font=small}
\caption{Link Heterogeneity in Different Network Settings}
\vspace{-3mm}
\begin{threeparttable}
\begin{tabular}{lccc|ccc}
\toprule
\multirow{2}{*}{Setting} & \multicolumn{3}{c}{Latency (ms)} & \multicolumn{3}{c}{Bandwidth (Mbit/s)} \\ 
\cmidrule(lr){2-4} \cmidrule(lr){5-7}
& Best & Worst & \% Diff & Best & Worst & \% Diff \\ 
\midrule
LAN   & 0.178 & 0.304 & 70.8\%  & 4790\tnote{a} & 4940\tnote{a} & 3.1\%  \\ 
MAN   & 0.383 & 0.953 & 148.8\% & 2540 & 2230  & 13.9\% \\ 
CMAN & 1.088 & 2.394 & 120\% & 1550\tnote{a} & 137\tnote{a}   & 1031\% \\  
WAN1  & 34.772 & 192.515 & 453.6\% & 868 & 142 & 511.3\% \\ 
WAN2  & 91.451 & 276.253 & 202.1\% & 341 & 108 & 215.7\% \\ 
Mixed & 34.799 & 276.256 & 693.9\% & 824 & 108 & 663.0\% \\ 
\bottomrule
\end{tabular}
\begin{tablenotes}
\item[a] Bandwidth is limited by cloud providers' policies.
\end{tablenotes}
\end{threeparttable}
\label{tab:node_location}
\end{table}
%-------------------------------

These exemplary real-world network settings would significantly benefit from a protocol that assigns all latency-critical communications to the link with the best latency between the parties, while directing the bulk of the messages through links with higher available bandwidth. Most existing protocols achieve the same or similar communication complexity for each party~\cite{FCW:KerLauRan16,CCS:AFLNO16,EC:FLNW17,SP:ABFLLN17,USENIX:Dalskov0K21}, focusing on reducing the total communication complexity---for instance, from two elements per party (six in total)~\cite{FCW:KerLauRan16} to one element per party (three in total)~\cite{CCS:AFLNO16} in the 3PC setting. 
However, in heterogeneous network environments like those described, designing an MPC protocol with varying round and communication complexities for each party could yield even greater performance improvements. This way, the protocol is less affected by the weaker network links in a given setting.

%-------------------------------
\subsubsection*{\textbf{Computational Complexity Matters}}
\begin{figure}[htb!]
    \centering
    \includegraphics[width=0.47\textwidth]{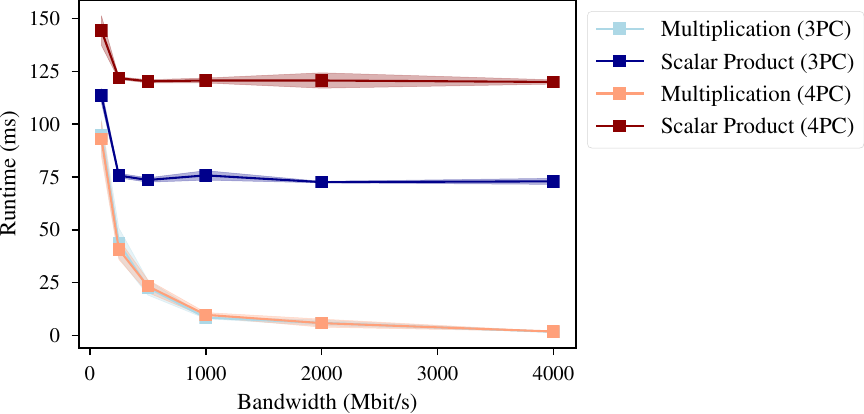}
    \vspace{-3mm}
    \captionsetup{font=small}
    \caption{Comparing Runtimes for Multiplications and Scalar Product with Varying Bandwidths}
    \label{fig:dot_band}
    \vspace{-4mm}
\end{figure}

%-------------------------------
As MPC deployments span a wider range of application domains, such as privacy-preserving machine learning~\cite{SP:NC23}, some MPC operations are becoming \-compu\-tation-bound rather than communication-bound. To demonstrate this, we implemented two of the most popular 3PC~\cite{CCS:AFLNO16} and 4PC protocols~\cite{USENIX:Dalskov0K21}, comparing the runtime of regular multiplications with scalar products of the same output size across varying bandwidths. Figure \ref{fig:dot_band} shows that while multiplications are often bottlenecked by communication in most settings, the runtime of scalar multiplications shows no significant benefit from bandwidths above 250 Mbit/s. Thus, these applications are not bounded by available bandwidth but rather are bottlenecked by local computations.      

%-------------------------------
\subsubsection*{\textbf{Limitations of Existing Implementations}}
%-------------------------------
Araki et al. \cite{CCS:AFLNO16} demonstrated that implementing a semi-honest 3PC protocol in a homogeneous network setting can achieve a throughput of seven billion AND gates per second. However, the implementation of~\cite{CCS:AFLNO16} has not been published, and open-source implementations do not come close to that level of throughput. For instance, on our test setup, the popular open-source library MP-SPDZ \cite{CCS:Keller20} achieves a throughput of less than ten million independent AND gates per second using the same 3PC protocol. To investigate whether this limitation also applies to other implementations, we measure the throughput of $\Z{32}$ and AND gates per second for several state-of-the-art frameworks. \tabref{otherFrameworks} shows the results of our comparison. 

We measured the throughput of replicated secret sharing protocols in the honest-majority setting for MP-SPDZ~\cite{CCS:Keller20} and ABY3~\cite{CCS:MohRin18}. MOTION~\cite{TOPS:MOTION} and MPyC~\cite{TPMPC:Sch18} provide only $n$-party computation protocols secure against a dishonest majority, which naturally achieve lower throughput. Where available, we used each  framework's concurrency features. For instance, we utilized MP-SPDZ's \texttt{@multithread} instruction and vectorized array multiplication operators. Although ABY3 does not natively support multithreading, we implemented multithreading on top of their framework for our measurements. \tabref{otherFrameworks} presents both ``external" measurements obtained manually by timing the start and end of an executable, and ``internal" measurements provided by most frameworks out of the box. We found that all tested existing frameworks achieve only several thousand to a few million gates per second on a 25 Gbit/s network, significantly below the theoretical capabilities expected given the available network bandwidth. Notably, the frameworks do not achieve $32$ times higher throughput for AND gates compared to $\Z{32}$ multiplication gates, despite the latter require $32$ times more data to be sent. This indicates that existing frameworks struggle to accelerate Boolean computations. 

Given the analysis above, we make a case for designing efficient MPC protocols that address identified bottlenecks related to network heterogeneity and computational complexity. To demonstrate our protocols' theoretical contributions but moreover to ensure that the progress in MPC protocol design over the last years is reflected in practice, we also make a case for novel implementations that find ways to accelerate the basic building blocks required by nearly every MPC protocol.      
%============================
\section{Preliminaries}
\label{sec:preliminaries}
%============================
In this section, we provide the preliminaries, including notations and sharing semantics. Our protocols are designed to operate over $\ell$-bit rings of the form $\Z{\ell}$ and also support the Boolean ring, denoted by $\Z{}$. Additionally, the protocols use function-dependent preprocessing~\cite{ASIACRYPT:GordonR018,ACNS:BenNieOmr19,JC:KPPS23} for efficiency, similar to works like ASTRA~\cite{CCSW:CCPS19} and Tetrad~\cite{NDSS:KotiPRS22}. 

%--------------------------------
\subsection{Notations}
\label{sec:notations}
%--------------------------------
We use $\Partyset$ to denote the set of parties and $\Party{i}$ to denote the $i$th party. While $\Partyset = \{\Party{0}, \Party{1}, \Party{2}\}$ denotes the parties in our 3PC setting, $\Partyset = \{\Party{0}, \Party{1}, \Party{2}, \Party{3}\}$ denotes the parties in our 4PC protocols. Moreover, $\Partysubset{\Phi}$ denotes a subset of $\Partyset$ comprising of parties in the set $\Phi$. For instance, $\Partysubset{\Phi}$ or simply $\Partysubset{i,j}$ indicates the set of parties $\Phi = \{\Party{i}, \Party{j}\}$. Similarly, we use $x_{\Phi}$ or simply $x_{i,j}$ to denote the value $x$ possessed by all parties in $\Phi = \{\Party{i}, \Party{j}\}$. 
We use $\csec$ to denote the computational security parameter which is set to $128$ in our protocols.

%--------------------------------
\paragraph{Sharing Semantics}
\label{sec:semantics}
%--------------------------------
We use the \emph{masked evaluation} technique~\cite{C:LPSY15,CCS:WanRanKat17a,ACNS:BenNieOmr19} in our protocols, where each secret $x \in \Z{\ell}$ is associated with a mask $\lv{x}{}$ and the masked value $\mv{x}{}$ such that $\mv{x}{} = x + \lv{x}{}$. The mask $\lv{x}{}$ is \emph{input-independent} and thus all the operations involving only $\lv{x}{}$ can be carried out in the preprocessing phase, while $\mv{x}{}$ is the \emph{input-dependent} share. 
Additionally, we use $\mverify{x}{} = x + \lv{x}{} + \lv{x}{*}$ to denote a doubly masked value which represents the secret $x$ masked using two independent masks $\lv{x}{}$ and $\lv{x}{*}$ (cf.~\secref{4PC} for details). 

%--------------------------------
\paragraph{Secret-Sharing Schemes}
\label{sec:sharing-schemes}
%--------------------------------
We use two different sharing schemes throughout our protocols, and their overview is provided below. 
\begin{enumerate}
    %------------
    \item $\sqr{\cdot}$-sharing: A value $x \in \Z{\ell}$ is $\sqr{\cdot}$-shared among $\Partysubset{\Phi}$, if each $\Party{i} \in \Partysubset{\Phi}$ holds $x^i$ such that $\sum_{i} x^i = x$.
    %------------
    \item $\shr{\cdot}$-sharing: A value $x \in \Z{\ell}$ is $\shr{\cdot}$-shared among $\Partysubset{\Phi}$, if parties in $\Partysubset{\Phi}$ hold $\mv{x}{}$ and $\sqr{\lv{x}{}}$ such that $\mv{x}{} = x + \lv{x}{}$.
    %------------
\end{enumerate}
The exact definitions of the above schemes differ slightly between the 3PC and 4PC protocols, as discussed in their respective sections. 

Additionally, $\shr{\cdot}^B$ represents Boolean sharing, where addition and multiplication are replaced by XOR and AND gates, respectively. Similarly, $\shr{\cdot}^A$ denotes arithmetic sharing. The superscript is omitted when the type of sharing is clear from the context.

%--------------------------------
\paragraph{Messages and Verification}
%--------------------------------
A value computed by $\Partysubset{\Phi}$ is denoted by $\cval{\Phi}{}$ and for multiple values, the $j$th value is denoted by $\cval{\Phi}{j}$. Similarly, a message communicated by $\Partysubset{\Phi}$ is denoted by $\msg{\Phi}{}$ and for multiple messages, the $j$th message is denoted by $\msg{\Phi}{j}$.
The notation $\Partyverify{i}$ indicates that any trailing computations or communications performed by $\Party{i}$ are used only for verifying messages from other parties and do not add any delays in the main protocol execution. In other words, this notation indicates that these specific computations and communications are not latency-critical and do not contribute to the round complexity of the protocol.

%--------------------------------
\subsection{Generating Shared Random Values}
\label{sec:generate-random}
%--------------------------------
To improve communication efficiency, our protocols utilize the $\FuncSRNG$ functionality~(\figref{funcsrng}), which allows a subset of parties~$\Partysubset{\Phi}$ to generate fresh random values without interaction. Protocol $\piSRNG$~(\figref{srng}) shows the instantiation of $\FuncSRNG$ in our protocols. We refer to this as sampling using a shared random value generator~(SRNG), where random values are generated with the help of a pseudorandom function~(PRF). Additionally, the protocol assumes a shared-key setup ($\Funcsetup$) is established at the beginning, which is the case with most existing protocols~\cite{CCS:AFLNO16,CCSW:CCPS19,NDSS:ChaRacSur20,USENIX:Dalskov0K21}.

\begin{protocolbox}{$\piSRNG(\Partysubset{\Phi},\key{\Phi}, \counter{\Phi}, \ell') \rightarrow \randval{\Phi} \in \Z{\ell'}$}{Generating shared random values}{fig:srng}
	%-----
	\justify
	\algoHeadacmart{One-time Setup:} 
    Parties in $\Partysubset{\Phi}$ invoke $\Funcsetup$ to obtain a unique shared key $\key{\Phi} \in \Z{\csec}$. Parties initialize a counter $\counter{\Phi} \in \mathbb{N}$ to $0$ and a buffer stack $\buffer{\Phi} \leftarrow \emptyset$ with $\buffertop$ denoting the index of top element.
	\justify
	\algoHeadacmart{Procedure:} 
	Let $PRF: \Z{\csec} \times \mathbb{N} \rightarrow \Z{\csec}$ be a pseudorandom function. To generate a new random value in $\Z{\ell'}$, each $\Party{} \in \Partysubset{\Phi}$ does the following:
	\begin{enumerate}[itemsep=0mm]
		%-----
        \item If $|\buffer{\Phi}| < \ell'$, compute $r_{\csec} = PRF(\key{\Phi}, \counter{\Phi})$ to obtain $\csec$ random bits and add them to the buffer $\buffer{\Phi}$. Set $\counter{\Phi} \leftarrow \counter{\Phi} + 1$.
		%-----	
		\item Pop $\ell'$ bits $b_{\buffertop-\ell',\dots,\buffertop}$ from $\buffer{\Phi}$ and compute $\randval{\Phi} = \sum_{i=0}^{\ell'} b_i \cdot 2^i$.
		%-----
	\end{enumerate}      
\end{protocolbox}

%--------------------------------
\subsection{Compare-View Functionality}
\label{sec:compare-view}
%--------------------------------
To achieve malicious security in our 4PC protocols, each party needs to verify the correctness of the messages it receives. To enable this, the parties have access to a Compare-View functionality, similar to the joint-message passing in SWIFT~\cite{USENIX:KPPS21} and the \textsf{jsnd} primitive in Tetrad~\cite{NDSS:KotiPRS22}. Protocol $\piCompareView$ in~\figref{compareview} instantiates this functionality.

Let $\{v_1, \dots, v_n\}$ denote a set of values obtained by the parties in $\Partysubset{\Phi}$ through the messages received during the protocol execution and/or local computation. Protocol $\piCompareView$ ensures that all parties in $\Partysubset{\Phi}$ will either have the same set of values or will return $\abort$ if there is an inconsistency. To achieve this, they compare a single hash of their concatenated views of $\{v_1, \dots, v_n\}$.

\begin{protocolbox}{$\piCompareView(\Partysubset{\Phi}, \{v_1, \dots, v_n\}) \rightarrow \bool$}{Verifying the correctness of received values}{fig:compareview}
	%-----
	\justify
	\algoHeadacmart{Assumptions:} 
    Each party in $\Partysubset{\Phi}$ obtains a set of fixed-length values $\{v_1, \dots, v_n\}$ during the protocol execution. Parties in $\Partysubset{\Phi}$ have access to a collision-resistant hash function $\Hash$ and $||$ denotes concatenation.
	\justify
	\algoHeadacmart{Procedure:} 
	\begin{enumerate}[itemsep=0mm]
		%-----
		\item Parties in $\Partysubset{\Phi}$ compute and exchange $h_V = \Hash(v_1||v_{2}||\dots||v_n)$.
        %-----	
        \item If the received hashes are consistent with the computed hash, they return $\true$ and $\continue$ with the protocol. Otherwise, they return $\false$ and $\abort$ the protocol.
		%-----
	\end{enumerate} 
\end{protocolbox}

%============================
\section{$\threepcname$: 3PC Protocol}
\label{sec:3PC}
%============================
This section details our 3PC protocol over rings, namely $\threepcname$, which is secure against up to one semi-honest corruption among the computing parties $\Partyset = \{\Party{0}, \Party{1}, \Party{2}\}$. Similar to ASTRA~\cite{CCSW:CCPS19}, our protocol requires the communication of one ring element in the preprocessing phase and two elements in the online phase per multiplication. If preferred, the preprocessing phase can also be executed during the online phase. 

%---------------------------------------
\subsection{Sharing Semantics}
\label{sec:3pc-sharing-semantics}
%---------------------------------------
\makeparafit
We detail the sharing semantics of our protocol based on masked evaluation and comparing it to ASTRA in~\tabref{3pc-secret-sharing}. 
Similar to ASTRA, in our scheme for a secret $x$, each online party $\Party{1}$ and $\Party{2}$ will possess one share of the mask $\lv{x}{}$. However, the parties' input-dependent shares are masked differently: In ASTRA, both parties hold the same input-dependent share $\mv{x}{}$ corresponding to~$x$ being masked by $\lv{x}{}$. In our scheme, $\Party{1}$ holds $\mv{x}{2}$ and $\Party{2}$ holds $\mv{x}{1}$, which correspond to~$x$ being masked with one of the shares of~$\lv{x}{}$.
Although the amount of information possessed by the parties in our protocol is the same as in ASTRA, this sharing mainly improves online computation, as discussed later in this section. Additionally, we introduce the notion of \emph{persistent shares}, which represent the actual values each party should store in memory during the protocol execution.

\vspace{-1mm}
\begin{table}[htb!]
\centering
\small
\captionsetup{font=small}
\caption{$\threepcname$: 3PC Secret Sharing.}
\label{tab:3pc-secret-sharing}
\vskip -0.1in
    \begin{tabular}{rccc}
        \toprule
        & Party & ASTRA~\cite{CCSW:CCPS19} & $\threepcname$ (3PC)\\ 
        \midrule
        \multirow{3}{*}{\makecell[r]{Sharing\\Semantics\\$\sh{x}$}}
        & $\Party{0}$ & $\lv{x}{1}, \lv{x}{2}$ & $\lv{x}{1}, \lv{x}{2}$ \\
        & $\Party{1}$ & $\mv{x}{}, \lv{x}{1}$ & $\mv{x}{2}, \lv{x}{1}$ \\
        & $\Party{2}$ & $\mv{x}{}, \lv{x}{2}$ & $\mv{x}{1}, \lv{x}{2}$ \\
        \midrule
        \multirow{3}{*}{\makecell[r]{Persistent\\Shares}}
        & $\Party{0}$ & $\lv{x}{}$ & $\lv{x}{1}, \lv{x}{2}$ \\
        & $\Party{1}$ & $\mv{x}{}, \lv{x}{1}$ & $\mv{x}{2}, \lv{x}{1}$ \\
        & $\Party{2}$ & $\mv{x}{}, \lv{x}{2}$ & $\mv{x}{1}, \lv{x}{2}$ \\
        \midrule
        \multicolumn{2}{r}{\multirow{2}{*}{Correlation}} & $\lv{x}{} = \lv{x}{1} + \lv{x}{2}$ & $\mv{x}{1} = x + \lv{x}{1}$ \\
        & & $\mv{x}{} = x + \lv{x}{}$ & $\mv{x}{2} = x + \lv{x}{2}$ \\
        \bottomrule
    \end{tabular}
    \vspace{-2mm}
\end{table}

Note that our sharing scheme is \emph{linear}. Thus, given public constants $\alpha, \beta, \gamma$ and secret-shares $\sh{x}, \sh{y}$, parties can locally compute the shares of  $\sh{\alpha x + \beta y + \gamma}$.

%---------------------------------------
\subsection{Input Sharing and Output Reconstruction}
\label{sec:3pc-input-output}
%---------------------------------------
To allow party $\Party{I}$ to secret share a value $x \in \Z{\ell}$, we modify the shared key setup functionality~($\Funcsetup$) so that $\Party{I}$ receives PRF keys for both masks $\lv{x}{1}, \lv{x}{2}$. This enables $\Party{I}$ to compute all the shares and send them to the respective parties.

To reconstruct a secret shared value $\sh{x}$ towards $\Party{O}$, $\Party{2}$ sends $\mv{x}{1}$ and $\Party{0}$ sends $\lv{x}{1}$ to $\Party{O}$. Similarly, for a public reconstruction among $\Partyset$, $\Party{0}$ sends $\lv{x}{1}$ to $\Party{2}$, while $\Party{2}$ sends $\lv{x}{2}$ to $\Party{1}$ and $\mv{x}{1}$ to $\Party{0}$. 
Each party then holds two of the three shares and can compute $x = \mv{x}{1} - \lv{x}{1}  = \mv{x}{2} - \lv{x}{2}$.

%---------------------------------------
\subsection{Multiplication}
\label{sec:3pc-multiplication}
%---------------------------------------
To multiply two secret-shared values $\sh{a}$ and $\sh{b}$, the 3PC protocol in ASTRA involves the online parties locally computing an additive sharing of the masked output $\mv{c}{}$. Then, they exchange these shares to reconstruct the final result. However, we approach our protocol from a different perspective and aim to eliminate the errors in each party's local computation by using messages from other parties.

\makeparafit
The sharing semantics of our 3PC protocols are designed so that $\Party{1}$ and $\Party{2}$ can communicate to obtain a masked version of the output $c = ab$ with an \emph{input-independent} error. In the preprocessing phase, $\Party{0}$ prepares a message for $\Party{2}$ to correct this input-independent error. This ensures that all parties obtain valid and masked shares of $c$ according to our sharing semantics. 
Protocol $\piMult$~(\figref{piMult3PC}) outlines the formal steps in our multiplication protocol, detailed below.

\begin{protocolbox}{$\piMult(\sh{a}, \sh{b}) \rightarrow \sh{c}$}{$\threepcname$: 3PC multiplication protocol}{fig:piMult3PC}
	%----
	\justify 
	\algoHeadacmart{Preprocessing:}
	\begin{enumerate}[itemsep=0mm]
		%------
		\item Sample random values using $\piSRNG$:
		\begin{align*}
            &\Partysubset{0,1}: \lv{c}{1}, r_{0,1} &
            &\Partysubset{0,2}: \lv{c}{2}
		\end{align*}
		\item Locally compute:
		\begin{align*}
            &\Party{0}: \msgset{0}{} = \lv{a}{2} \lv{b}{2} - (\lv{a}{1} - \lv{a}{2}) (\lv{b}{1} - \lv{b}{2}) + r_{0,1} 
		\end{align*}
        \item Communicate:
		\begin{align*}
            &\Party{0} \rightarrow \Party{2}: \msgset{0}{} 
		\end{align*}
		%------
		%------
	\end{enumerate}
	\justify
    \algoHeadacmart{Online:}
	\begin{enumerate}[itemsep=0mm]
		%-------
		\item Locally compute:
		\begin{align*}
            &\Party{1}: \valset{1}{} = \mv{a}{2} \lv{b}{1} + \mv{b}{2} \lv{a}{1} + r_{0,1} &
            \Party{2}: \valset{2}{} = \mv{a}{1} \mv{b}{1} + \msgset{0}{}
		\end{align*}
		%------
        \item Communicate:
		\begin{align*}
            &\Party{1} \rightarrow \Party{2}: \msgset{1}{} = \valset{1}{} - \lv{c}{1}  &
            \Party{2} \rightarrow \Party{1}: \msgset{2}{} = \valset{2}{} + \lv{c}{2} 
		\end{align*}
		%-------
		\item Locally compute:
		\begin{align*}
            &\Party{1}: \mv{c}{2} = \msgset{2}{} - \valset{1}{} &
            \Party{2}: \mv{c}{1} = \valset{2}{} - \msgset{1}{}
		\end{align*}
		%------
	\end{enumerate}     
\end{protocolbox}
%------------------

% \medskip
Given $\sh{a}, \sh{b}$, note that $\Party{2}$ can compute $\mv{a}{1}\cdot\mv{b}{1}$, which is equivalent to computing
\begin{equation}
    \mv{a}{1}\cdot\mv{b}{1} = (a + \lv{a}{1})(b + \lv{b}{1})= ab + a \lv{b}{1} + b \lv{a}{1} + \lv{a}{1} \lv{b}{1},
\end{equation}
thus obtaining the output $c = ab$ with an input-dependent error $a \lv{b}{1} + b \lv{a}{1}$ and an input-independent error $\lv{a}{1} \lv{b}{1}$. Our goal is to correct these errors using messages from $\Party{0}$ and $\Party{1}$ while obliviously inserting the mask $\lv{c}{1}$, such that $\Party{2}$ obtains $\mv{c}{1} = ab + \lv{c}{1}$. In a similar fashion, $\Party{1}$ should obtain its input-dependent share $\mv{c}{2}$.

%--------------------------
\paragraph{Preprocessing Phase}
%--------------------------
Using their pre-shared keys, the parties locally sample masks for the output ($\lv{c}{1}, \lv{c}{2}$) and a random pad $r_{0,1}$ to mask the intermediate message. $\Party{0}$ sends the message $\msgset{0}{} = \lv{a}{1} \lv{b}{2} + \lv{a}{2} \lv{b}{1} - \lv{a}{1} \lv{b}{1} + r_{0,1}$ to $\Party{2}$. This message serves the purpose of correcting the input-independent error when $\Party{1}$ and $\Party{2}$ communicate during the online phase. Note that the mask $r_{0,1}$ ensures that $\Party{2}$ cannot infer any values from $\Party{0}$'s message.

%--------------------------
\paragraph{Online Phase}
%--------------------------
$\Party{2}$ uses $\msgset{1}{} = \mv{a}{2} \lv{b}{1} + \mv{b}{2} \lv{a}{1} + r_{0,1} - \lv{c}{1}$ from $\Party{1}$ to eliminate the input-dependent error $a \lv{b}{1} + b \lv{a}{1}$ from its computation. However, this results in a larger input-independent error compared to $\lv{a}{1} \lv{b}{1}$. Note that $\Party{0}$ holds all input-independent shares and can therefore help $\Partysubset{1,2}$ correcting any input-independent error. Hence, this error is removed using message $\msgset{0}{}$ obtained from $\Party{0}$ during the preprocessing phase. $\Party{2}$ obtains:
\begin{align*}
     \mv{c}{1} &= \valset{2}{} - \msgset{1}{} \\
     &= (ab + a \lv{b}{1} + b \lv{a}{1} + \lv{a}{1} \lv{b}{1} + \msgset{0}{}) - (\valset{1}{} - \lv{c}{1})  \\
     &= (ab + \lv{c}{1}) + \msgset{0}{} - (\lv{a}{1} \lv{b}{2} + \lv{a}{2} \lv{b}{1} - \lv{a}{1} \lv{b}{1} + r_{0,1}) = ab + \lv{c}{1}
\end{align*}

The steps for $\Party{1}$ are similar, except that it locally computes $\mv{a}{2} \lv{b}{1} + \mv{b}{2} \lv{a}{1} + r_{0,1} = a \lv{b}{1} + b \lv{a}{1} + \lv{a}{1} \lv{b}{2} + \lv{a}{2} \lv{b}{1} + r_{0,1}$ and uses the message from $\Party{2}$ to obtain the correct share $\mv{c}{2} = ab + \lv{c}{2}$. The correctness of our protocol is detailed in \secref{A3C}.

%---------------------------------------
\subsection{Discussions}
\label{sec:3pc-discussions}
%---------------------------------------
Our 3PC multiplication protocol $\piMult$~(cf.~\figref{piMult3PC}) in $\threepcname$ has the same computational complexity as that of ASTRA~\cite{CCSW:CCPS19}. However, as shown in~\tabref{comp_full} in~\secref{others}, we first eliminate one multiplication in ASTRA with a straightforward optimization. Our approach lets us then shift some local computation of ASTRA from online to preprocessing, leading to computational gains in the online phase. 

The security of our semi-honest 3PC, $\threepcname$, can be proven using a real-world/ideal-world simulation paradigm~\cite{Goldreich2004,EPRINT:Lindell16}, using a simplified variant of the simulation strategy discussed in~\secref{simulation-strategy}. Due to its similarity to ASTRA, we omit a formal security proof for it. 

Regarding bandwidth and latency requirements for the three network links among $\Partysubset{0,1,2}$, our 3PC protocol tolerates high latency between $\Partysubset{0,2}$ and $\Partysubset{0,1}$ as they do not communicate in the online phase. $\Partysubset{1,2}$ on the other hand need to synchronously evaluate the circuit while waiting for each other's messages after each communication round. Our protocol also tolerates low bandwidth between $\Partysubset{0,1}$ as they do not communicate in both phases. (cf.~\figref{3PC_OFF} and \figref{3PC_ON} for details).

%---------------------------------------
\subsubsection*{\textbf{Malicious security}}
\label{sec:3pc-malicious}
%---------------------------------------
Instead of improving the security of $\threepcname$ to tolerate malicious corruption, we chose a 4PC setting for malicious security due to the following reasons. Maliciously secure 3PC protocols over rings mainly uses one of two approaches: Distributed Zero-Knowledge Proofs (DZKP)~\cite{C:BBCGI19} and triple sacrificing~\cite{SP:BSSSY24}. The DZKP approach achieves sublinear communication complexity at the expense of increased computational complexity, whereas the triple sacrificing approach maintains low computational complexity but necessitates a higher communication overhead. For instance, 3PC maliciously secure  protocols in Replicated~\cite{EC:FLNW17,SP:ABFLLN17}, ASTRA~\cite{CCSW:CCPS19} and SWIFT~\cite{USENIX:KPPS21,NDSS:PatSur20} require total communications of $21$, $25$ and $12$ ring elements per multiplication, respectively, when utilizing the triple sacrificing technique~\cite{SP:BSSSY24}. Using DZKP~\cite{C:BBCGI19} reduces the communication requirement to $6$ ring elements, but with significantly higher local computational demands. 

\makeparafit
On the other hand, maliciously secure 4PC schemes such as Fantastic Four~\cite{USENIX:Dalskov0K21} and Tetrad~\cite{NDSS:KotiPRS22} require only $6$ and $5$ communication elements per multiplication, respectively, and eliminate the need for these expensive primitives. By exploiting the additional redundancy of parties' shares, these protocols utilize a joint-send primitive that is similar to our Compare-View functionality introduced in \secref{preliminaries}. Therefore, we use our 3PC protocol $\threepcname$ as the foundation and have designed a maliciously secure 4PC protocol, $\fourpcname$, based on it. The details about $\fourpcname$ are presented next.

%============================
\section{$\fourpcname$: 4PC Protocol}
\label{sec:4PC}
%============================
In this section, we detail our 4PC protocol over rings, namely $\fourpcname$, which is secure against up to one malicious corruption among the computing parties $\Partyset = \{\Party{0}, \Party{1}, \Party{2}, \Party{3}\}$. Our protocol requires communicating two ring elements in the preprocessing phase and three elements in the online phase per multiplication, similar to Tetrad~\cite{NDSS:KotiPRS22}. 
The protocol builds upon the 3PC protocol~$\threepcname$ described in~\secref{3PC}, incorporating the necessary redundancy to verify all messages sent between the parties. However, the core difference between our protocol and Tetrad lies in our use of two independent masks to hide the secret, which are distributed carefully across the parties, as discussed in~\secref{4pc-sharing-semantics}. 
Regarding latency requirements of the six network links among $\Partyset$, only the link between $\Partysubset{1,2}$ contains latency-critical communication. During the online phase, $\Partysubset{1,2}$ need to wait for each other's messages before they can synchronously proceed with the next gate in the circuit. All other links are either only utilized in the preprocessing phase, or only utilized for verification. If a party requires messages only for verification, it can delay processing of these messages to the end of the protocol without affecting the progress of the other parties as they evaluate the circuit. 
Regarding bandwidth requirements, our protocol tolerates three low bandwidth links between the parties as these are not utilized in both phases. 
Additionally, we present a variant of our protocol optimized for heterogeneous network settings. This variant increases the number of unutilized links to four and therefore performs well in settings even when the majority of parties share weak network links. The details are provided in~\secref{4pc-multiplication-hetero}.
%---------------------------------------
\subsection{Sharing Semantics}
\label{sec:4pc-sharing-semantics}
%---------------------------------------
\tabref{4pc-secret-sharing} details the sharing semantics of our 4PC protocol and compares it with Tetrad. 
The sharing semantics of our 4PC protocol extend those of the 3PC protocol by introducing redundancy to verify messages exchanged among the parties. 
At a high level, $\Party{0}$ holds an input-dependent share to help verify the communication between $\Partysubset{1,2}$ using messages from $\Party{3}$. Additionally, $\Party{3}$ helps $\Partysubset{1,2}$ verify messages sent by $\Party{0}$.

\begin{table}[htb!]
\centering
\small
\captionsetup{font=small}
\caption{$\fourpcname$: 4PC Secret Sharing.}
\label{tab:4pc-secret-sharing}
\vskip -0.1in
    \begin{tabular}{rccc}
        \toprule
        & Party & Tetrad~\cite{NDSS:KotiPRS22} & $\fourpcname$\\ 
        \midrule
        \multirow{4}{*}{\makecell[r]{Sharing\\Semantics\\$\sh{x}$}}
        & $\Party{0}$ & $\mv{x}{}, \lv{x}{1}, \lv{x}{2}$ & $\mw{x}{}, \lv{x}{1}, \lv{x}{2}$ \\
        & $\Party{1}$ & $\mv{x}{}, \lv{x}{0}, \lv{x}{1}$ & $\mv{x}{}, \lw{x}{}, \lv{x}{1}$ \\
        & $\Party{2}$ & $\mv{x}{}, \lv{x}{0}, \lv{x}{2}$ & $\mv{x}{}, \lw{x}{}, \lv{x}{2}$ \\
        & $\Party{3}$ & $\lv{x}{0}, \lv{x}{1}, \lv{x}{2}$ & $\lw{x}{}, \lv{x}{1}, \lv{x}{2}$ \\
        \midrule
        \multirow{4}{*}{\makecell[r]{Persistent\\Shares}}
        & $\Party{0}$ & $\mv{x}{}, \lv{x}{1}, \lv{x}{2}$ & $\mw{x}{}, \lv{x}{}$ \\
        & $\Party{1}$ & $\mv{x}{}, \lv{x}{0}, \lv{x}{1}$ & $\mv{x}{}, \lv{x}{1}$ \\
        & $\Party{2}$ & $\mv{x}{}, \lv{x}{0}, \lv{x}{2}$ & $\mv{x}{}, \lv{x}{2}$ \\
        & $\Party{3}$ & $\lv{x}{0}, \lv{x}{1}, \lv{x}{2}$ & $\lw{x}{}, \lv{x}{}$ \\
        \midrule
        \multicolumn{2}{r}{\multirow{3}{*}{Correlation}} 
        & $\lv{x}{} = \lv{x}{0} + \lv{x}{1} + \lv{x}{2}$ & $\lv{x}{} = \lv{x}{1} + \lv{x}{2}$ \\
        & & $\mv{x}{} = x + \lv{x}{}$ & $\mv{x}{} = x + \lv{x}{}$ \\
        & &  & $\mw{x}{} = x + \lw{x}{}$ \\
        \bottomrule
    \end{tabular}
    \begin{itemize}[leftmargin=5mm]
        \renewcommand\labelitemi{}
        \item $*$ indicates share not correlated to $\sh{\lv{x}{}}$ and used only for verification. 
    \end{itemize}
    \vspace{-1mm}
\end{table}

Tetrad uses a single mask $\lv{x}{}$ to hide the secret $x$ and distributes the additive shares of $\lv{x}{}$ among the parties to create redundancy. Specifically, all the input-independent shares $\lv{x}{0}, \lv{x}{1}, \lv{x}{2}$ in Tetrad are correlated to mask the identical input-dependent share $\mv{x}{}$ held by $\Partysubset{0,1,2}$.
In contrast, our protocol uses two independent masks, $\lv{x}{}$ and $\lw{x}{}$, and correspondingly two masked values, $\mv{x}{}$ and $\mw{x}{}$. This enables each party to obtain different input-dependent shares, $\mv{x}{}$ and $\mw{x}{}$, similar to our 3PC scheme. This approach reduces the correlation between input-independent shares, resulting in each party needing to store fewer shares in memory~(cf. persistent shares in~\tabref{4pc-secret-sharing}). Additionally, it allows us to reduce the number of basic instructions per multiplication gate by more than half compared to Tetrad, as shown in~\tabref{comp_score}.
Note that no single party's share reveals anything about $x$, but holding two distinct shares suffices to obtain $x$.

%---------------------------------------
\subsection{Input Sharing and Output Reconstruction}
\label{sec:4pc-input-output}
%---------------------------------------
To securely share a value $x \in \Z{\ell}$ in the presence of a malicious adversary, the input party $\Party{I}$ obtains the PRF keys corresponding to the masks $\lv{x}{1}$, $\lv{x}{2}$ and $\lw{x}{}$, similar to the 3PC scheme. $\Party{I}$ then computes and sends $\mverify{x}{} = x + \lv{x}{1} + \lv{x}{2} + \lw{x}{}$ to $\Partysubset{0,1,2}$. The parties compare their views of $\mverify{x}{}$ using the compare-view functionality (cf.~$\piCompareView$ in~\figref{compareview}) and locally convert it to their respective input-dependent share by subtracting the corresponding masks they generated together with $\Party{I}$.

To reconstruct a secret $\sh{x}$ towards output party $\Party{O}$, $\Party{0}$ sends $\lv{x}{}$ and $\Party{2}$ sends $\mv{x}{}$ to $\Party{O}$. $\Party{O}$ then uses $\piCompareView$ to compare their views of $\lv{x}{}$ and $\mv{x}{}$  with $\Party{3}$ and $\Party{1}$, respectively.  Since one party in each pair of $\{\Party{1}, \Party{2}\}$ and $\{\Party{0}, \Party{3}\}$ is honest, $\Party{O}$ will either obtain the correct $x$ or $\abort$ the protocol. 

The formal protocols for input sharing~($\piShare$) and output reconstruction with abort~($\piRec$) are provided in~\secref{additional}, along with a \emph{fair}~\cite{JC:CohLin17} reconstruction protocol~($\piFairRec$), which guarantees that if the adversary receives an output, the honest parties do as well.

%---------------------------------------
\subsection{Multiplication}
\label{sec:4pc-multiplication}
%---------------------------------------
\makeparafit
Protocol $\piMult$~(\figref{piMult4PC}) outlines the formal steps in our multiplication protocol, detailed below. To multiply two secret-shared values, $\sh{a}$ and $\sh{b}$, the parties proceed similarly to 3PC multiplication~(cf.~\secref{3pc-multiplication}). The goal is to eliminate errors in each party's local computation by using messages from other parties. 
These messages are carefully designed to ensure that if $\Party{j}$ computes a value $v$ using a message sent by $\Party{i}$, another party $\Party{k}$ also obtains $v$, either through local computation or with the help of a set of verified messages.
This allows the use of the compare-view functionality~(cf.~$\piCompareView$ in~\figref{compareview}) to achieve malicious security with abort.

%--------------------------
\paragraph{Preprocessing Phase}
%--------------------------
During this phase, parties non-interactively compute all the input-independent shares of $c$, specifically $\lv{c}{1}$, $\lv{c}{2}$, and $\lw{c}{}$, using $\piSRNG$~(cf.~\figref{srng}). As in our 3PC, $\Party{0}$ then sends a message $\msgset{0}{}$ to $\Party{2}$, which aims to eliminate the input-independent error during the communication between $\Party{1}$ and $\Party{2}$ in the online phase. This message also ensures the correct mask $\lv{c}{}$ is inserted into their input-dependent shares. Additionally, $\Party{0}$ needs to obtain an input-dependent share $\mw{c}{} = ab + \lw{c}{}$, which it uses to verify $\Partysubset{1,2}$'s communication in the online phase. For this purpose, $\Party{3}$ computes and sends a message $\msgset{3}{}$ to eliminate the input-independent error of $\Party{0}$'s computation during the online phase. This way, $\Party{0}$ receives an input-dependent share of $c$ without interacting with $\Partysubset{1,2}$, which is utilized for verification.

\smallskip
\begin{protocolbox}{$\piMult(\sh{a}, \sh{b}) \rightarrow \sh{c}$}{$\fourpcname$: 4PC multiplication protocol}{fig:piMult4PC}
	%----
	\justify 
	\algoHeadacmart{Preprocessing:}
	\begin{enumerate}[itemsep=0mm]
		%------
		\item Sample random values using $\piSRNG$:
		\begin{align*}
            &\Partysubset{0,1,3}: r_{0,1,3}, \lv{c}{1} &
            &\Partysubset{0,2,3}: \lv{c}{2} &
            &\Partysubsetverify{1,2,3}: r_{1,2,3}, \lw{c}{}
		\end{align*}
		\item Locally compute:
		\begin{align*}
            &\Party{0},\Partyverify{3}: \lv{c}{} = \lv{c}{1} + \lv{c}{2}  \\
            &\Party{0},\Partyverify{3}: \msgset{0,3}{} = \lv{c}{} + \lv{a}{} \lv{b}{} + r_{0,1,3}  \\
            &\Partyverify{3}: \msgset{3}{} = \lv{a}{} (\lv{b}{} - \lw{b}{}) - \lv{b}{} \lw{a}{} - \lw{c}{} + r_{1,2,3}    
		\end{align*}
        \item Communicate:
		\begin{align*}
            &\Party{0} \rightarrow \Party{2}: \msgset{0,3}{} &
            &\Partyverify{3} \rightarrow \Partyverify{0}: \msgset{3}{} 
		\end{align*}
		%------
		%------
	\end{enumerate}
	\justify
    \algoHeadacmart{Online:}
	\begin{enumerate}[itemsep=0mm]
		%-------
		\item Locally compute:
		\begin{align*}
            &\Partyverify{0}: \valset{0}{} = \mw{a}{} \lv{b}{} + \mw{b}{} \lv{a}{} &
            &\Partysubset{1,2}: \valset{1,2}{} = \mv{a}{} \mv{b}{}\\
            &\Party{1}: \msgset{1}{} = \mv{a}{} \lv{b}{1} + \mv{b}{} \lv{a}{1} + r_{0,1,3} &
            &\Party{2}: \msgset{2}{} = \mv{a}{} \lv{b}{2} + \mv{b}{} \lv{a}{2} - \msgset{0,3}{} \\
            &\Partysubsetverify{1,2}: \msgset{1,2}{} = \valset{1,2}{} + r_{1,2,3} \\
            &\Party{1}: \valset{1}{} = \valset{1,2}{} - \msgset{1}{} &
            &\Party{2}: \valset{2}{} = \valset{1,2}{} - \msgset{2}{} 
		\end{align*}
		%------
        \item Communicate:
		\begin{align*}
            &\Party{1} \rightarrow \Party{2}: \msgset{1}{} &
            &\Party{2} \rightarrow \Party{1}: \msgset{2}{} &
            &\Partyverify{2} \rightarrow \Partyverify{0}: \msgset{1,2}{} 
		\end{align*}
		%-------
		\item Locally compute:
		\begin{align*}
            &\Party{1}: \mv{c}{} = \valset{1}{} - \msgset{2}{}  &
            &\Party{2}: \mv{c}{} = \valset{2}{} - \msgset{1}{}  \\
            &\Partyverify{0}: \mw{c}{} = \msgset{1,2}{} - (\valset{0}{} + \msgset{3}{}) \\
            &\Partysubsetverify{1,2}: \mverify{c}{} = \mv{c}{} + \lw{c}{} &
            &\Partyverify{0}: \mverify{c}{} = \mw{c}{} + \lv{c}{} 
        \end{align*}
        \item Compare views using $\piCompareView$:
        \begin{align*}
            &\Partysubsetverify{0,1}: \msgset{1,2}{} &
            &\Partysubsetverify{2,3}: \msgset{0,3}{} &
            &\Partysubsetverify{0,1,2}: \mverify{c}{}
		\end{align*}
		%------
	\end{enumerate}
\end{protocolbox}
%------------------

%--------------------------
\paragraph{Online Phase}
%--------------------------
During the online phase, $\Party{1}$ and $\Party{2}$ locally compute $\mv{a}{} \mv{b}{} = ab + a \lv{b}{} + b \lv{a}{} + \lv{a}{} \lv{b}{}$. They then obtain their share $\mv{c}{} = ab + \lv{c}{}$ by removing the error terms using the messages $\msgset{1}{}$ and $\msgset{2}{}$ that they exchange. Note that both these messages are masked with $r_{0,1,3}$ and $\lv{c}{}$ (contained in $\msgset{0,3}{}$), to prevent leakage of information. For correctness, note that:
\begin{align*}
    \msgset{1}{} 
    &= \mv{a}{} \lv{b}{1} + \mv{b}{} \lv{a}{1} + r_{0,1,3} \\
    &= a \lv{b}{1} + b \lv{a}{1} + \lv{a}{} \lv{b}{1} + \lv{a}{1} \lv{b}{} + r_{0,1,3} 
    \\[1mm]
    \msgset{2}{} 
    &= \mv{a}{} \lv{b}{2} + \mv{b}{} \lv{a}{2} - \msgset{0,3}{} \\
    &= a \lv{b}{2} + b \lv{a}{2} + \lv{a}{} \lv{b}{2} + \lv{a}{2} \lv{b}{} - \msgset{0,3}{}  
    \\[1mm]
    \msgset{1}{} + \msgset{2}{} 
    &= a \lv{b}{} + b \lv{a}{} + 2 \lv{a}{} \lv{b}{} + r_{0,1,3} - \msgset{0,3}{} 
    % \\[1mm]
    % \mv{a}{} \mv{b}{} &= ab + a \lv{b}{} + b \lv{a}{} + \lv{a}{} \lv{b}{} 
\end{align*}

Thus, the input-dependent error $a \lv{b}{} + b \lv{a}{}$ in $\mv{a}{} \mv{b}{}$ matches the input-dependent error in $\msgset{1}{} + \msgset{2}{}$. Moreover, the input-independent error $\lv{a}{} \lv{b}{} - r_{0,1,3}$ can be eliminated using $\msgset{0,3}{} = \lv{c}{} + \lv{a}{} \lv{b}{} + r_{0,1,3}$ from $\Party{0}$, while simultaneously inserting the mask~$\lv{c}{}$. Using this insight, $\Party{1}$ and $\Party{2}$ can obtain their share as 
\begin{align*}
    \mv{c}{} &= \mv{a}{} \mv{b}{} - (\msgset{1}{} + \msgset{2}{})\\ 
             &= ab - \lv{a}{} \lv{b}{} - r_{0,1,3} + \msgset{0,3}{}
             = ab + \lv{c}{}
\end{align*}

To verify $\Partysubset{1,2}$'s communication we just described, $\Party{0}$ also needs to obtain an input-dependent share in our 4PC protocol.
To obtain its share $\mw{c}{} = ab + \lw{c}{}$, $\Party{0}$ begins by locally computing $\valset{0}{} = \mw{a}{} \lv{b}{} + \mw{b}{} \lv{a}{}$. To eliminate the errors from $\valset{0}{}$ and derive $\mw{c}{}$, it uses the message $\msgset{1,2}{}$ received from $\Party{2}$ during the online phase and $\msgset{3}{}$ from $\Party{3}$ during preprocessing. The final share is $\mw{c}{} = \msgset{1,2}{} - (\valset{0}{} + \msgset{3}{})$. For correctness, note that:
\begin{align*}
    \msgset{1,2}{} 
    &= \mv{a}{} \mv{b}{} + r_{1,2,3} 
    = ab + a \lv{b}{} + b \lv{a}{} + \lv{a}{} \lv{b}{} + r_{1,2,3} 
    \\[1mm]
    \valset{0}{} 
    &= \mw{a}{} \lv{b}{} + \mw{b}{} \lv{a}{} = a \lv{b}{} + b \lv{a}{} + \lw{a}{} \lv{b}{} + \lv{a}{} \lw{b}{} 
    \\[1mm]
    \msgset{3}{} 
    &= \lv{a}{} (\lv{b}{} - \lw{b}{}) -  \lw{a}{}\lv{b}{} - \lw{c}{} + r_{1,2,3} 
\end{align*}

%---------------------------------------
\subsubsection*{Verifying Communication} 
%---------------------------------------
To ensure the correctness of the overall protocol, parties need to ensure that they have received correct messages. Each party does this by comparing their received messages with another party who can compute the same message in a different way, using the compare-view functionality~(cf.~$\piCompareView$ in~\figref{compareview}). For instance, to verify $\msgset{0,3}{}$ received from $\Party{0}$ during preprocessing, $\Party{2}$ compares its view of $\msgset{0,3}{}$ with $\Party{3}$, who can compute the same message locally. Similarly, $\Party{0}$ and $\Party{1}$ can jointly verify $\msgset{1,2}{}$. 

For the remaining messages $\msgset{3}{}$, $\msgset{1}{}$, and $\msgset{2}{}$, we observe that a single check of $\mverify{c}{} = ab + \lv{c}{} + \lw{c}{}$ by the parties in $\Partysubset{0,1,2}$ is sufficient. If $\Party{3}$'s message $\msg{3}{}$ is incorrect, then $\Party{1}$'s and $\Party{2}$'s correct views will differ from $\Party{0}$'s corrupted view. Similarly, if either $\Party{1}$'s message $\msgset{1}{}$ to $\Party{2}$ or $\Party{2}$'s message $\msgset{2}{}$ to $\Party{1}$ is incorrect, $\Party{0}$'s correct view will differ from their corrupted views of $\mverify{c}{}$. Since our protocol tolerates up to one corrupted party, only one of these cases can occur. Therefore, the parties have successfully verified all messages exchanged during the multiplication protocol.

Since the message $\msgset{1,2}{}$ from $\Party{2}$ to $\Party{0}$ during the online phase is used only for verification, it can be delayed for all the gates by a constant factor (say 300ms) without affecting the protocol's throughput in the amortized sense. Consequently, $\Partysubset{0,2}$ can operate over a high-latency link, even if the evaluated circuit has a high multiplicative depth. Messages used in this manner can also be considered part of a constant-round post-processing phase.
In our protocol, the parties achieve low computational complexity by reusing the calculated terms across messages, verification, and obtaining shares.
The correctness of our protocol and the verification of messages are detailed in \secref{A4C}.

%---------------------------------------
\subsubsection*{Post-processing routine} 
%---------------------------------------
Here, we detail the flow of the protocol when $\Party{0}$ has an arbitrary delay. Consider a circuit with $n$ consecutive multiplication gates. Let's assume that $\Partysubset{0,3}$ have completed the preprocessing phase. During the online phase, $\Party{1}$ and $\Party{2}$ interactively execute steps (1) to (3) of $\piMult$~(cf.~\figref{piMult4PC}) for each multiplication gate sequentially. 
However, all messages from $\Party{2}$ to $\Party{0}$ corresponding to the $n$ gates are delayed until the end of the protocol and sent in one round. Upon receiving the messages, $\Party{0}$ performs its local computations for all $n$ gates. Following this, $\Party{0}$ compares its views with other parties using the compare-view ~(cf.~$\piCompareView$ in~\figref{compareview}) functionality. Meanwhile, $\Party{2}$ and $\Party{3}$ execute~$\piCompareView$ on the message $\msgset{0,3}{}$ for all $n$ gates. 

While this provides the intuition, the parties are not required to execute the preprocessing, online, and postprocessing phases sequentially. Instead, they can simultaneously execute all phases. In this context, communication and computation marked by $\Partysubsetverify{}{}$ are non-blocking, indicating they are not latency-critical. Conversely, all other computation and communication are blocking, necessitating at least one party to wait for another party's communication and computation before proceeding with the protocol. By interleaving all three phases, the total runtime of the protocol is minimized in practice. This approach is utilized in our implementation~(cf.~\secref{bench-scaling} for details) to reduce the total runtime compared to online-only protocols like Fantastic Four \cite{USENIX:Dalskov0K21}. Since the preprocessing phase relies only on local computations, the online phase is unlikely to be blocked due to dependencies from the preprocessing phase in this interleaved processing model.

%---------------------------------------
\subsection{Multiplication in Heterogeneous Networks}
\label{sec:4pc-multiplication-hetero}
%---------------------------------------
The 4PC multiplication protocol $\piMult$ in~\figref{piMult4PC} tolerates three links with low bandwidth and five links with high latency between the parties~(cf.~\figref{4PC_OFF} and \figref{4PC_ON} in \secref{hom_het}). 
Here, we provide a variant of the multiplication protocol, $\piMultH$~(cf.~\figref{piMult4PCH}), tolerates four links with low bandwidth by shifting the communication of multiple messages to the same link. This way, this variant is resilient to network settings where the majority of the links between the parties are weak as demonstrated in \figref{bdw}.

At a high level, we shift the communication in the network link between $\Partysubset{0,3}$ to the already utilized link between $\Partysubset{0,2}$. Specifically, we replace the message $\msgset{3}{}$ from $\Party{3}$ to $\Party{0}$ in the preprocessing phase of $\piMult$ with another message $\msgset{1,2}{2}$ that $\Party{2}$ sends to $\Party{0}$ during the online phase. $\Party{0}$ then verifies the communication between $\Party{1}$ and $\Party{2}$ using $\msgset{1,2}{2}$ instead of $\msgset{3}{}$. This modification has the added advantage that $\Party{3}$ does not need to communicate with any other party during circuit evaluation, allowing it to have an arbitrarily weak network links to all other parties. A downside compared to the base protocol is that $\Party{2}$ needs to send two messages to $\Party{0}$ which halves the throughput on that network link. 

They key difference of $\piMultH$ compared to $\piMult$ is as follows: During the preprocessing phase, $\Party{3}$ computes $\valset{0,3}{} = \lv{a}{} (\lv{b}{} - \lw{b}{}) - \lv{b}{} \lw{a}{} - \lv{c}{} + r_{1,2,3}$ but does not send it to $\Party{0}$. During the online phase, $\Party{1}$ and $\Party{2}$ proceed similarly to $\piMult$ to obtain $\mv{c}{}$ but they also compute two messages, $\msgset{1,2}{1}$ and $\msgset{1,2}{2}$, which $\Party{2}$ sends to $\Party{0}$. $\Party{0}$ then utilizes $\msgset{1,2}{1}$ to compute its share $\mw{c}{}$ and uses $\msgset{1,2}{2}$ to verify $\Partysubset{1,2}$'s communication.  

%---------------------------------------
\subsubsection*{Verifying Communication} 
%---------------------------------------
To help $\Party{1}$ and $\Party{2}$ verify the correctness of their exchanged messages $\msgset{1}{}$ and $\msgset{2}{}$, $\Party{0}$ proceeds as follows: 
$\Party{0}$ computes $\valset{0,3}{} = \msgset{1,2}{2} - \valset{0}{}$ and compares its view with~$\Party{3}$, who computed it locally during preprocessing. If the views are consistent, $\Party{0}$ confirms that $\msgset{1,2}{2} = \msgset{1}{} + \msgset{2}{} + r_{1,2,3}$, implying $\msgset{1}{} + \msgset{2}{}$ is correct. Given that at most one of $\Party{1}$ or $\Party{2}$ can be corrupt, the correctness of $\msgset{1}{} + \msgset{2}{}$ implies that both $\msgset{1}{}$ and $\msgset{2}{}$ are correct. Additionally, $\Party{0}$ compares its view of $\msgset{1,2}{2}$ with $\Party{1}$ for consistency, while $\Party{2}$ ensures the correctness of $\msgset{0,3}{}$ received from $\Party{0}$ by comparing it with $\Party{3}$. Assuming that both checks succeed, one can
verify the correctness of $\valset{0,3}{} = \msgset{1,2}{2} - \valset{0}{}$ by noting the following:
\begin{align*}
    \msgset{1,2}{2} 
    &= \msgset{1}{} + \msgset{2}{} + r_{1,2,3} 
    = a \lv{b}{} + b \lv{a}{} + \lv{a}{} \lv{b}{} - \lv{c}{} + r_{1,2,3} 
    \\[1mm]
    \valset{0}{} 
    &= \mw{a}{} \lv{b}{} + \mw{b}{} \lv{a}{}
    = a \lv{b}{} + b \lv{a}{} + \lv{a}{} \lw{b}{} + \lv{b}{} \lw{a}{}
    \\[1mm]
    \valset{0,3}{} 
    &= \msgset{1,2}{2} - \valset{0}{}
    = \lv{a}{} \lv{b}{} - \lv{a}{} \lw{b}{} - \lv{b}{} \lw{a}{} - \lv{c}{} + r_{1,2,3}
\end{align*}
 
Finally, to verify correctness of $\msgset{1,2}{1}$ sent by $\Party{2}$, $\Party{0}$ performs a consistency check with $\Party{1}$. This will ensure that $\Party{0}$ obtained the correct share $\mw{c}{}$. 

The correctness of our protocol and the verification of messages are detailed in \secref{A4CH}. 
The security of $\fourpcname$ is proved using real-world/ideal-world simulation paradigm~\cite{Goldreich2004,EPRINT:Lindell16} and the details are provided in~\secref{security-4PC}.

\begin{protocolbox}{$\piMultH(\sh{a}, \sh{b}) \rightarrow \sh{c}$}{$\fourpcname$: Heterogeneous 4PC multiplication protocol}{fig:piMult4PCH}
	%----
	\justify 
	\algoHeadacmart{Preprocessing:}
	\begin{enumerate}[itemsep=0mm]
		%------
		\item Sample random values using $\piSRNG$:
		\begin{align*}
            &\Partysubset{0,1,3}: r_{0,1,3}, \lv{c}{1} &
            &\Partysubset{0,2,3}: \lv{c}{2} &
            &\Partysubsetverify{1,2,3}: r_{1,2,3}, \lw{c}{} 
		\end{align*}
		\item Locally compute:
		\begin{align*}
            &\Party{0},\Partyverify{3}: \lv{c}{} = \lv{c}{1} + \lv{c}{2}  \\
            &\Party{0},\Partyverify{3}: \msgset{0,3}{} = \lv{c}{} + \lv{a}{} \lv{b}{} + r_{0,1,3}  \\
            &\Partyverify{3}: \valset{0,3}{} = \lv{a}{} (\lv{b}{} - \lw{b}{}) - \lv{b}{} \lw{a}{} - \lv{c}{} + r_{1,2,3} 
		\end{align*}
        \item Communicate:
		\begin{align*}
            &\Party{0} \rightarrow \Party{2}: \msgset{0,3}{} 
		\end{align*}
		%------
		%------
	\end{enumerate}
	\justify
    \algoHeadacmart{Online:}
	\begin{enumerate}[itemsep=0mm]
		%-------
		\item Locally compute:
		\begin{align*}
            &\Party{1}: \msgset{1}{} = \mv{a}{} \lv{b}{1} + \mv{b}{} \lv{a}{1} + r_{0,1,3} &
            &\Party{2}: \msgset{2}{} = \mv{a}{} \lv{b}{2} + \mv{b}{} \lv{a}{2} - \msgset{0,3}{} \\
            &\Party{1}: \valset{1}{} = \mv{a}{} \mv{b}{} - \msgset{1}{} &
            &\Party{2}: \valset{2}{} = \mv{a}{} \mv{b}{} - \msgset{2}{}\\
            &\Partyverify{0}: \valset{0}{} = \mw{a}{} \lv{b}{} + \mw{b}{} \lv{a}{} 
		\end{align*}
		%------
        \item Communicate:
		\begin{align*}
            &\Party{1} \rightarrow \Party{2}: \msgset{1}{} &
            &\Party{2} \rightarrow \Party{1}: \msgset{2}{}
		\end{align*}
		%-------
		\item Locally compute:
		\begin{align*}
            &\Party{1}: \mv{c}{} = \valset{1}{} - \msgset{2}{}  &
            &\Party{2}: \mv{c}{} = \valset{2}{} - \msgset{1}{}  \\
            &\Partysubsetverify{1,2}: \msgset{1,2}{1} = \mv{c}{} + \lw{c}{} &
            &\Partysubsetverify{1,2}: \msgset{1,2}{2} = \msgset{1}{} + \msgset{2}{} + r_{1,2,3}
        \end{align*}
        %------
        \item Communicate:
		\begin{align*}
            &\Partyverify{2} \rightarrow \Partyverify{0}: \msgset{1,2}{1} &
            &\Partyverify{2} \rightarrow \Partyverify{0}: \msgset{1,2}{2}
		\end{align*}
        %-------
		\item Locally compute:
		\begin{align*}
            &\Partyverify{0}: \valset{0,3}{} = \msgset{1,2}{2} - \valset{0}{}  &
            &\Partyverify{0}: \mw{c}{} = \msgset{1,2}{1} - \lv{c}{}
        \end{align*}
        %-------
        \item Compare views using $\piCompareView$:
        \begin{align*}
            &\Partysubsetverify{0,1}: \msgset{1,2}{1}, \msgset{1,2}{2} &
            &\Partysubsetverify{0,3}: \valset{0,3}{} &
            &\Partysubsetverify{2,3}: \msgset{0,3}{}
		\end{align*}
		%------
	\end{enumerate}
\end{protocolbox}
%------------------

%============================
\section{Additional Protocols}
\label{sec:add}
%============================
So far, we have seen the details of our protocols over $\ell$-bit rings. However, to use our protocols for real-world applications, it requires support for Fixed-Point Arithmetic~(FPA)~\cite{FC:CatrinaS10,SP:MohZha17} and mixed circuits~\cite{USENIX:PSSY21,CCS:MohRin18,NDSS:ChaRacSur20}. We summarize our contributions in this direction below and the formal details are provided in~\secref{other-protocols}.
Note that these protocols use the same network links between the parties as our multiplication protocol. Therefore, they maintain their high tolerance to weak network links which we empirically verify in \secref{app-eval}. 

\begin{enumerate}[wide, labelwidth=!, labelindent=10pt, parsep=2pt]
    %------------
    \item Truncation~(\secref{other-truncation}): Protocols using Fixed-Point Arithmetic require a truncation operation to prevent overflows during computation and maintain precision~\cite{FC:CatrinaS10}. Probabilistic truncation~\cite{SP:MohZha17} is one such method, which takes a share $\shr{x} = (\mv{x}{}, \lv{x}{})$ and outputs its truncated version $\ptrunc{\shr{x}} = \left(\lfloor \frac{\mv{x}{}}{2^t} \rfloor, \lfloor \frac{\lv{x}{}}{2^t} \rfloor\right)$. With high probability, $\ptrunc{\shr{x}} \approx \lfloor \frac{x}{2^t} \rfloor$. Here $t$ denotes the number of fractional bits in the FPA representation. 
    
    Similar to ASTRA~\cite{CCSW:CCPS19,MPCLeague} and Tetrad \cite{NDSS:KotiPRS22}, probabilistic truncation can be incorporated to our multiplication protocols at no additional communication cost and inherits the computational improvements from our multiplication protocols. 
    %------------
    \item Matrix Multiplication~(\secref{other-matrix}): Similar to existing works~\cite{CCSW:CCPS19,NDSS:KotiPRS22,USENIX:Dalskov0K21}, our multiplication protocols can be extended trivially to handle matrix multiplications, where the addition and multiplication operations are replaced with its matrix counterparts. Moreover, dot product can be viewed as a special case of matrix multiplication.
    %------------
    \item Comparisons~(\secref{other-comparisons}): Comparisons are a key element in many applications, including non-linear activation functions like ReLU. In the context of rings, comparisons are typically performed by examining the most significant bit~\cite{CCS:MohRin18,CCSW:CCPS19}: if this bit is $1$, the value is negative; if it is $0$, the value is non-negative. 
    %------------
    \item Arithmetic to Boolean Conversion~(\secref{other-A2B}): Given the arithmetic sharing of $x \in \Z{\ell}$, this protocol generates the Boolean sharing of all the $\ell$ bits of $x$. 
    %------------
    \item Bit to Arithmetic Conversion~(\secref{other-bit2A}): Given the Boolean sharing of a bit $a^{\bitb}$, denoted by $\sh{a^{\bitb}}^B$, the protocol generates the arithmetic equivalent shares $\sh{a^{\bitb}}^A$.
    %------------
\end{enumerate}
%============================
\section{MPC in Various Network Settings}
\label{sec:hom_het}
%============================
In this section, we analyse the network links among the parties in our 3PC and 4PC protocols from the view point of latency and bandwidth. For this analysis, we distinguish three different categories of network links between the parties.
\begin{enumerate}[wide, labelwidth=!, labelindent=10pt, parsep=2pt]
    %-----------------
    \item \emph{Category~I:} This category includes links used only for setting up pre-shared keys between parties or exchanging hashes for verification at protocol's end, not for the bulk of communication; e.g., link between $\Partysubset{0,1}$ in our protocols. Low bandwidth or high latency on these links does not significantly impact performance.
    %-----------------
    \item \emph{Category~II:} This category involves links among parties that depend on each other's messages to jointly evaluate a circuit. An example is the network link between $\Partysubset{1,2}$. These parties must interactively evaluate each layer of the circuit and repeatedly wait for each other's incoming communication. Hence, a circuit with a multiplicative depth of $k$ needs at least $k$ sequential chunks of messages exchanged between these parties. Low bandwidth or high latency on these links significantly reduces performance.
    %-----------------
    \item \emph{Category~III:} This category covers links only required during a constant-round preprocessing phase or verification. An example is the link between $\Partysubset{0,2}$. $\Party{0}$ only sends messages to $\Party{2}$ in the preprocessing phase. These messages can be received in a single chunk, and delaying them only adds a constant overhead to the protocol's execution time irrespective of the circuit's multiplicative depth. In $\fourpcname$, $\Party{2}$ also sends messages to $\Party{0}$ for verification purposes. In contrast to the link between $\Partysubset{1,2}$, there is no bidirectional dependency associated with $\Party{0}$ receiving $\Party{2}$'s messages: $\Party{0}$ can compute all its messages to other parties based solely on its input-independent shares during the preprocessing. Therefore, low bandwidth on these links reduces performance noticeably, but high latency does not.
    %-----------------
\end{enumerate}

\figref{comm_het} illustrates the resulting link requirements between the parties following these three different cases, where links of Category~II are denoted by bold arrows ($\boldsymbol{\rightarrow}$), while links of Category~III are denoted by dotted arrows ($\dashrightarrow$). The remaining links between the parties are of Category~I and therefore mainly underutilized. 

\vspace{-2mm}
%-------------------------------
\begin{figure}[htb!]
%-----------------------------------------------
\begin{center}
%-----------------------------------------------
\hrule
\smallskip
\textbf{\footnotesize $\threepcname$: 3PC Multiplication}\\[0.5mm]
\hrule
\vspace{1mm}%\vskip\baselineskip 
%-----------------------------------------------
\begin{subfigure}{.225\textwidth}
      \let\shipout\myshipout

\makeatletter
\newcommand*{\encircled}[1]{\relax\ifmmode\mathpalette\@encircled@math{#1}\else\@encircled{#1}\fi}
\newcommand*{\@encircled@math}[2]{\@encircled{$\m@th#1#2$}}
\newcommand*{\@encircled}[1]{%
\tikz[baseline,anchor=base]{\node[draw,circle,outer sep=0pt,inner sep=.2ex] {#1};}} \makeatother

\begin{tikzpicture}
    \tikzset{
        arr/.style={-latex, line width=.75pt},
        carr/.style={-latex, line width=2pt},
        cor/.style={color=black},
        box/.style={rectangle, draw, color=black},
        circ/.style={circle, draw, color=black}
    }

    \node[circ] (p1) at (0,0) {$P_0$};
    \node[circ] (p2) at (4,0) {$P_1$};
    \node[circ] (p3) at (2,-2) {$P_2$};

        \path[-{Stealth[length=3mm, width=2mm]}, dashed] (p1) edge node [above] {$1$} (p3);

\end{tikzpicture}
    \vspace{-1mm}  
    \captionsetup{font=scriptsize}   
    \caption{Preprocessing~(Base)}
  \label{fig:3PC_OFF}
\end{subfigure}%
\begin{subfigure}{.225\textwidth}
     \hfill 
      \let\shipout\myshipout

\makeatletter
\newcommand*{\encircled}[1]{\relax\ifmmode\mathpalette\@encircled@math{#1}\else\@encircled{#1}\fi}
\newcommand*{\@encircled@math}[2]{\@encircled{$\m@th#1#2$}}
\newcommand*{\@encircled}[1]{%
\tikz[baseline,anchor=base]{\node[draw,circle,outer sep=0pt,inner sep=.2ex] {#1};}} \makeatother

\begin{tikzpicture}
    \tikzset{
        arr/.style={-latex, line width=.75pt},
        carr/.style={-latex, line width=2pt},
        cor/.style={color=black},
        box/.style={rectangle, draw, color=black},
        circ/.style={circle, draw, color=black}
    }

    \node[circ] (p1) at (0,0) {$P_0$};
    \node[circ] (p2) at (4,0) {$P_1$};
    \node[circ] (p3) at (2,-2) {$P_2$};

        \path[{Stealth[length=3mm, width=2mm]}-{Stealth[length=3mm, width=2mm]}, line width=1.5pt] (p2) edge node [above] {$1$} (p3);

\end{tikzpicture}
    \vspace{-1mm}  
    \captionsetup{font=scriptsize}   
    \caption{Online~(Base)}
  \label{fig:3PC_ON}
\end{subfigure}%
%-----------------------------------------------

%-----------------------------------------------
\vspace{1mm}%\vskip\baselineskip 
\hrule
\smallskip
\textbf{\footnotesize $\fourpcname$: 4PC Multiplication}\\[0.5mm]
\hrule
\vspace{1mm}%\vskip\baselineskip  
%-----------------------------------------------

%-----------------------------------------------
\begin{subfigure}{.225\textwidth}
      \let\shipout\myshipout

\makeatletter
\newcommand*{\encircled}[1]{\relax\ifmmode\mathpalette\@encircled@math{#1}\else\@encircled{#1}\fi}
\newcommand*{\@encircled@math}[2]{\@encircled{$\m@th#1#2$}}
\newcommand*{\@encircled}[1]{%
\tikz[baseline,anchor=base]{\node[draw,circle,outer sep=0pt,inner sep=.2ex] {#1};}} \makeatother

\begin{tikzpicture}
    \tikzset{
        arr/.style={-latex, line width=.75pt},
        carr/.style={-latex, line width=2pt},
        cor/.style={color=black},
        box/.style={rectangle, draw, color=black},
        circ/.style={circle, draw, color=black}
    }

    \node[circ] (p1) at (0,0) {$P_0$};
    \node[circ] (p4) at (4,0) {$P_3$};
    \node[circ] (p3) at (0,-2) {$P_2$};
    \node[circ] (p2) at (4,-2) {$P_1$};

        \path[-{Stealth[length=3mm, width=2mm]}, dashed] (p1) edge node [right] {$1$} (p3);
        \path[-{Stealth[length=3mm, width=2mm]}, dashed] (p4) edge node [above] {$1$} (p1);

\end{tikzpicture}
    \vspace{-1mm}  
    \captionsetup{font=scriptsize}   
    \caption{Preprocessing~(Base)}
  \label{fig:4PC_OFF}
\end{subfigure}%
\begin{subfigure}{.225\textwidth}
     \hfill 
      \let\shipout\myshipout

\makeatletter
\newcommand*{\encircled}[1]{\relax\ifmmode\mathpalette\@encircled@math{#1}\else\@encircled{#1}\fi}
\newcommand*{\@encircled@math}[2]{\@encircled{$\m@th#1#2$}}
\newcommand*{\@encircled}[1]{%
\tikz[baseline,anchor=base]{\node[draw,circle,outer sep=0pt,inner sep=.2ex] {#1};}} \makeatother

\begin{tikzpicture}
    \tikzset{
        arr/.style={-latex, line width=.75pt},
        carr/.style={-latex, line width=2pt},
        cor/.style={color=black},
        box/.style={rectangle, draw, color=black},
        circ/.style={circle, draw, color=black}
    }

    \node[circ] (p1) at (0,0) {$P_0$};
    \node[circ] (p4) at (4,0) {$P_3$};
    \node[circ] (p3) at (0,-2) {$P_2$};
    \node[circ] (p2) at (4,-2) {$P_1$};

        \path[-{Stealth[length=3mm, width=2mm]}, dashed] (p3) edge node [right] {$1$} (p1);
        \path[{Stealth[length=3mm, width=2mm]}-{Stealth[length=3mm, width=2mm]}, line width=1.5pt] (p2) edge node [above] {$1$} (p3);

\end{tikzpicture}
    \vspace{-1mm}  
    \captionsetup{font=scriptsize}   
    \caption{Online~(Base)}
  \label{fig:4PC_ON}
\end{subfigure}%
%-----------------------------------------------

%-----------------------------------------------
\vspace{1mm}%\vskip\baselineskip   
%-----------------------------------------------
\begin{subfigure}{.225\textwidth}
     \let\shipout\myshipout

\makeatletter
\newcommand*{\encircled}[1]{\relax\ifmmode\mathpalette\@encircled@math{#1}\else\@encircled{#1}\fi}
\newcommand*{\@encircled@math}[2]{\@encircled{$\m@th#1#2$}}
\newcommand*{\@encircled}[1]{%
\tikz[baseline,anchor=base]{\node[draw,circle,outer sep=0pt,inner sep=.2ex] {#1};}} \makeatother

\begin{tikzpicture}
    \tikzset{
        arr/.style={-latex, line width=.75pt},
        carr/.style={-latex, line width=2pt},
        cor/.style={color=black},
        box/.style={rectangle, draw, color=black},
        circ/.style={circle, draw, color=black}
    }

    \node[circ] (p1) at (0,0) {$P_0$};
    \node[circ] (p4) at (4,0) {$P_3$};
    \node[circ] (p3) at (0,-2) {$P_2$};
    \node[circ] (p2) at (4,-2) {$P_1$};

        \path[-{Stealth[length=3mm, width=2mm]}, dashed] (p1) edge node [right] {$1$} (p3);

\end{tikzpicture}
     \captionsetup{font=scriptsize}
     \vspace{-1mm}
    \caption{Preprocessing~(Heterogeneous)}
  \label{fig:4PC_OFFH}
\end{subfigure}%
\begin{subfigure}{.225\textwidth}
     \hfill 
      \let\shipout\myshipout

\makeatletter
\newcommand*{\encircled}[1]{\relax\ifmmode\mathpalette\@encircled@math{#1}\else\@encircled{#1}\fi}
\newcommand*{\@encircled@math}[2]{\@encircled{$\m@th#1#2$}}
\newcommand*{\@encircled}[1]{%
\tikz[baseline,anchor=base]{\node[draw,circle,outer sep=0pt,inner sep=.2ex] {#1};}} \makeatother

\begin{tikzpicture}
    \tikzset{
        arr/.style={-latex, line width=.75pt},
        carr/.style={-latex, line width=2pt},
        cor/.style={color=black},
        box/.style={rectangle, draw, color=black},
        circ/.style={circle, draw, color=black}
    }

    \node[circ] (p1) at (0,0) {$P_0$};
    \node[circ] (p4) at (4,0) {$P_3$};
    \node[circ] (p3) at (0,-2) {$P_2$};
    \node[circ] (p2) at (4,-2) {$P_1$};

        \path[-{Stealth[length=3mm, width=2mm]}, dashed] (p3) edge node [right] {$2$} (p1);
        \path[{Stealth[length=3mm, width=2mm]}-{Stealth[length=3mm, width=2mm]}, line width=1.5pt] (p2) edge node [above] {$1$} (p3);

\end{tikzpicture}
    \captionsetup{font=scriptsize}
    \vspace{-1mm}
    \caption{Online~(Heterogeneous)}
  \label{fig:4PC_ONH}
\end{subfigure}%
%-----------------------------------------------

    \vspace{-2mm}
    \captionsetup{font=small}
    \caption{Communication pattern during multiplication}
    \label{fig:comm_het} 
\end{center}
%-----------------------------------------------
\begin{flushleft}
\rule{0in}{0.5em}\footnotesize 
-- (Bandwidth-critical) Dashed arrows denote constant communication rounds.\\ 
-- (Bandwidth \& Latency-critical) Bold arrows denote communication rounds linear in the circuit's multiplicative depth. 
\end{flushleft}
%-----------------------------------------------
\vspace{-5mm}
\end{figure}
%-------------------------------

%---------------------------------------
\subsubsection*{\textbf{Heterogeneous Network Settings}}
%---------------------------------------
When designing a protocol for heterogeneous network settings, our goal is to maximize the use of Category~I links while minimizing the use of Category~II links. Consequently, in our 3PC and heterogeneous 4PC protocols, only one link between the parties is latency-critical, and only two links are bandwidth-critical.

In our heterogeneous 4PC protocol, $\Party{3}$ only needs to use Category~I links that are not performance-critical. This means that any network configuration that performs well with our semi-honest 3PC protocol will also perform well in a malicious setting, as long as the parties can agree on a fourth participant. However, this protocol variant requires $\Party{2}$ to send two elements of communication per multiplication gate in one direction as part of verification, which effectively throttles communication on that link by a factor of two.
Therefore, this variant performs better during the actual computation, but may delay the computation's verification in settings where the available bandwidth between parties $\Partysubset{0,2}$ is less than two times higher than the bandwidth between $\Partysubset{1,2}$. Given these considerations, parties can decide which protocol to choose.

%Therefore, this variant is strictly better in scenarios where the available bandwidth between parties $\Partysubset{0,2}$ is at least two times higher than the bandwidth between $\Partysubset{1,2}$. In this case, the link between $\Partysubset{1,2}$ is the bottleneck of the protocol. 

%---------------------------------------
\subsubsection*{\textbf{Homogeneous Network Settings}}
%---------------------------------------
While minimizing link dependencies is beneficial in heterogeneous network settings where network link properties between parties differ, we show how our protocols can be adapted to network settings where this is not a primary concern. In homogeneous network settings, where all parties have similar bandwidth and latency, an efficient protocol distributes its communication complexity evenly across all links. This approach has been previously used for load-balancing MPC protocols~\cite{NDSS:KotiPRS22,EPRINT:LuZR23}. 

Any $n$-party protocol can be converted into a protocol optimized for homogeneous network settings by running $n!$ independent circuits in parallel. In each evaluation, the parties select a unique permutation of their roles in the protocol. For example, consider a 3PC protocol with parties $\Party{i}$, $\Party{j}$, and $\Party{k}$. There are six unique permutations to assign the roles of $\Partysubset{0,1,2}$ to $\Party{i}$, $\Party{j}$, and $\Party{k}$.

For every protocol using $l$ elements of global communication, the number of messages per circuit remains the same, but all communication channels are now utilized equally. We refer to the technique of switching player assignments during joint evaluation of MPC workloads as \emph{Split-Roles}.  \figref{comm_hom} illustrates the resulting communication between nodes when using our 3PC and 4PC protocols in a homogeneous network setting when utilizing this technique. To optimize their communication pattern for a given network setting, the parties can also customize the set of permutations depending on which links should be utilized more and which ones less.

%-------------------------------
\begin{figure}[htb!]
\begin{center}
%-----------------------------------------------
\hrule
\smallskip
\textbf{\footnotesize $\threepcname$: 3PC Multiplication}\\[0.5mm]
\hrule
\vspace{0.5mm}%\vskip\baselineskip  
%-----------------------------------------------

%-----------------------------------------------
\begin{subfigure}{.225\textwidth}
      \let\shipout\myshipout

\makeatletter
\newcommand*{\encircled}[1]{\relax\ifmmode\mathpalette\@encircled@math{#1}\else\@encircled{#1}\fi}
\newcommand*{\@encircled@math}[2]{\@encircled{$\m@th#1#2$}}
\newcommand*{\@encircled}[1]{%
\tikz[baseline,anchor=base]{\node[draw,circle,outer sep=0pt,inner sep=.2ex] {#1};}} \makeatother

\begin{tikzpicture}
    \tikzset{
        arr/.style={-latex, line width=.75pt},
        carr/.style={-latex, line width=2pt},
        cor/.style={color=black},
        box/.style={rectangle, draw, color=black},
        circ/.style={circle, draw, color=black}
    }

    \node[circ] (p1) at (0,0) {$P_0$};
    \node[circ] (p2) at (4,0) {$P_1$};
    \node[circ] (p3) at (2,-2) {$P_2$};

        \path[{Stealth[length=3mm, width=2mm]}-{Stealth[length=3mm, width=2mm]}, dashed] (p1) edge node [above] {$\frac{1}{6}$} (p2);
        \path[{Stealth[length=3mm, width=2mm]}-{Stealth[length=3mm, width=2mm]}, dashed] (p2) edge node [above] {$\frac{1}{6}$} (p3);
        \path[{Stealth[length=3mm, width=2mm]}-{Stealth[length=3mm, width=2mm]}, dashed, draw=none] (p1) edge node [above] {$\frac{1}{6}$} (p3);

\end{tikzpicture}
     \vspace{-1mm}
     \captionsetup{font=scriptsize}
     \caption{Preprocessing ~(Homogeneous)}
  \label{fig:3PC_OFF_HOM}
\end{subfigure}%
\begin{subfigure}{.225\textwidth}
     \hfill 
      \let\shipout\myshipout

\makeatletter
\newcommand*{\encircled}[1]{\relax\ifmmode\mathpalette\@encircled@math{#1}\else\@encircled{#1}\fi}
\newcommand*{\@encircled@math}[2]{\@encircled{$\m@th#1#2$}}
\newcommand*{\@encircled}[1]{%
\tikz[baseline,anchor=base]{\node[draw,circle,outer sep=0pt,inner sep=.2ex] {#1};}} \makeatother

\begin{tikzpicture}
    \tikzset{
        arr/.style={-latex, line width=.75pt},
        carr/.style={-latex, line width=2pt},
        cor/.style={color=black},
        box/.style={rectangle, draw, color=black},
        circ/.style={circle, draw, color=black}
    }

    \node[circ] (p1) at (0,0) {$P_0$};
    \node[circ] (p2) at (4,0) {$P_1$};
    \node[circ] (p3) at (2,-2) {$P_2$};

        \path[{Stealth[length=3mm, width=2mm]}-{Stealth[length=3mm, width=2mm]}, line width=1.5pt] (p1) edge node [above] {$\frac{1}{3}$} (p2);

        \path[{Stealth[length=3mm, width=2mm]}-{Stealth[length=3mm, width=2mm]}, line width=1.5pt] (p2) edge node [above] {$\frac{1}{3}$} (p3);

        \path[{Stealth[length=3mm, width=2mm]}-{Stealth[length=3mm, width=2mm]}, line width=1.5pt] (p1) edge node [above] {$\frac{1}{3}$} (p3);

\end{tikzpicture}
    \vspace{-1mm}
    \captionsetup{font=scriptsize}
    \caption{Online~(Homogeneous)}
  \label{fig:3PC_ON_HOM}
\end{subfigure}%
%-----------------------------------------------

%-----------------------------------------------
\vspace{1mm}%\vskip\baselineskip  
\hrule
\smallskip
\textbf{\footnotesize $\fourpcname$: 4PC Multiplication}\\[0.5mm]
\hrule
\vspace{0.5mm}%\vskip\baselineskip    
%-----------------------------------------------

%-----------------------------------------------
\begin{subfigure}{.225\textwidth}
      \let\shipout\myshipout

\makeatletter
\newcommand*{\encircled}[1]{\relax\ifmmode\mathpalette\@encircled@math{#1}\else\@encircled{#1}\fi}
\newcommand*{\@encircled@math}[2]{\@encircled{$\m@th#1#2$}}
\newcommand*{\@encircled}[1]{%
\tikz[baseline,anchor=base]{\node[draw,circle,outer sep=0pt,inner sep=.2ex] {#1};}} \makeatother

\begin{tikzpicture}
    \tikzset{
        arr/.style={-latex, line width=.75pt},
        carr/.style={-latex, line width=2pt},
        cor/.style={color=black},
        box/.style={rectangle, draw, color=black},
        circ/.style={circle, draw, color=black}
    }

    \node[circ] (p1) at (0,0) {$P_0$};
    \node[circ] (p4) at (4,0) {$P_3$};
    \node[circ] (p3) at (0,-2) {$P_2$};
    \node[circ] (p2) at (4,-2) {$P_1$};

        \path[{Stealth[length=3mm, width=2mm]}-{Stealth[length=3mm, width=2mm]}, dashed] (p1) edge node [above] {} (p2);
        \path[{Stealth[length=3mm, width=2mm]}-{Stealth[length=3mm, width=2mm]}, dashed] (p1) edge node [right] {$\frac{1}{6}$} (p3);
        \path[{Stealth[length=3mm, width=2mm]}-{Stealth[length=3mm, width=2mm]}, dashed] (p1) edge node [above] {$\frac{1}{6}$} (p4);
        \path[{Stealth[length=3mm, width=2mm]}-{Stealth[length=3mm, width=2mm]}, dashed] (p2) edge node [above] {$\frac{1}{6}$} (p3);
        \path[{Stealth[length=3mm, width=2mm]}-{Stealth[length=3mm, width=2mm]}, dashed] (p2) edge node [right] {$\frac{1}{6}$} (p4);
        \path[{Stealth[length=3mm, width=2mm]}-{Stealth[length=3mm, width=2mm]}, dashed] (p3) edge node [above] {$\frac{1}{6}$} (p4);

\end{tikzpicture}
    \vspace{-1mm}
    \captionsetup{font=scriptsize}
    \caption{Preprocessing ~(Homogeneous)}
  \label{fig:4PC_HOM_OFF}
\end{subfigure}%
\begin{subfigure}{.25\textwidth}
     \hfill 
      \let\shipout\myshipout

\makeatletter
\newcommand*{\encircled}[1]{\relax\ifmmode\mathpalette\@encircled@math{#1}\else\@encircled{#1}\fi}
\newcommand*{\@encircled@math}[2]{\@encircled{$\m@th#1#2$}}
\newcommand*{\@encircled}[1]{%
\tikz[baseline,anchor=base]{\node[draw,circle,outer sep=0pt,inner sep=.2ex] {#1};}} \makeatother

\begin{tikzpicture}
    \tikzset{
        arr/.style={-latex, line width=.75pt},
        carr/.style={-latex, line width=2pt},
        cor/.style={color=black},
        box/.style={rectangle, draw, color=black},
        circ/.style={circle, draw, color=black}
    }

    \node[circ] (p1) at (0,0) {$P_0$};
    \node[circ] (p4) at (4,0) {$P_3$};
    \node[circ] (p3) at (0,-2) {$P_2$};
    \node[circ] (p2) at (4,-2) {$P_1$};

        \path[{Stealth[length=3mm, width=2mm]}-{Stealth[length=3mm, width=2mm]}, line width=1.5pt] (p1) edge node [above] {} (p2);
        \path[{Stealth[length=3mm, width=2mm]}-{Stealth[length=3mm, width=2mm]}, line width=1.5pt] (p1) edge node [right] {$\frac{1}{4}$} (p3);
        \path[{Stealth[length=3mm, width=2mm]}-{Stealth[length=3mm, width=2mm]}, line width=1.5pt] (p1) edge node [above] {$\frac{1}{4}$} (p4);
        \path[{Stealth[length=3mm, width=2mm]}-{Stealth[length=3mm, width=2mm]}, line width=1.5pt] (p2) edge node [above] {$\frac{1}{4}$} (p3);
        \path[{Stealth[length=3mm, width=2mm]}-{Stealth[length=3mm, width=2mm]}, line width=1.5pt] (p2) edge node [right] {$\frac{1}{4}$} (p4);
        \path[{Stealth[length=3mm, width=2mm]}-{Stealth[length=3mm, width=2mm]}, line width=1.5pt] (p3) edge node [above] {$\frac{1}{4}$} (p4);

\end{tikzpicture}
    \vspace{-1mm}
    \captionsetup{font=scriptsize}
    \caption{Online~(Homogeneous)}
  \label{fig:4PC_HOM_ON}
\end{subfigure}%
%-----------------------------------------------

    \vspace{-2mm}
    \captionsetup{font=small}
    \caption{Communication pattern using Split-Roles}
    \label{fig:comm_hom} 
\end{center}
%-----------------------------------------------
\begin{flushleft}
\rule{0in}{0.5em}\footnotesize 
Bandwidth-critical: Dashed arrows denote constant communication rounds.\\ 
Bandwidth \& Latency-critical: Bold arrows denote communication rounds linear in the circuit's multiplicative depth.
\end{flushleft}
%-----------------------------------------------
\vspace{-4mm}
\end{figure}
%-------------------------------

%---------------------------------------
\subsubsection*{\textbf{Other Network Settings}}
%---------------------------------------
Our protocols offer high flexibility in optimizing communication patterns for different network settings, thanks to their high tolerance for weak network links and their use of replicated sharing semantics. The parties can utilize their replicated secret sharing to adjust their communication pattern on a per-message basis. 

When parties hold one input-dependent and one input-independent share, any outgoing or incoming message can be shifted between parties without introducing any additional communication costs. For instance, in our 3PC and 4PC protocols, $\Party{1}$'s and $\Party{2}$'s computation and communication pattern can be easily adjusted such that $\Party{0}$ sends its message in the preprocessing phase to $\Party{1}$ instead of $\Party{2}$. By selecting the percentage of messages to send to a specific party, $\Party{0}$ can granularly adjust the utilization of different network links based on available bandwidth.
%============================
\section{Implementation \& Benchmarking}
\label{sec:benchmarking}
%============================
We implemented our protocols and related state-of-the-art ones in C++. Specifically, we implemented two other state-of-the-art protocols for each category, namely the 3PC protocols ASTRA~\cite{CCSW:CCPS19} and Replicated~\cite{CCS:AFLNO16} and the 4PC protocols Fantastic Four~\cite{USENIX:Dalskov0K21} and Tetrad~\cite{NDSS:KotiPRS22}. One protocol in each category offers function-dependent preprocessing (ASTRA, Tetrad), while the other does not (Replicated, Fantastic Four).

All results in this section are based on a test setup of 3 to 4 nodes. Each node is connected with a 25 Gbit/s duplex link to each other node and equipped with a 32-core AMD EPYC 7543 processor. The round-trip latency between nodes is 0.3 ms.
If not stated otherwise, we do not use a separate preprocessing phase but perform all preprocessing operations during the online phase.

This section is divided into three main parts:
\begin{enumerate}[wide, labelwidth=!, labelindent=10pt, parsep=2pt]
    %-----------------
    \item In \secref{bench-scaling}, we demonstrate how to scale MPC protocols to achieve a throughput of billions of gates per second under standard conditions. We propose a series of optimizations that can be universally applied to any MPC protocol. 
    %-----------------
    \item \secref{bench-bottlenecks} highlights the improvements of our protocols in practical scenarios, including real-world network settings with different latencies and bandwidths across parties. This part addresses and overcomes the practical bottlenecks of MPC discussed in \secref{Bottlenecks}.
    %-----------------
%    \item Finally, we present the benchmarking results for AES circuit evaluations using our protocols in \secref{bench-AES}.
    %-----------------
\end{enumerate}

%---------------------------------------------
\subsection{Scaling MPC to Billions of Gates}
\label{sec:bench-scaling}
%---------------------------------------------
Our framework integrates a comprehensive set of universal optimizations: Bitslicing, Vectorization, multiprocessing, hardware instructions like \texttt{VAES} and \texttt{SHA}, and adjustable message buffering. In this section, we show how stacking up these optimizations allows us to scale MPC workloads to billions of gates per second. Additionally, our results confirm our finding that existing open-source frameworks struggle to achieve a throughput of more than a few million gates per second~(cf.~\secref{Bottlenecks}). 

\makeparafit
Different MPC protocols typically require at least some of the following components to evaluate a non-linear gate:  Encrypted communication channels, cryptographically secure random number generators and hash functions, as well as multiple elementary operations. Additionally, MPC workloads are highly sensitive to network bandwidth and latency. Therefore, implementations benefit significantly from accelerating these basic instructions and optimizing network communication. We focus on these aspects next.

%-----------------------------------------------
\subsubsection*{\textbf{Accelerating Basic Instructions}}
%-----------------------------------------------
As MPC protocols require the same kind of instructions repeatedly for each gate, we use the Single Instruction Multiple Data~(SIMD) approach to accelerate them by using wider register sizes. For example, we can use the \texttt{AVX-2} instruction set to perform eight 32-bit additions, 256 1-bit logic gates, or two 128-bit AES rounds in parallel using a single instruction on a 256-bit register. For SSL-encrypted communication, we rely on the OpenSSL library. Notably, the throughput of the cryptographic instructions is, on average, only 5-10 times slower than the throughput of the non-cryptographic instructions thanks to available hardware instructions such as \texttt{VAES} and \texttt{SHA}.

%---------------------------------------
\paragraph{Bitslicing and Vectorization}
%---------------------------------------
The key idea of Bitslicing is that computing a bitwise logical operation on an $m$-bit register effectively operates like $m$ parallel Boolean conjunctions, each processing a single bit~\cite{PLDI:MercadierD19}. Thus, Bitslicing can accelerate single-bit operations such as AND or XOR, provided that the bits to be operated on are packed properly. For example, instead of performing a single 1-bit XOR operation across two bits, we could perform 32 XOR operations on 32-bit data if the bits are packed into a single 32-bit variable. Furthermore, one can exploit hardware instruction sets such as \texttt{AVX-2} to pack 256 bits and compute 256 XOR operations in parallel. To support Bitslicing on various CPU architectures, we utilize the header files provided by the Usuba Bitslicing compiler~\cite{PPOPP:MercadierDLM18}.

%--------------------------------
\begin{figure}[htbp]
    \begin{subfigure}[b]{0.45\textwidth}
        \centering
        \includegraphics[width=\linewidth, trim = 0cm 5cm 0cm 5cm, clip]{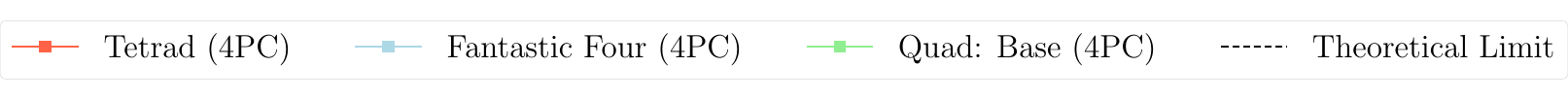}
    \end{subfigure}
    \centering
    \begin{subfigure}[b]{0.225\textwidth}
        \centering
        \includegraphics[width=\textwidth]{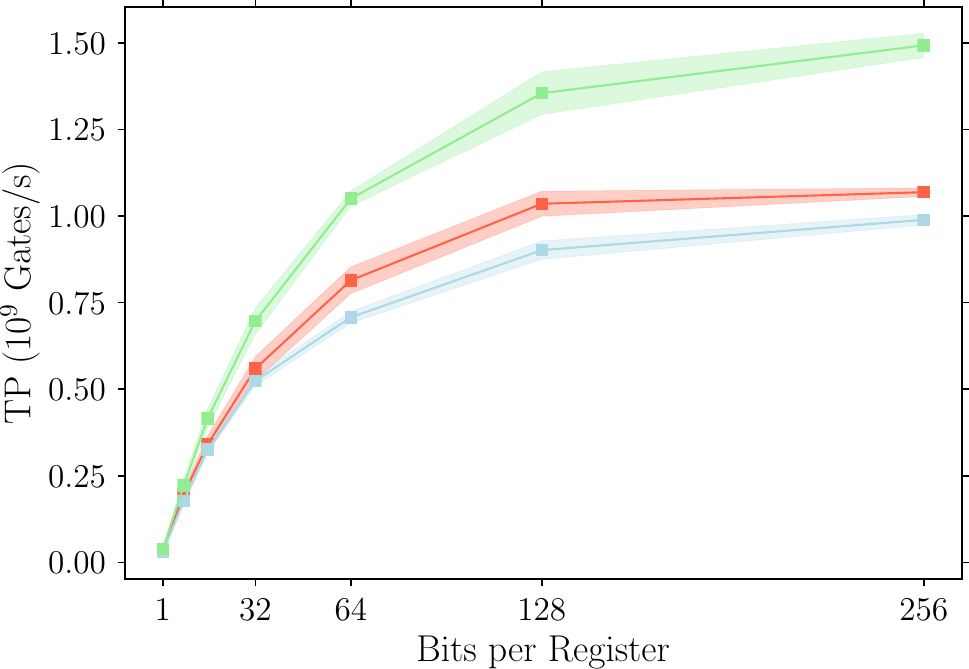}
    \end{subfigure}
    \hfill
    \begin{subfigure}[b]{0.225\textwidth}
        \centering
        \includegraphics[width=\textwidth]{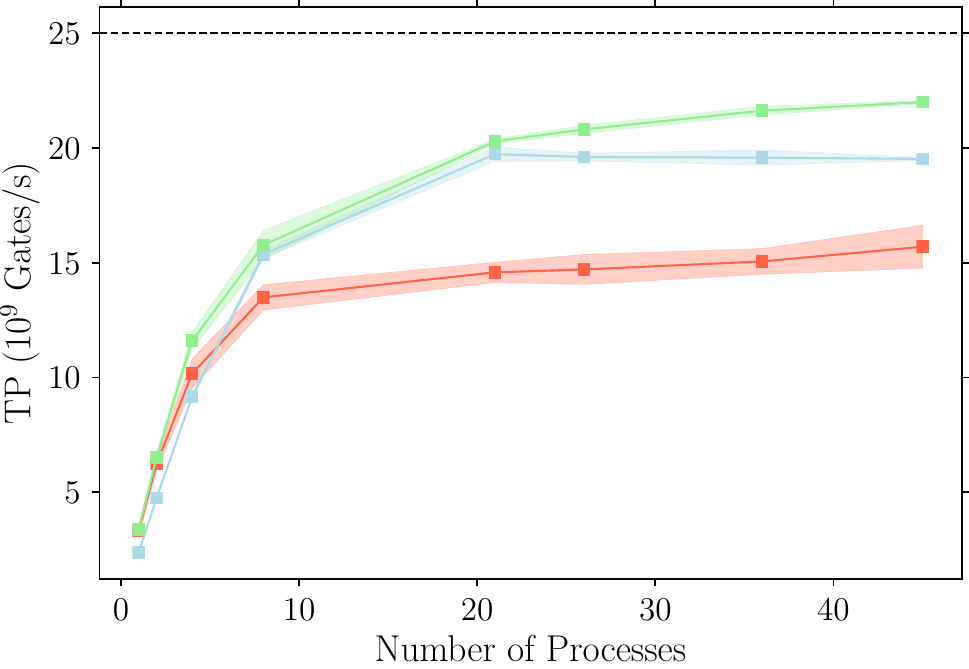}
    \end{subfigure}
    \vspace{-3mm}
    \captionsetup{font=small}
    \caption{Throughput when utilizing Bitslicing and Multiprocessing\label{fig:bitmult}}
\end{figure}
%--------------------------------
\makeparafit
\figref{bitmult}~(left) shows the throughput of the 4PC protocols when utilizing Bitslicing with increasing bit widths to evaluate AND gates. As we increase the bit width, we effectively utilize hardware parallelism on a single core to process a batch of 256 Boolean operations using a single instruction. Consequently, the throughput measured on our test setup increases over 100 times when performing 256 AND gates in parallel on an \texttt{AVX-2} register, compared to using a single Boolean variable and performing one instruction for each input. 

%--------------------------------
\begin{table}[htbp]
\captionsetup{font=small}
\caption{Throughput for protocols without using Split-Roles}
\label{tab:throughput}
\centering
\small
\vspace{-3mm}
\begin{threeparttable}
\begin{tabular}{clcc}
\hline
\multirow{2}{*}{Setting} & \multirow{2}{*}{Protocol} & Billion Gates/s & Theoretical \\
& & ($\sigma \pm \mu$) & Limit (\%) \\
\hline
\multirow{3}{*}{3PC} & Replicated & 24.57 $\pm$ 0.13 & 98.28 $\pm$ 0.51 \\
& ASTRA & 23.91 $\pm$ 0.12 & 95.64 $\pm$ 0.50 \\
& $\threepcname$ & 23.81 $\pm$ 0.38 & 95.24 $\pm$ 1.50 \\
\hline
\multirow{3}{*}{4PC} & Fantastic Four & 18.93 $\pm$ 0.44 & 75.71 $\pm$ 1.75 \\
& Tetrad & 15.18 $\pm$ 0.27 & 60.72 $\pm$ 1.10 \\
& $\fourpcname$ & 22.04 $\pm$ 0.05 & 88.16 $\pm$ 0.20 \\
\hline
\end{tabular}
\end{threeparttable}
\end{table}

%--------------------------------

Various MPC workloads, such as privacy-preserving machine learning, or user authentication as motivated by~\cite{CCS:AFLNO16}, operate on batches of data and can benefit significantly from utilizing multiple cores. \figref{bitmult}~(right) and \tabref{throughput} show that when combining Bitslicing with multiprocessing, our implementations achieve a throughput of more than 15 billion AND gates per second across all protocols. These results are within $61\% - 98\%$ of the theoretical limit of 25 billion AND gates per second that we can achieve on a 25-Gbit/s network without using Split-roles. The remaining gap in throughput is likely due to networking overhead incurred when sending and receiving messages with multiple threads using conventional sockets.

%-----------------------------------------------
\subsubsection*{\textbf{Optimizing Network Communication}}
%-----------------------------------------------
Optimizing local computations for MPC workloads is only effective until the network bottlenecks further performance improvements. In the following lines, we present techniques to fully utilize the available network bandwidth between parties. 

%------------------------------
\paragraph{Buffering}
%------------------------------
When evaluating a circuit, the parties must exchange a certain number of messages in each communication round. A party can either send each message as soon as it is computed, or instead buffer a set of messages and send them all at once. Our measurements showed a 50 times difference in throughput between an ideal and worst-case buffer size. On our test setup, buffering between $0.3$MB and $3$MB of messages led to the highest throughput.

%------------------------------
\paragraph{Split-Roles}
%------------------------------
As introduced in \secref{hom_het}, Split-Roles refers to parties switching player assignments across MPC workloads to perform load balancing.
Using Split-Roles, all messages are equally distributed between the parties, and the available network bandwidth is fully utilized. For instance, on a 25-Gbit/s network, we can theoretically achieve a throughput of 50 billion AND gates per second by utilizing Split-Roles with a 3PC protocol that requires three elements of global communication. We can increase the throughput further by executing a 3PC protocol with four parties, essentially creating a 4PC protocol. This way, we can achieve a theoretical throughput of 100 billion AND gates per second on a 25-Gbit/s network as the total number of links between the parties doubles. \tabref{throughput_split} shows the throughput of the implemented protocols when utilizing Split-Roles along with our other tweaks.

%\vspace{-2mm}
%------------------------------

\begin{table}[htbp]
\captionsetup{font=small}
\caption{Throughput for protocols when using Split-Roles}
\label{tab:throughput_split}
\centering
\small
\vspace{-3mm}
\begin{threeparttable}
\begin{tabular}{clll}
\hline
\multirow{2}{*}{Setting} & \multirow{2}{*}{Protocol} & Billion Gates/s & Theoretical \\
& & ($\sigma \pm \mu$) & Limit (\%) \\
\hline
\multirow{5}{*}{3PC SH} & Replicated & $36.16 \pm 0.35$ & $72.32 \pm 0.70$\ \\
 & ASTRA & $45.81 \pm 0.47$ & $91.62 \pm 0.94$\ \\
 & $\threepcname$ & $44.04 \pm 0.71$ & $88.08 \pm 1.42$\ \\ \cline{2-4}
 & ASTRA (Online) & $58.16 \pm 4.07$ & $77.55 \pm 5.43$\ \\
 & $\threepcname$ (Online) & $61.95 \pm 2.71$ & $82.60 \pm 3.61$\ \\ \hline
\multirow{5}{*}{4*PC SH\tnote{a}} & Replicated & $64.50 \pm 0.72$ & $64.50 \pm 0.72$\ \\
 & ASTRA & $73.81 \pm 1.37$ & $73.81 \pm 1.37$\ \\
 & $\threepcname$ & $70.96 \pm 1.42$ & $70.96 \pm 1.42$\ \\ \cline{2-4}
 & ASTRA (Online) & $90.93 \pm 3.92$ & $60.62 \pm 2.61$\ \\
 & $\threepcname$ (Online) & $88.21 \pm 7.22$ & $58.81 \pm 4.81$\ \\ \hline
\multirow{5}{*}{4PC MAL} & Fantastic Four & $26.92 \pm 0.07$ & $44.87 \pm 0.11$\ \\
 & Tetrad & $32.30 \pm 0.97$ & $43.07 \pm 1.29$\ \\
 & $\fourpcname$ & $38.29 \pm 0.81$ & $51.05 \pm 1.08$\ \\ \cline{2-4}
 & Tetrad (Online) & $47.05 \pm 1.44$ & $47.05 \pm 1.44$\ \\
 & $\fourpcname$ (Online) & $59.39 \pm 4.92$ & $59.39 \pm 4.92$\ \\ \hline
 TTP & Trusted Third Party  & 540.99 $\pm$ 20.57 & - \\ \hline
\end{tabular}
      \begin{tablenotes}
            \item SH refers to semihonest, MAL refers to malicious
            \item[a] 4*PC SH refers to executing a 3PC protocol on four nodes
        \end{tablenotes}
\end{threeparttable}
\end{table}

%------------------------------

%\vspace{-3mm}
%------------------------------
\paragraph{Online Phase}
%------------------------------
Most protocols we implemented offer a preprocessing phase that can be detached from the online phase. \tabref{throughput_split} shows the throughput of the implemented protocols when considering both phases and when only considering the Online Phase. We additionally compare the throughput of the Online Phase to the throughput a Trusted Third Party (TTP) can achieve on the same hardware. Notably, the throughput when utilizing a TTP is less than one order of magnitude higher than the secure alternatives when utilizing all aforementioned optimizations.

%---------------------------------------------
\subsubsection*{\textbf{AES Benchmark}}
%---------------------------------------------
AES is a common benchmark for assessing the performance of MPC frameworks and protocols. Araki et al. \cite{CCS:AFLNO16} have achieved the highest AES throughput so far, with 1.3 million 128-bit AES blocks per second. To test whether our tweaks on the throughput of raw $AND$ and multiplication gates translate to more complex circuits, we benchmark the throughput of 128-bit AES blocks based on the open-source AES circuit of \cite{PPOPP:MercadierDLM18} using the implemented protocols . 
%As the basis for the AES circuit, we utilize the optimized AES circuit proposed by USUBA \cite{PPOPP:MercadierDLM18}. 
We perform over 90 million AES blocks in parallel using all optimizations introduced in this section. 

%------------------------------
\begin{table}[htb!]
\captionsetup{font=small}
\caption{Throughput in million AES blocks per second}
\label{table:aes}
\centering
\captionsetup{font=small}
\small
\vspace{-3mm}
\begin{threeparttable}
\begin{tabular}{cllll}
\hline
\multirow{2}{*}{Setting} & \multirow{2}{*}{Protocol} & Million Blocks/s & Theoretical \\
& & ($\sigma \pm \mu$) & Limit (\%) \\
\hline
\multirow{5}{*}{3PC} & Replicated  & 5.18 $\pm$ 0.06 & 54.71\ $\pm$ 0.59\ \\
& ASTRA & 6.03 $\pm$ 0.32 & 63.69\ $\pm$ 3.35\ \\
& $\threepcname$~ & 5.13 $\pm$ 0.06 & 54.16\ $\pm$ 0.68\ \\ \cline{2-4}
& ASTRA (Online) & 6.69 $\pm$ 0.21 & 47.12\ $\pm$ 1.46\ \\
& $\threepcname$ (Online) & 6.69 $\pm$ 0.11 & 47.07\ $\pm$ 0.77\ \\ 
\hline
\multirow{5}{*}{4PC} & Fantastic Four & 2.11 $\pm$ 0.06 & 18.73\ $\pm$ 0.53\ \\
& Tetrad & 2.57 $\pm$ 0.03 & 22.81\ $\pm$ 0.23\ \\
& $\fourpcname$~ & 3.63 $\pm$ 0.06 & 32.25\ $\pm$ 0.53\ \\ \cline{2-4}
& Tetrad (Online) & 2.86 $\pm$ 0.04 & 15.34\ $\pm$ 0.22\ \\
& $\fourpcname$ (Online) & 4.15 $\pm$ 0.07 & 22.23\ $\pm$ 0.36\ \\ 
\hline
& Trusted Third Party  & 17.21 $\pm$ 0.05 & - \\ \hline
\end{tabular}
\end{threeparttable}
\end{table}

%------------------------------

Table \ref{table:aes} shows the throughput in AES blocks per second. While our protocols cannot saturate the network to the same degree as for raw $AND$ gates, we can still achieve four times higher throughput than previous work using the same 3PC protocol as \cite{CCS:AFLNO16}. The 4PC protocols achieve around half the performance of the 3PC protocols in our implementation. Depending on the computational complexity of the underlying protocol, we are able to saturate between $15\%$ and $64\%$ of our 25 Gbit/s connection.

%---------------------------------------------
\subsection{Overcoming MPC Bottlenecks}
\label{sec:bench-bottlenecks}
%---------------------------------------------
Our protocols excel especially in two scenarios: real-world (heterogeneous) network settings, and computational extensive tasks. We simulate heterogeneous network settings by using Linux traffic control (tc) to restrict the bandwidth and latency between nodes. 

%---------------------------------------
\subsubsection*{\textbf{Latency Constrained Workloads}}
\begin{figure*}[htb!]
    \centering
    \begin{subfigure}[b]{0.9\textwidth}
        \centering
        \includegraphics[width=\linewidth, trim = 0cm 4.5cm 0cm 4.5cm, clip]{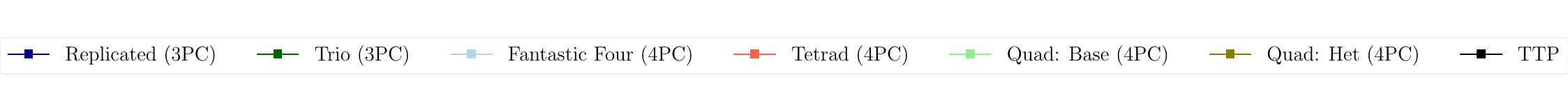}
    \end{subfigure}
    \hspace{0.01\textwidth}
    \begin{subfigure}[b]{0.3\textwidth}
        \centering
  \includegraphics[width=0.95\linewidth, trim = 0cm 0cm 6.3cm 0cm, clip]{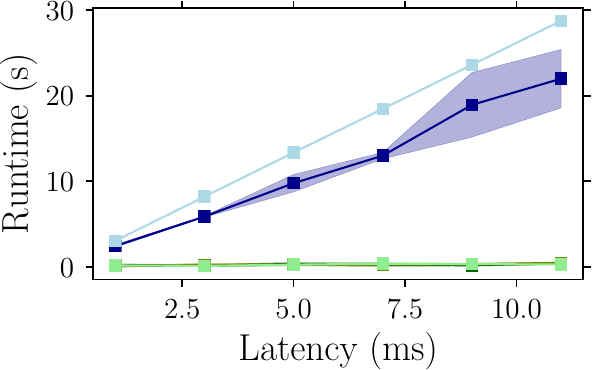}
        \vspace{-1mm}
        \captionsetup{font=footnotesize}
        \caption{AES Runtime: Restricting Latency between $\frac{5}{6}$ of all links}
        \label{fig:AES_lat}
    \end{subfigure}
    \hspace{0.01\textwidth}
    \begin{subfigure}[b]{0.3\textwidth}
        \centering
        \includegraphics[width=0.9\linewidth, trim = 0cm 0cm 6.3cm 0cm, clip]{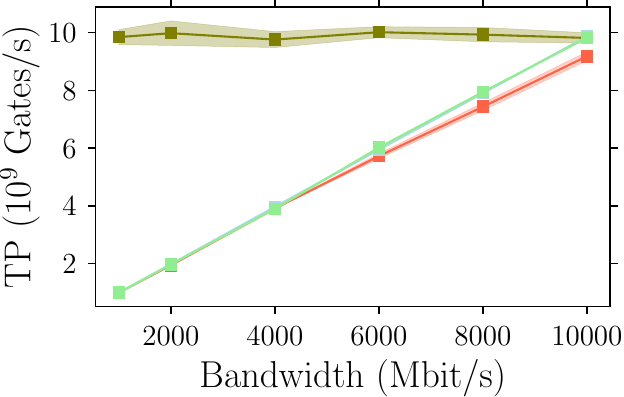}
        \vspace{-1mm}
        \captionsetup{font=footnotesize}
        \caption{Multiplication Throughput (TP): Restricting bandwidths between $\frac{2}{3}$ of all links}
        \label{fig:bdw}
    \end{subfigure}
    \hspace{0.01\textwidth}
    \begin{subfigure}[b]{0.3\textwidth}
        \centering
        \includegraphics[width=0.92\linewidth, trim = 0cm 0cm 6.3cm 0cm, clip]{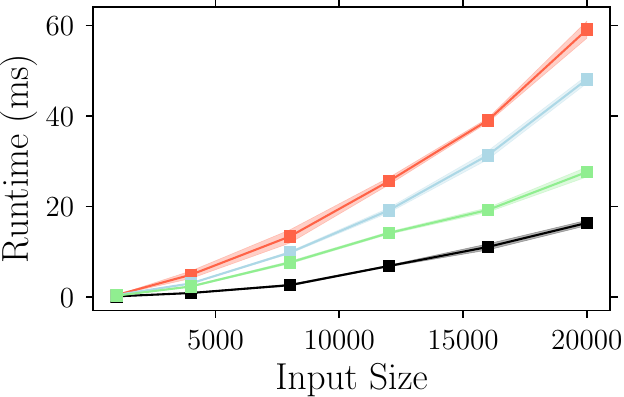}
        \vspace{-1mm}
        \captionsetup{font=footnotesize}
        \caption{Vector-Matrix-Product Runtime against TTP Baseline }
        \label{fig:dot}
    \end{subfigure}
    \vspace{-2mm}
    \captionsetup{font=small}
    \caption{Settings where our protocols excel}
    \label{fig:sweetspots}
    \vspace{-2mm}
\end{figure*}

%---------------------------------------
MPC workloads that process a small number of sequential operations are typically bottlenecked by the network latency. 
Figure \ref{fig:AES_lat} shows that when evaluating a single AES block, our protocols are marginally affected when throttling the network links between a majority of the parties.

To demonstrate the utility of our protocols for real-world latency-restricted workloads, we compare the runtime of our protocols when encrypting an AES block against a baseline of the Fantastic Four and Replicated protocols. 
Table \ref{table:workload_real} shows the protocols' runtime if we apply the real-world latencies measured in the WAN1 setting (c.f. Table \ref{tab:node_location}) ranging from 34.7ms to 192.5ms.
The results show that the 455\% difference between the best and the worst link measured in Table \ref{tab:node_location} closely translates to the real-world advantage of 451\% of our 4PC protocol to the baseline.

%When we however simply allow one link between the parties to be unaffected by the link throttling, figures \ref{fig:mult_latp} and \ref{fig:fixed_mult_latp} show, that our protocols are only marginally affected by the restriction compared to the baseline. 

%Our A2B and Bit2A setup protocols do not require any latency-critical communication between the parties and are thus classified by us as 0-round protocols. Figures \ref{fig:a2bit_lat} and \ref{fig:bit2a_lat} show that even if we restrict the latency of all links between parties, these are marginally affected compared to the one-round baseline. Note that these are only the setup phases required before performing an arithmetic XOR or Boolean addition to complete the share conversion.

%------------------------------
%\input{combined_figures/latency_restrictions_new}
%\input{combined_figures/latency_restrictions_real}
%------------------------------

%Beyond these basic primitives, we also show the advantage of relying only on a few links for the more complex AES circuit. Figure \ref{fig:AES_lat} shows that our protocols are significantly faster than Replicated 3PC and Fantastic Four in latency-restricted settings when the network latency between the parties differs.

%------------------------------
%------------------------------

%---------------------------------------
\subsubsection*{\textbf{Bandwidth Constrained Workloads}}
%---------------------------------------

MPC workloads that process a large batch of independent operations are typically bottlenecked by the network bandwidth. 
Figure \ref{fig:bdw} shows that even if we restrict the bandwidth of $\frac{2}{3}$ of all links in our setup when evaluating a large number of multiplications, our 4PC variation optimized for heterogeneous settings is mostly unaffected by the restriction.

Our real-world bandwidth-restricted workload consists of two million parallel AES computations. Table \ref{table:workload_real} shows that if we apply the real-world network properties measured in WAN1 ranging from 142 Mbit/s to 868 Mbit/s of available network bandwidth between parties, our 4PC protocol achieves more than four times higher throughput compared to the baseline. 
%When we restrict the bandwidth between $\frac{1}{3}$ of the nodes in our setup, our 3PC protocol still achieves the same throughput of approx. 24 billion $AND$ gates per second, as measured in the unrestricted setting. This is due to the link between $\Partysubset{0,1}$ not being utilized at all in the multiplication protocol.
%Figure \ref{fig:bdw} shows that even if we restrict the bandwidth of $\frac{2}{3}$ of the links in our setup, our 4PC variation optimized for heterogeneous settings still achieves a throughput of approx. 10 billion $AND$ gates per second. The figure shows that while the bandwidth restriction affects all other protocols, it does not affect the throughput of our heterogeneous protocol.

%---------------------------------------
\subsubsection*{\textbf{Compute Constrained Workloads}}
%---------------------------------------
MPC workloads that process large matrix multiplication or dot products are typically bottlenecked by local computation. 
Our protocols excel at computationally demanding tasks due to their reduced number of basic instructions compared to related work. ABY3 \cite{CCS:MohRin18} first discovered that the communication complexity to evaluate a dot product of arbitrary size in the semi-honest replicated 3PC setting is that of a single multiplication. Thus, sufficiently large dot products become inevitably computation-bound. 
%To benchmark the performance of dot products, we compute the product of a vector of size $n$ with a matrix of size $n \times n$, resulting in $n$ dot products of size $l=n$.%
To benchmark the performance of dot products, we compute a batch of large vector-matrix products with vector sizes between 1,000 and 20,000 elements.
A vector-matrix product is, for instance, required in privacy-preserving machine learning when evaluating fully connected layers.

Figure \ref{fig:dot} shows that in ideal network settings, our 4PC protocol is two times faster when evaluating large dot products than Tetrad and Fantastic Four. Furthermore, our Trusted-Third-Party implementation is less than two times faster than our 4PC protocol on the same hardware.
Table \ref{table:workload_real} shows that in the real-world WAN setting, this translates to up to 33\% faster runtimes for an input size of 20,000 elements and a batch size of 64.

\begin{table}[htb!]
\captionsetup{font=small}
\caption{Runtime in seconds for the three different real-world workloads in WAN1 network setting}
\label{table:workload_real}
\centering
\captionsetup{font=small}
\small
\vspace{-3mm}
\begin{threeparttable}
\begin{tabular}{lrrr}
\hline
Protocol & Lat.\tnote{a} ($\sigma \pm \mu$) & Bdw.\tnote{b} ($\sigma \pm \mu$) & Comp.\tnote{c} ($\sigma \pm \mu$) \\
\hline
Replicated  & 81.17 $\pm$ 0.00 & 103.00 $\pm$ 0.29 & 1.02 $\pm$ 0.13 \\
$\threepcname$ & 43.91 $\pm$ 0.03 & 45.39 $\pm$ 0.12 & 0.87 $\pm$ 0.04 \\ 
\hline
Fantastic Four & 243.49 $\pm$ 0.08 & 255.69 $\pm$ 0.92 & 1.82 $\pm$ 0.08 \\
$\fourpcname$ & 44.17 $\pm$ 0.05 & 60.07 $\pm$ 2.16 & 1.37 $\pm$ 0.12 \\ 
\hline
\end{tabular}
\begin{tablenotes}
\item[a] Workload contains a single AES block.
\item[b] Workload contains two million parallel AES blocks.
\item[c] Workload contains 64 vector-matrix products, 20,000 elements. 
\end{tablenotes}
\end{threeparttable}
\end{table}

\subsubsection*{\textbf{AES Benchmark in Various Network Settings}}

Finally, we benchmark the throughput of our protocols against the Replicated 3PC and Fantastic Four 4PC protocols by using the AES workload from Table \ref{table:aes} with two million blocks in the CMAN, WAN1, and Mixed real-world network settings (c.f. Table \ref{tab:node_location}). The results show that in all of these settings, our protocols benefit from their heterogeneous properties, enabling them to outperform the others between two to eight times depending on the level of network heterogeneity.

\begin{table}[htb!]
\captionsetup{font=small}
\caption{Throughput in thousand AES blocks per second for different network configurations}
\label{table:aes_real}
\centering
\captionsetup{font=small}
\small
\vspace{-3mm}
\begin{threeparttable}
\begin{tabular}{lrrr}
\hline
Protocol & CMAN ($\sigma \pm \mu$) & WAN1 ($\sigma \pm \mu$) & Mixed ($\sigma \pm \mu$) \\
\hline
Replicated  & 25.77 $\pm$ 0.00 & 19.88 $\pm$ 0.06 & 13.78 $\pm$ 0.12 \\
$\threepcname$~ & 189.81 $\pm$ 0.88 & 45.12 $\pm$ 0.12 & 44.80 $\pm$ 0.44 \\
\hline
Fantastic Four & 25.63 $\pm$ 0.01 & 8.01 $\pm$ 0.03 & 5.36 $\pm$ 0.01 \\
$\fourpcname$~ & 83.52 $\pm$ 0.00 & 34.09 $\pm$ 1.22 & 43.98 $\pm$ 0.87 \\
\hline
\end{tabular}
\end{threeparttable}
\end{table}

\vspace{-1mm}
%============================
\section{Conclusion}
\label{sec:Conclusion}
%============================
In this work, we introduced novel three-party and four-party computation protocols designed to achieve high throughput across various network settings. Leveraging outsourced MPC computations, our protocols enable efficient MPC for any number of input parties. Our open-source implementation demonstrates the capability to evaluate billions of gates per second, even in scenarios where inter-party links exhibit high latency and limited bandwidth. This underscores MPC's ability to manage intensive workloads across real-world settings characterized by varied computational node capabilities. 
Our benchmarks highlight the need for optimizing both communication and computational aspects of MPC protocols as well as enhancements in MPC implementations to bridge the performance gap with Trusted Third Party setups.

Looking forward, an interesting direction for future research involves exploring optimizations tailored for heterogeneous network conditions in different honest-majority scenarios. Moreover, given the high throughput our implementation achieves while processing Boolean and arithmetic gates, future research could extend these techniques to handle more complex tasks, such as privacy-preserving machine learning applications that involve large-scale dot product evaluations and non-linear activation functions.
%============================
\subsection*{Acknowledgements}
%============================
\ifpublicversion
The authors Yongqin Wang and Murali Annavaram would like to acknowledge the support by Defense Advanced Research Projects Agency (DARPA) under Contract Nos. HR001120C0088, NSF award number  2224319, REAL@USC-Meta center, and VMware gift. The views, opinions, and/or findings expressed are those of the author(s) and should not be interpreted as representing the official views or policies of the Department of Defense or the U.S. Government. The author Hossein Yalame would like to acknowledge the support of the CRYPTECS project, which is funded by the Federal Ministry of Education and Research under Grant Agreement No. 16KIS1441.
\else
We used ChatGPT4 to revise the text to correct typos and grammatical errors.
\fi
%----------------------------

\bibliographystyle{ACM-Reference-Format}
\bibliography{ref_cryptobibshort,ref_myreferencesshort}

%\clearpage
\appendix
%--------------------------
%============================
\section{Correctness}
\label{sec:Computations}
%============================
In this section, we explain the correctness of our 3PC~(cf.~\secref{3PC}) and 4PC~(cf.~\secref{4PC}) multiplication protocols by unfolding the computations, showing that each party receives a valid share of the output $c = ab$. Additionally, for our 4PC protocol, which provides malicious security with abort against an adversary $\Adv$ corrupting at most one party, we show that each message sent by $\Adv$ can be verified by a set of honest parties using $\piCompareView$~(cf.~\figref{compareview}), ensuring that any inconsistency will be detected.

%----------------------------
\subsection{$\threepcname$: 3PC Protocol}
\label{sec:A3C}
%----------------------------
In our 3PC multiplication protocol $\piMult$~(cf.~\figref{piMult3PC}), $\Party{0}$ along with $\Party{1}$/$\Party{2}$ locally samples their share of mask $\lv{c}{}$ during the preprocessing by invoking $\piSRNG$~(cf.~\figref{srng}) using the shared PRF keys.  
To see the correctness of $\Party{1}$'s computation during the online phase, note that:
\begin{align*}
    \underline{\text{Party }}&\underline{\Party{1}\text{ in }\piMult~\text{(cf.~\figref{piMult3PC})}:} \\
    \mv{c}{2} &= \msgset{2}{} - \valset{1}{} \\
            &= \mv{a}{1} \mv{b}{1} + \msgset{0}{} + \lv{c}{2} - (\mv{a}{2} \lv{b}{1} + \mv{b}{2} \lv{a}{1} + r_{0,1}) \\
            &= ab + a \lv{b}{1} + b \lv{a}{1} + \lv{a}{1} \lv{b}{1} + \lv{c}{2} + \msgset{0}{} \\
            &~~\hspace*{3mm}~~- a \lv{b}{1} - b \lv{a}{1} - \lv{a}{1} \lv{b}{2} - \lv{a}{2} \lv{b}{1} - r_{0,1}  \\
            &= ab + \lv{a}{1} \lv{b}{1} - \lv{a}{1} \lv{b}{2} - \lv{a}{2} \lv{b}{1} - r_{0,1} + \lv{c}{2} + \msgset{0}{} \\
            &= ab + \lv{a}{1} \lv{b}{1} - \lv{a}{1} \lv{b}{2} - \lv{a}{2} \lv{b}{1} - r_{0,1} + \lv{c}{2} + \lv{a}{2} \lv{b}{2}\\
            &~~\hspace*{4mm}~~- (\lv{a}{1} \lv{b}{1} - \lv{a}{2} \lv{b}{1} - \lv{a}{1} \lv{b}{2} + \lv{a}{2} \lv{b}{2}) + r_{0,1} \\
            &= ab + \lv{c}{2}  
\end{align*}
Similarly, $\Party{2}$ computes the following:
\begin{align*}
    \underline{\text{Party }}&\underline{\Party{2}\text{ in }\piMult~\text{(cf.~\figref{piMult3PC})}:} \\
    \mv{c}{1} &= \valset{2}{} - \msgset{1}{} \\
        &= \mv{a}{1} \mv{b}{1} + \msgset{0}{} - (\mv{a}{2} \lv{b}{1} + \mv{b}{2} \lv{a}{1} + r_{0,1} - \lv{c}{1}) \\
            &= ab + a \lv{b}{1} + b \lv{a}{1} + \lv{a}{1} \lv{b}{1} + \lv{c}{1} + \msgset{0}{} \\
            &~~\hspace*{3mm}~~- a \lv{b}{1} - b \lv{a}{1} - \lv{a}{1} \lv{b}{2} - \lv{a}{2} \lv{b}{1} - r_{0,1}  \\
            &= ab + \lv{a}{1} \lv{b}{1} - \lv{a}{1} \lv{b}{2} - \lv{a}{2} \lv{b}{1} - r_{0,1} + \lv{c}{1} + \msgset{0}{} \\
            &= ab + \lv{a}{1} \lv{b}{1} - \lv{a}{1} \lv{b}{2} - \lv{a}{2} \lv{b}{1} - r_{0,1} + \lv{c}{1} + \lv{a}{2} \lv{b}{2}\\
            &~~\hspace*{3mm}~~- (\lv{a}{1} \lv{b}{1} - \lv{a}{2} \lv{b}{1} - \lv{a}{1} \lv{b}{2} + \lv{a}{2} \lv{b}{2}) + r_{0,1} \\
            &= ab + \lv{c}{1}  
\end{align*}

%----------------------------
\subsection{$\fourpcname$: 4PC Protocol}
\label{sec:A4C}
%----------------------------
Similar to the 3PC protocol, parties locally sample the shares of masks $\lv{c}{}$ and $\lw{c}{}$ during the preprocessing of $\piMult$~(cf.~\figref{piMult4PC}). Note that $\Party{3}$ has no local computation in the online phase except the execution of compare-view functionality with $\Party{2}$. The computation of $\Partysubset{0,1,2}$ during the online phase is as follows:

\begin{align*}
    \underline{\text{Party }}&\underline{\Party{0}\text{ in }\piMult~\text{(cf.~\figref{piMult4PC})}:} \\
    \mw{c}{} &= \msgset{1,2}{} - (\valset{0}{} + \msgset{3}{}) \\
        &= \valset{1,2}{} + r_{1,2,3} - (\mw{a}{} \lv{b}{} + \mw{b}{} \lv{a}{}) - \msgset{3}{} \\
        &= \mv{a}{} \mv{b}{} - (\mw{a}{} \lv{b}{} + \mw{b}{} \lv{a}{}) - \msgset{3}{} + r_{1,2,3} \\
        &= ab + a \lv{b}{} + b \lv{a}{} + \lv{a}{} \lv{b}{} - a \lv{b}{} - b \lv{a}{} \\
        &~~\hspace*{4mm}~~- \lw{a}{} \lv{b}{} - \lw{b}{} \lv{a}{} - \msgset{3}{} + r_{1,2,3} \\
        &= ab + \lv{a}{} \lv{b}{} - \lw{a}{} \lv{b}{} - \lw{b}{} \lv{a}{}  - \msgset{3}{} + r_{1,2,3} \\
        &= ab + \lv{a}{} \lv{b}{} - \lw{a}{} \lv{b}{} - \lw{b}{} \lv{a}{} - \lv{a}{} (\lv{b}{} - \lw{b}{}) + \lv{b}{} \lw{a}{} + \lw{c}{} \\
        &= ab + \lw{c}{} 
\end{align*}

\begin{align*}
    \underline{\text{Party }}&\underline{\Party{1}\text{ in }\piMult~\text{(cf.~\figref{piMult4PC})}:} \\
    \mv{c}{} &= \valset{1}{} - \msgset{2}{}  = \mv{a}{} \mv{b}{} - \msgset{1}{} - \msgset{2}{} \\
        &= \mv{a}{} \mv{b}{} - ( \mv{a}{} \lv{b}{1} + \mv{b}{} \lv{a}{1} + r_{0,1,3}) \\
        &~~\hspace*{9mm}~~- (\mv{a}{} \lv{b}{2} + \mv{b}{} \lv{a}{2} - \msgset{0,3}{}) \\
        &= \mv{a}{} \mv{b}{} - (\mv{a}{} \lv{b}{} + \mv{b}{} \lv{a}{} + r_{0,1,3} - \msgset{0,3}{}) \\
        &= ab + a \lv{b}{} + b \lv{a}{} + \lv{a}{} \lv{b}{} - a \lv{b}{} - b \lv{a}{} \\
        &~~\hspace*{4mm}~~- \lv{a}{} \lv{b}{} - \lv{a}{} \lv{b}{} - r_{0,1,3} + \msgset{0,3}{}\\
        &= ab - \lv{a}{} \lv{b}{} - r_{0,1,3} + \msgset{0,3}{} = ab + \lv{c}{}~~\hspace*{15mm}~~ 
\end{align*}

\begin{align*}
    \underline{\text{Party }}&\underline{\Party{2}\text{ in }\piMult~\text{(cf.~\figref{piMult4PC})}:} \\
    \mv{c}{} &= \valset{2}{} - \msgset{1}{} = \mv{a}{} \mv{b}{} - \msgset{1}{} - \msgset{2}{} = ab + \lv{c}{}~\hspace*{1mm}~
\end{align*}

%----------------------------------------
\subsubsection*{Correctness of messages}
\label{sec:A4C-CV}
%----------------------------------------
While the equations above showed that the parties correctly compute their shares, the correctness is conditioned on obtaining the correct messages during the protocol execution. To show the correctness of the messages communicated during the protocol (denoted by $\msgset{\cdot}{}$ in $\piMult$), we reduce each such message to an instance of compare-view functionality instantiated using $\piCompareView$ protocol. We use $\msgerror{\cdot}{}{}$ to denote a potentially corrupted message with an error~$\error{}$.

We consider the cases of adversary $\Adv$ corrupting each parties separately and the reductions are shown below:
\begin{description}[itemsep=3mm]
    %===========================================
    \item[\underline{Case: $\Adv = \Party{0}$}]
    %--------
    \item 
    \begin{tabular}{lp{10cm}}
    Corrupted message: & $\msgerror{0,3}{}{} = \msgset{0,3}{} + \error{}$ \\
    Reduction: & $\piCompareView(\Partysubsetverify{2,3}, \msgset{0,3}{})$ 
    \end{tabular}
    %--------
    %===========================================
    \item[\underline{Case: $\Adv = \Party{1}$}]
    %--------
    \item
    \begin{tabular}{lp{10cm}}
    Corrupted message: & $\msgerror{1}{}{} = \msgset{1}{} + \error{}$ \\
    Reduction: & $\piCompareView(\Partysubsetverify{0,1,2}, \mverify{c}{})$ \\[2mm]
    \end{tabular}
    %--------
    If $e \neq 0$, \Party{2} obtains $\mverify{c}{} - \error{}$ and $\piCompareView(\Partysubsetverify{0,1,2}, \mverify{c}{})$ fails.
    %--------
    %===========================================
    \item[\underline{Case: $\Adv = \Party{2}$}]
    %--------
    \item
    \begin{tabular}{lp{10cm}}
    Corrupted message: & $\msgerror{2}{}{1} = \msgset{2}{} + \error{1}$, $\msgerror{1,2}{}{2} = \msgset{1,2}{} + \error{2}$ \\
    Reduction: & $\piCompareView(\Partysubsetverify{0,1,2}, \mverify{c}{})$, $\piCompareView(\Partysubsetverify{0,1}, \msgset{1,2}{})$ \\[2mm]
    \end{tabular}
    %--------
    If $\error{2}{} \neq 0$, $\piCompareView(\msgset{1,2}{},\Partysubsetverify{0,1})$ fails. If $\error{2}{} = 0$ but $\error{1}{} \neq 0$, \Party{1} obtains $\mverify{c}{} - \error{1}{}$ and $\piCompareView(\mverify{c}{},\Partysubsetverify{0,1,2})$ fails.  Hence, any errors $\error{1}{} \neq 0 \lor \error{2}{} \neq 0$ leads to $\abort$.\\[0.5mm]
    %--------
    %===========================================
    \item[\underline{Case: $\Adv = \Party{3}$}]
    %--------
    \item
    \begin{tabular}{lp{10cm}}
    Corrupted message: & $\msgerror{3}{}{} = \msgset{3}{} + \error{}$ \\
    Reduction: & $\piCompareView(\Partysubsetverify{0,1,2}, \mverify{c}{})$ \\[2mm]
    \end{tabular}
    %--------
    If $e \neq 0$, \Party{0} obtains $\mverify{c}{} - \error{}$ and $\piCompareView(\Partysubsetverify{0,1,2}, \mverify{c}{})$ fails.
    %--------
    %===========================================
\end{description}
    
%----------------------------
\subsection{Heterogeneous 4PC Protocol}
\label{sec:A4CH}
%----------------------------
Similar to analysis of $\piMult$ in~\secref{A4C}, parties locally sample the shares of $\lv{c}{}$ and $\lw{c}{}$ during the preprocessing of $\piMultH$~(cf.~\figref{piMult4PCH}). Also, $\Party{3}$ has no local computation in the online phase except invocations of $\piCompareView$ with $\Party{0}$ and $\Party{2}$. The computation of $\Partysubset{0,1,2}$ during the online phase is as follows:
\begin{align*}
    \Party{0}: \mw{c}{} &= \msgset{1,2}{1} - \lv{c}{} = \mv{c}{} + \lw{c}{} - \lv{c}{} = ab + \lw{c}{} \\
    \Party{1}: \mv{c}{} &= \valset{1,2}{} - \msgset{2}{} = \mv{a}{} \mv{b}{} - \msgset{1}{} - \msgset{2}{} = ab + \lv{c}{} \\
    \Party{2}: \mv{c}{} &= \valset{1}{} - \msgset{1}{} = \mv{a}{} \mv{b}{} - \msgset{1}{} - \msgset{2}{} = ab + \lv{c}{} 
\end{align*}

%----------------------------------------
\subsubsection*{Correctness of messages}
\label{sec:A4CH-CV}
%----------------------------------------
To ensure that all corrupted messages~$\msgerror{\cdot}{}{}$ are caught during verification, we reduce the correctness of messages to an instance of $\piCompareView$, similar to $\piMult$ in~\secref{A4C-CV}.

\begin{description}
    %===========================================
    \item[\underline{Case: $\Adv = \Party{0}$}]
    %--------
    \item 
    \begin{tabular}{lp{10cm}}
    Corrupted message: & $\msgerror{0,3}{}{} = \msgset{0,3}{} + \error{}$ \\
    Reduction: & $\piCompareView(\Partysubsetverify{2,3}, \msgset{0,3}{})$ \\[2mm]
    \end{tabular}
    %--------
    %===========================================
    \item[\underline{Case: $\Adv = \Party{1}$}]
    %--------
    \item
    \begin{tabular}{lp{10cm}}
    Corrupted message: & $\msgerror{1}{}{} = \msgset{1}{} + \error{}$ \\
    Reduction: & $\piCompareView(\Partysubsetverify{0,3}, \valset{0,3}{})$ \\[3mm]
    \end{tabular}
    %--------
    If $\error{} \neq 0$, $\Party{2}$ obtains $\msgerror{1,2}{2}{} = \msgset{1,2}{2} + \error{}$ and sends it to $\Party{0}$, who computes $\valerror{0,3}{}{}$ based on $\msgerror{1,2}{2}{}$. The following equation shows that $\piCompareView(\Partysubsetverify{0,3}, \valset{0,3}{})$ fails in this case.
    \begin{align*}
        \valerror{0,3}{}{} 
        &= \msgerror{1,2}{2}{} - \valset{0}{} \\
        &= a \lv{b}{} + b \lv{a}{} + \lv{a}{} \lv{b}{} - \lv{c}{} + r_{1,2,3} + \error{}  - \mw{a}{} \lv{b}{} - \mw{b}{} \lv{a}{}  \\
        &= \lv{a}{} \lv{b}{} - \lw{a}{} \lv{b}{} - \lw{b}{} \lv{a}{} + r_{1,2,3} - \lv{c}{} + \error{}  = \valset{0,3}{} + \error{} 
    \end{align*}
    %--------
    %===========================================
    \item[\underline{Case: $\Adv = \Party{2}$}]
    %--------
    \item
    \begin{tabular}{lp{10cm}}
    Corrupted message: 
    &$\msgerror{2}{}{1} = \msgset{2}{} + \error{1}$\\
    &$\msgerror{1,2}{1}{2} = \msgset{1,2}{1} + \error{2}$ \\
    &$\msgerror{1,2}{2}{3} = \msgset{1,2}{2} + \error{3}$ \\
    Reduction: 
    & $\piCompareView(\Partysubsetverify{0,3}, \valset{0,3}{})$\\ &$\piCompareView(\Partysubsetverify{0,1}, \msgset{1,2}{1})$, $\piCompareView(\Partysubsetverify{0,1}, \msgset{1,2}{2})$
    \\[3mm]
    \end{tabular}
    %--------
    If $\error{1}{} \neq 0$, honest $\Party{1}$ obtains $\msgset{1,2}{1} - \error{1}$ and $\msgset{1,2}{2} - \error{1}$. In that case, any assignment other than $\error{1} = \error{2} = \error{3}$ leads to abort due to $\piCompareView(\Partysubsetverify{0,1}, \msgset{1,2}{1})$ or $\piCompareView(\Partysubsetverify{0,1}, \msgset{1,2}{2})$. Assigning $\error{1} = \error{2} = \error{3} \neq 0$ leads to $\Party{0}$ obtaining a corrupted $\valset{0,3}{}$. Hence, $\piCompareView(\Partysubsetverify{0,3}, \valset{0,3}{})$ fails. \\
    If $\error{1}{} = 0$, $\Party{1}$ obtains $\msgset{1,2}{1}$ and $\msgset{1,2}{2}$. Then, $\error{2}{} \neq 0$ or $\error{3}{} \neq 0$ leads to abort due to $\piCompareView(\Partysubsetverify{0,1}, \msgset{1,2}{1})$ or $\piCompareView(\Partysubsetverify{0,1}, \msgset{1,2}{2})$.\\[0.5mm]
    %--------%===========================================
    \item[\underline{Case: $\Adv = \Party{3}$}]
    %--------
    \item $\Party{3}$ does not communicate any message of the form ~$\msgerror{\cdot}{}{}$.
    %===========================================
\end{description}
%============================
\section{Additional Details}
\label{sec:additional}
%============================
This section provides the formal details for the protocols discussed in \secref{4PC} and \secref{add}. We begin with the protocols for input sharing and output reconstruction in our 4PC protocol $\fourpcname$. Next, we show how to extend the properties of our protocols—such as high tolerance to weak network links, low computational complexity, and best-known communication complexity—to other protocols like multiplication with truncation, comparison, and bit conversions, as discussed in \secref{add}.

%----------------------
\subsection{$\fourpcname$: Handling Inputs and Outputs in 4PC}
\label{sec:4PC-IO-protocols}
%----------------------
Protocol $\piShare$ in~\figref{piShare} provides the details for input sharing in our 4PC protocol $\fourpcname$, where $\Party{I}$ holds the secret input $x \in \Z{\ell}$ to be shared among $\Partysubset{0,1,2,3}$. 

\begin{protocolbox}{$\piShare(\Partysubset{I},x) \rightarrow \sh{a}$}{$\fourpcname$: 4PC Secret Sharing}{fig:piShare}
	%=========================
	\justify 
	\algoHeadacmart{Setup:} 
	\begin{enumerate}[itemsep=0mm]
		%-------
        \item Invoke $\Funcsetup$ to obtain the following PRF keys:
		\begin{align*}
            &\Partysubset{I,0,1,3}: \key{I,0,1,3}  &   \Partysubset{I,0,2,3}: \key{I,0,2,3} & &\Partysubset{I,1,2,3}: \key{I,1,2,3} 
		\end{align*}
        %-------
    \end{enumerate}
    %=========================
    \justify
    \algoHeadacmart{Preprocessing:} 
    \begin{enumerate}[itemsep=0mm]
        %-------
		\item Sample random values using $\piSRNG$~(\figref{srng}):
		\begin{align*}            
            &\Partysubset{I,1,2,3}^V: \lw{x}{}  &
            &\Partysubset{I,0,1,3}: \lv{x}{1} &
            &\Partysubset{I,0,2,3}: \lv{x}{2}  
		\end{align*}
        %-------
		\item Locally compute:
		\begin{align*}
            &\Partysubset{I,0,3}^V: \lv{x}{} = \lv{x}{1} + \lv{x}{2}   
		\end{align*}
        %-------
	\end{enumerate}
    %=========================
	\justify
    \algoHeadacmart{Online:} 
	\begin{enumerate}[itemsep=0mm]
        %-------
        \item Communicate:
		\begin{align*}
            &\Partysubset{I} \rightarrow \Partysubset{0,1,2}: \mverify{x}{} = x + \lv{x}{} + \lw{x}{} 
		\end{align*}
        %-------            
        \item Compare views using $\piCompareView$~(\figref{compareview}):
        \begin{align*}
            &\Partysubsetverify{0,1,2}: \mverify{x}{} 
		\end{align*}
		%-------
		\item Locally compute:
		\begin{align*}
            &\Partyverify{0}: \mw{x}{} = \mverify{x}{} -  \lv{x}{}
            &\Partysubset{1,2}: \mv{x}{} = \mverify{x}{} - \lw{x}{} 
        \end{align*}
		%------
	\end{enumerate}
    %=========================
    \hrule\smallskip
    \small \centering \textit{$\Partysubset{I}$ denotes the input party holding secret $x \in \Z{\ell}$.}
\end{protocolbox}
%------------------

Protocol $\piRec$ in~\figref{piRec} allows the parties in $\Partysubset{0,1,2,3}$ to reconstruct a shared secret $\sh{x}$ towards a designated party $\Party{O}$. This protocol trivially extends to a public reconstruction among $\Partysubset{0,1,2,3}$, where each party in $\Partysubset{0,1,2,3}$ is assigned the role of $\Party{O}$. 

\begin{protocolbox}{$\piRec(\Partysubset{O},\sh{x}) \rightarrow x$}{$\fourpcname$: 4PC Reconstruction}{fig:piRec}
    %=========================
	\justify 
    \algoHeadacmart{Postprocessing:} 
    \begin{enumerate}[itemsep=0mm]
      %-------
      \item Communicate:
      \begin{align*} 
        &\Party{2} \rightarrow \Party{O}: \mv{x}{} &
        &\Party{0} \rightarrow \Party{O}: \lv{x}{}    
      \end{align*}
      %-------
      \item Locally compute:
	  \begin{align*}
        &\Party{O}: x = \mv{x}{} - \lv{x}{} 
      \end{align*}
      %-------
      \item Compare views using $\piCompareView$~(\figref{compareview}):
      \begin{align*}
        &\Partysubset{O,1}^V: \mv{x}{}
        &\Partysubset{O,3}^V: \lv{x}{} 
      \end{align*}
      %-------
    \end{enumerate}
	%=========================
    \hrule\smallskip
    \small \centering \textit{$\Party{O}$ denotes the output party to obtain the secret $x$. }
\end{protocolbox}
%------------------

In the reconstruction protocol $\piRec$~(cf.~\figref{piRec}) discussed above, an adversary $\Adv$ can selectively $\abort$ for some $\Party{O}$ by sending incorrect messages during $\piCompareView$ within $\piRec$. To ensure fairness during a public reconstruction, we provide a fair reconstruction variant $\piFairRec$ in~\figref{piFairRec}, similar to Tetrad~\cite{NDSS:KotiPRS22}.

In $\piFairRec$, we let the parties $\Partysubset{0,1,2,3}$ store the actual secret-shares for all the output wires, instead of storing only the persistent shares~(cf.~\tabref{4pc-secret-sharing}). This modification provides the advantage that each of the following values---$\mverify{x}{}$, $\lw{x}{}$, $\lv{x}{1}$, $\lv{x}{2}$---will be held by three out of the four parties. Given that there is at most one corruption, this guarantees a honest majority of each of these values among the parties and we leverage the same. 

Similar to Tetrad, each party maintains an \emph{aliveness} bit indicating whether some of the verification prior to reconstruction failed or not. Parties mutually exchange this bit and accepts the majority. If the agreement is $\continue$, parties proceed with sending the values to $\Party{O}$ who computes output as $x = \mverify{x}{} - \lw{x}{} - \lv{x}{1} - \lv{x}{2}$.

\begin{protocolbox}{$\piFairRec(\Partysubset{O},\sh{x}) \rightarrow x$}{$\fourpcname$: 4PC Fair Reconstruction}{fig:piFairRec}
    %=========================
	\justify 
    \algoHeadacmart{Postprocessing:} 
    \begin{enumerate}[itemsep=0mm]
      %-------
      \item $\Partysubset{0,1,2,3}$ sets their aliveness bit $\bitb = 0$, if any of their verification during $\piCompareView$~(\figref{compareview}) fails. Else, set $\bitb = 1$.
      %-------
      \item $\Partysubset{0,1,2,3}$ mutually exchanges their 
      bit $\bitb$ and $\continue$, if there is a majority with $\bitb = 1$. Else, $\abort$.
      %-------
      \item Communicate:
      \begin{align*} 
        &\Partysubset{0,1,2} \rightarrow \Party{O}: \mverify{x}{} &
        &\Partysubset{1,2,3} \rightarrow \Party{O}: \lw{x}{} &\\
        &\Partysubset{0,1,3} \rightarrow \Party{O}: \lv{x}{1} &
        &\Partysubset{0,2,3} \rightarrow \Party{O}: \lv{x}{2} 
      \end{align*}
      %-------
      \item Locally compute using the majority values:
	  \begin{align*}
        &\Party{O}: x = \mverify{x}{} - \lw{x}{} - \lv{x}{1} - \lv{x}{2} 
      \end{align*}
      %-------
    \end{enumerate}
	%=========================
    \hrule\smallskip
    \small \centering \textit{$\Party{O}$ denotes the output party to obtain the secret $x$. }
\end{protocolbox}
%------------------

%----------------------
\subsection{Other Protocols}
\label{sec:other-protocols}
%----------------------
This section covers the formal details for the protocols discussed in~\secref{add}. The costs for additional protocols are shown in table \ref{tab:additional_protocols}.

\begin{table*}[htbp]
\small
\begin{threeparttable}[htbp]
\centering
\captionsetup{font=small}
\caption{Communication complexity of additional protocols}
\label{tab:additional_protocols}
\begin{tabular}{lllll}
\hline
Protocol & Scheme & Preprocessing & Online & Rounds \\ \hline

\multirow{2}{*}{Multiplication + Truncation} 
& 3PC & 1 & 2 & 1 \\ %\cline{2-5} 
& 4PC & 2 & 3 & 1 \\ \hline

\multirow{2}{*}{Arithmetic to Boolean (protocol-specific primitive) $\boldsymbol{+}$ Ripple Carry Adder\tnote{a}} 
& 3PC & $1 \boldsymbol{+}  (\ell - 1)$ & $0 \boldsymbol{+}  (2\ell - 1)$ & $0 \boldsymbol{+}  (\ell - 1)$ \\ %\cline{2-5} 
& 4PC & $1 \boldsymbol{+}  2(\ell - 1)$ & $1 \boldsymbol{+}  (3\ell - 1)$ & $0 \boldsymbol{+}  (\ell - 1)$ \\ \hline

\multirow{2}{*}{Arithmetic to Boolean (protocol-specific primitive) $\boldsymbol{+}$ Parallel Prefix Adder\tnote{a}} 
& 3PC & $1 \boldsymbol{+} O(\ell \cdot \log_2 \ell)$ & $0 \boldsymbol{+}  2 \cdot O(\ell \cdot \log_2 \ell)$ & $0 \boldsymbol{+}  (\log_2 \ell + 1)$ \\ %\cline{2-5} 
& 4PC & $1 \boldsymbol{+}  2 \cdot O(\ell \cdot \log_2 \ell)$ & $1 \boldsymbol{+}  3 \cdot O(\ell \cdot \log_2 \ell)$ & 0 $\boldsymbol{+} (\log_2 \ell + 1)$ \\ \hline

\multirow{2}{*}{Bit to Arithmetic (protocol-specific primitive) $\boldsymbol{+}$ Arithmetic XOR\tnote{a}} 
& 3PC & $1 \boldsymbol{+} 1$ & $0 \boldsymbol{+} 2$ & $0 \boldsymbol{+} 1$ \\ %\cline{2-5} 
& 4PC & $1 \boldsymbol{+} 2$ & $1 \boldsymbol{+} 2$ & $0 \boldsymbol{+} 1$ \\ \hline
                                    
\end{tabular}
\begin{tablenotes}
 \small
 \item $^{\text{a}}$ Protocol-specific primitive (c.f. \secref{other-A2B} and \secref{other-bit2A}) $\boldsymbol{+}$ high-level circuit implemented with elementary gates.
 \item 
 \end{tablenotes}
\end{threeparttable}
\end{table*}

%----------------------
\subsubsection{Truncation}
\label{sec:other-truncation}
%----------------------

Let $\ftrunc{x}$ denote $\lfloor \frac{x}{2^t} \rfloor$. Let $\ptrunc{\mv{x}{}}$ denote probabilistic truncation as defined by \cite{SP:MohZha17},
To truncate-multiply two values $a$ and $b$ in our schemes scheme, $\Partyset$ need to obtain a share of $\ftrunc{\mv{c}{}} = \ftrunc{ab + \lv{c}{}}$ and $\ftrunc{\lv{c}{}}$ for a random mask $\lv{c}{}$.  
$\ftrunc{\mv{c}{}}$ and $\ftrunc{\lv{c}{}}$ then form a truncation pair where $\ftrunc{\mv{c}{}} - \ftrunc{\lv{c}{}} = \ptrunc{c}  \approx \ftrunc{c}$ with high probability.

%----------------------
\paragraph{3PC}
%----------------------
In the following explanation, we use three sequential steps to explain the intuition of our 3PC truncation protocol shown in w \ref{fig:piMult3T}: First, $\Party{0}$ calculates an input-independent error, similar to our general multiplication protocol. $\Party{0}$ then locally truncates the input-dependent error to obtain $\ftrunc{\lv{c}{}}$. Then, $\Party{0}$ shares $\ftrunc{\lv{c}{}}$ among $\Partyset$.  Second, $\Partysubset{1,2}$ exchange messages to obtain $ab + \lv{c}{}$ in the clear. They proceed to locally truncate ${ab + \lv{x}{}}$. To share $\ftrunc{ab + \lv{c}{}}$ among $\Partyset$, all parties simply set their input-independent shares to $0$. 
As the parties now hold both shares, they perform the final step by subtracting their local shares to obtain $\sh{\ptrunc{ab}} = \sh{\ftrunc{ab + \lv{c}{}}} - \sh{\ftrunc{\lv{c}{}}}$. While this three-step procedure explains the intuition of our formal protocol in Figure \ref{fig:piMult3T}, the actual protocol does not create the two shares explicitly but fuses them into a single one to achieve a more efficient construction.

\begin{protocolbox}{$\piMult{_{3T}}(\sh{a}, \sh{b}) \rightarrow \sh{\ptrunc{c}}$}{$\threepcname$: 3PC multiplication with truncation}{fig:piMult3T}
    
	%----
	\justify 
	
	\algoHeadacmart{Preprocessing:} 
	\begin{enumerate}[itemsep=0mm]
		%------
		\item Sample random values using $\piSRNG$:
		\begin{align*}
            &P_{0,1}: r_{0,1}, \lv{c}{1} &
            &P_{0,2}: r_{0,2} 
		\end{align*}
        
		\item Locally compute:
		\begin{align*}
            &\Party{0}: \lv{c}{2} = \msgset{0}{} = \ftrunc{(\lv{a}{1} - \lv{a}{2}) (\lv{b}{1} - \lv{b}{2}) - \lv{a}{2} \lv{b}{2} + r_{0,1} + r_{0,2}} - \lv{c}{1}   
		\end{align*}
        
        \item Communicate:
		\begin{align*}
            &\Party{0} \rightarrow \Party{2}: \msgset{0}{} 
		\end{align*}
		%------
		%------
	\end{enumerate}
	\justify
	
    \algoHeadacmart{Online:} 
	\begin{enumerate}[itemsep=0mm]
		%-------
		\item Locally compute:
		\begin{align*}
            &\Party{1}: \msgset{1}{} = \mv{a}{2} \lv{b}{1} + \mv{b}{2} \lv{a}{1} - r_{0,1} &
            \Party{2}: \msgset{2}{} = \mv{a}{1} \mv{b}{1} + r_{0,2} 
		\end{align*}
        
		%------
        \item Communicate:
		\begin{align*}
            &\Party{1} \rightarrow \Party{2}: \msgset{1}{} &
            \Party{2} \rightarrow \Party{1}: \msgset{2}{} 
		\end{align*}
        
		%-------
		\item Locally compute:
		\begin{align*}
            &\Party{1}: \mv{c}{2} = \ftrunc{\msgset{2}{} - \msgset{1}{}} - \lv{c}{1} &\Party{2}: \mv{c}{1} = \ftrunc{\msgset{2}{} - \msgset{1}{}} - \msgset{0}{}  \\
            &\Party{2}: \lv{c}{2} = \msgset{0}{}   & 
		\end{align*}
        
		%------
	\end{enumerate}     
\end{protocolbox}
%------------------

Nevertheless, one can see that step 2-3 in the preprocessing phase are mainly responsible for sharing the locally-truncated input-independent error among $\Partysubset{1,2}$. Steps 1 and 2 in the online phase are responsible for letting $\Partysubset{1,2}$ obtain 
$ab + \lv{c}{}$ in the clear.
Finally, in step 3, $\Partysubset{1,2}$ parties locally truncate $ab + \lv{c}{}$ and perform the subtraction using their fused shares.

%----------------------
\paragraph{4PC}
%----------------------
Truncation also comes for free in our 4PC schemes. To truncate-multiply a value in our 4PC schemes,
\Party{0} shares the locally truncated $\ftrunc{\lv{c}{}}$ with \Party{1} and \Party{2} similarly to our 3PC protocol. $\Party{3}$ verifies this message.

As in our 3PC protocol, $ \Party{1} $ and $ \Party{2} $ obtain $ ab + \lv{c}{} $ in the clear. To do so, they exchange $ \msgset{1}{} = \mv{a}{} \lv{b}{1} + \mv{b}{} \lv{a}{1} - r_{0,1,3} $ and $ \msgset{1,2}{} = \mv{a}{} \lv{b}{2} + \mv{b}{} \lv{a}{2} - r_{0,2,3} $. They locally compute $ \mv{a}{} \mv{b}{} - \msgset{1}{} - \msgset{1,2}{} $ to obtain $ \mv{c}{} = ab - \lv{a}{} \lv{b}{} + r_{0,1,3} + r_{0,2,3} = ab + \lv{c}{} $. They then locally truncate $ \mv{c}{} $. In our 4PC protocol, $ \Party{0} $ also needs to obtain an input-dependent share for verification purposes.Thus, $\Partysubset{2}$ sends a masked message $\msgset{1,2}{}$ (\msgset{1,2}{1} in the heterogeneous variant) to $\Party{0}$. Observe that $\msgset{1,2}{}$ already contains both truncation pairs.

The verification of exchanged messages is handled analogously to our general multiplication protocol. Figures \ref{fig:piMultiplication4_T} and \ref{fig:piMultiplicationH_T} implement the presented intuition in a computationally efficient way.

\begin{protocolbox}{$\piMult{_{4T}}(\vl{\sh{a}}, \vl{\sh{b}}) \rightarrow \sh{\ptrunc{c}}$}{$\fourpcname$: 4PC multiplication with truncation}{fig:piMultiplication4_T}
    
	%----
	\justify 
	
	\algoHeadacmart{Preprocessing:} 
	\begin{enumerate}[itemsep=0mm]
		%------
		\item Sample random values using $\piSRNG$:
		\begin{align*}
            &P_{0,1,3}: r_{0,1,3}, \lv{c}{1} &
            &P_{0,2,3}: r_{0,2,3} &
            &P_{1,2,3}^V: r_{1,2,3}, \lw{c}{} 
		\end{align*}
        
		\item Locally compute:
		\begin{align*}
            &\Party{0},\Partyverify{3}: \lv{c}{} = \ftrunc{r_{0,1,3} + r_{0,2,3} - \lv{a}{} \lv{b}{}}  \\
            &\Party{0},\Partyverify{3}: \msgset{0,3}{} = \lv{c}{} - \lv{c}{1}  \\
            &\Partyverify{3}: \msgset{3}{} = \lv{a}{} (\lv{b}{} - \lw{b}{}) - \lv{b}{} \lw{a}{} - r_{0,1,2} - r_{0,2,3} + r_{1,2,3}    
		\end{align*}
        
        \item Communicate:
		\begin{align*}
            &\Party{0} \rightarrow \Party{2}: \msgset{0,3}{} \\
            &\Partyverify{3} \rightarrow \Partyverify{0}: \msgset{3}{} 
		\end{align*}
        
		%------
		%------
	\end{enumerate}
	\justify
	
    \algoHeadacmart{Online:} 
	\begin{enumerate}[itemsep=0mm]
		%-------
		\item Locally compute:
		\begin{align*}
            &\Party{1}: \msgset{1}{} = \mv{a}{} \lv{b}{1} + \mv{b}{} \lv{a}{1} - r_{0,1,3} &
             &\Partysubset{1,2}: \valset{1,2}{1} = \mv{a}{} \mv{b}{} \\
            &\Party{2}: \msgset{2}{} = \mv{a}{} \lv{b}{2} + \mv{b}{} \lv{a}{2} - r_{0,2,3} &
            &\Party{2}: \lv{c}{2} = \msgset{0,3}{} \\
            &\Partyverify{0}: \valset{0,1,2}{} = \mw{a}{} \lv{b}{} + \mw{b}{} \lv{a}{} + \lv{a}{} \lv{b}{} + \msgset{3}{} 
		\end{align*}
        
		%------
        \item Communicate:
		\begin{align*}
            &\Party{1} \rightarrow \Party{2}: \msgset{1}{}&
            &\Party{2} \rightarrow \Party{1}: \msgset{2}{} 
		\end{align*}
		\item Locally compute:
		\begin{align*}       
            &\Partysubset{1,2}: \valset{1,2}{2} = \msgset{1}{} + \msgset{2}{} &
            &\Partysubset{1,2}: \mv{c}{} = \ftrunc{\valset{1,2}{1} - \valset{1,2}{2}}  \\
            &\Partysubsetverify{1,2}: \msgset{1,2}{} = \mv{c}{} + \lw{c}{} &
            &\Partysubsetverify{1,2}: \valset{0,1,2}{} = \valset{1,2}{2} + r_{1,2,3} 
        \end{align*}

    \item Communicate:
		\begin{align*}
            &\Partyverify{2} \rightarrow \Partyverify{0}: \msgset{1,2}{} 
		\end{align*}

    \item Locally compute:
		\begin{align*}
            &\Partyverify{0}: \mw{c}{} = \msgset{1,2}{} - \lv{c}{} 
        \end{align*}
        
        \item Compare views using $\piCompareView$:
        \begin{align*}
            &\Partysubsetverify{0,1}: \msgset{1,2}{} &
            &\Partysubsetverify{0,1,2}: \valset{0,1,2}{} &
            &\Partysubsetverify{2,3}: \msgset{0,3}{}
		\end{align*}

		%------
	\end{enumerate}
\end{protocolbox}
%------------------

\begin{protocolbox}{$\piMult{_{4HT}}(\vl{\sh{a}}, \vl{\sh{b}}) \rightarrow \sh{\ptrunc{c}}$}{$\fourpcname$: 4PC heterogeneous multiplication protocol with truncation}{fig:piMultiplicationH_T}
    
	%----
	\justify 
	
	\algoHeadacmart{Preprocessing:} 
	\begin{enumerate}[itemsep=0mm]
		%------
		\item Sample random values using $\piSRNG$:
		\begin{align*}
            &P_{0,1,3}: r_{0,1,3}, \lv{c}{1} &
            &P_{0,2,3}: r_{0,2,3} &
            &P_{1,2,3}^V: r_{1,2,3}, \lw{c}{} 
		\end{align*}
        
		\item Locally compute:
		\begin{align*}
            &\Party{0},\Partyverify{3}: \lv{c}{} = \ftrunc{r_{0,1,3} + r_{0,2,3} - \lv{a}{} \lv{b}{}}  \\
            &\Party{0},\Partyverify{3}: \msgset{0,3}{} = \lv{c}{} - \lv{c}{1}  \\
            &\Partyverify{3}: \valset{0,3}{} = \lv{a}{} (\lv{b}{} - \lw{b}{}) - \lv{b}{} \lw{a}{} - r_{0,1,2} - r_{0,2,3} + r_{1,2,3}    
		\end{align*}
        
        \item Communicate:
		\begin{align*}
            &\Party{0} \rightarrow \Party{2}: \msgset{0,3}{} 
		\end{align*}
        
		%------
		%------
	\end{enumerate}
	\justify
	
    \algoHeadacmart{Online:} 
	\begin{enumerate}[itemsep=0mm]
		%-------
		\item Locally compute:
		\begin{align*}
            &\Party{1}: \msgset{1}{} = \mv{a}{} \lv{b}{1} + \mv{b}{} \lv{a}{1} - r_{0,1,3} &
            &\Party{2}: \msgset{2}{} = \mv{a}{} \lv{b}{2} + \mv{b}{} \lv{a}{2} - r_{0,2,3} \\
            &\Party{1}: \valset{1}{} = \mv{a}{} \mv{b}{} - \msgset{1}{} &
            &\Party{2}: \valset{2}{} = \mv{a}{} \mv{b}{} - \msgset{1,2}{} \\
            &\Partyverify{0}: \valset{0}{} = \mw{a}{} \lv{b}{} + \mw{b}{} \lv{a}{} + \lv{a}{} \lv{b}{} &
            &\Party{2}: \lv{c}{2} = \msgset{0,3}{} 
		\end{align*}
        
		%------
        \item Communicate:
		\begin{align*}
            &\Party{1} \rightarrow \Party{2}: \msgset{1}{} &
            &\Party{2} \rightarrow \Party{1}: \msgset{2}{} 
		\end{align*}
        
		%-------
		\item Locally compute:
		\begin{align*}
            &\Party{1}: {\mv{c}{}}= \ftrunc{\valset{1}{} - \msgset{2}{}}  &
            &\Party{2}: {\mv{c}{}}= \ftrunc{\valset{2}{} - \msgset{1}{}}  \\
            &\Partysubsetverify{1,2}: \msgset{1,2}{1} = {\mv{c}{}} + \lw{c}{} &
            &\Partysubsetverify{1,2}: \msgset{1,2}{2} = \msgset{1}{} + \msgset{2}{} + r_{1,2,3} 
        \end{align*}

    \item Communicate:
		\begin{align*}
            &\Partyverify{2} \rightarrow \Partyverify{0}: \msgset{1,2}{1} &
            &\Partyverify{2} \rightarrow \Partyverify{0}: \msgset{1,2}{2} 
		\end{align*}

    \item Locally compute:
		\begin{align*}
            &\Partyverify{0}: \valset{0,3}{} = \msgset{1,2}{2} - \valset{0}{} &
            &\Partyverify{0}: \mw{c}{} = \msgset{1,2}{1} - \lv{c}{} 
        \end{align*}
        
        \item Compare views using $\piCompareView$:
        \begin{align*}
            &\Partysubsetverify{0,1}: \msgset{1,2}{1}, \msgset{1,2}{2} &
            &\Partysubsetverify{0,3}: \valset{0,3}{} &
            &\Partysubsetverify{2,3}: \msgset{0,3}{}
		\end{align*}

		%------
	\end{enumerate}     
\end{protocolbox}
%------------------

%----------------------
\subsubsection{Matrix Multiplication}
\label{sec:other-matrix}
%----------------------
As observed in several works~\cite{SP:MohZha17,CCS:MohRin18,NDSS:PatSur20,SP:BSSSY24}, a matrix multiplication $C = A \cdot B$ can be trivially computed with an existing MPC multiplication protocol that operates correctly on single values. Suppose the parties hold $\sh{A}$ and $\sh{B}$. In that case, they execute the protocol regularly but compute a local matrix addition for each addition operation and a local matrix multiplication for each multiplication operation. Let $|C|$ be the total number of elements in the output matrix $C$. For each call to $\piSRNG$ the parties need to sample  $|C|$ elements to ensure unique masks for every value. Every message in the protocol also becomes of size $|C|$. Since scalar products have an output size of $1$, the cost of a scalar multiplication is the same as the cost of a regular multiplication.  

%----------------------
\subsubsection{Comparisons}
\label{sec:other-comparisons}
%----------------------
To evaluate a comparison $\sh{a} > \sh{b}$ we use the following established sequence proposed by \cite{CCS:MohRin18}. First, the parties calculate $\sh{z} = \sh{b} - \sh{a}$. Note that the sign bit of $c$ is 1 if $a > b$, and $0$ otherwise. By converting $\sh{c}$ into a Boolean share, the parties can extract a share of its sign bit. Afterward, the parties can convert the share of the sign bit back to an arithmetic share to use the result of the comparison. 

Both conversions require a protocol-specific primitive followed by a protocol-agnostic high-level circuit that can be implemented using multiplication and addition gates in $\Z{}$ or $\Z{\ell}$. The protocol-specific primitives convert a single share from either the Boolean or the arithmetic domain to two separate shares in the target computation domain. These two separate shares then get combined into a single share of the target domain with the respective high-level circuit. In both our 3PC and 4PC schemes, the protocol-specific primitive for switching computation domains are part of non-latency critical communication and thus do not add to the round complexity. 

%----------------------
\subsubsection{Arithmetic to Boolean}
\label{sec:other-A2B}
%----------------------
To convert an arithmetic share $\sh{a}^A$ to a Boolean share $\sh{a}^B$, the parties compute Boolean shares of $b = \sh{a + \lv{a}{}}^B$ and $c = \sh{- \lv{a}{}}^B$. The parties then use a Boolean adder to compute $\sh{b}^B + \sh{c}^B = \sh{a + \lv{a}{}}^B + \sh{-\lv{a}{}]^B}$ to receive an XOR-sharing of $\sh{a}^B$. If the parties only wish to obtain a single bit of $\sh{a}$, they can use a carry-out adder instead of a Boolean adder. Figures \ref{fig:A2B3} and \ref{fig:A2B4} show the formal protocols for our 3PC and 4PC protocols respectively.

%----------------------
\paragraph{3PC}
%----------------------
To compute a share of $b^B = \sh{-\lv{a}{}}^B$, \Partysubset{0,1} sample $r_{0,1}$ and \Party{0} sends $\msgset{0}{} = \sh{-\lv{a}{}}^B \oplus r_{0,1}$ to \Party{2} in the preprocessing phase. The parties then define their shares as shown in Figure \ref{fig:A2B3}. In the online phase, each party locally computes a share of $b = \sh{a +\lv{a}{}}^B$. The parties proceed to compute $\sh{a}^B = \sh{a + \lv{a}{}}^B + \sh{-\lv{a}{}}^B$ using a Boolean adder.
To verify the correctness, observe that $\sh{-\lv{a}{}} = \mv{c}{1} \oplus \lv{c}{1}$ and $\sh{-\lv{a}{}} = \mv{c}{2} \oplus \lv{c}{2}$. Likewise, $\sh{a + \lv{a}{}} = \mv{b}{1} \oplus \lv{b}{1}$ and $\sh{a + \lv{a}{}} = \mv{b}{2} \oplus \lv{b}{2}$. Hence, the parties successfully created two Boolean shares $\sh{b}$ and $\sh{c}$ such that $\sh{a}^B = \sh{b}^B + \sh{c}^B$.

\begin{protocolbox}{$\piAtoB{_{,3PC}}(\vl{\sh{a}^A}) \rightarrow \sh{a}^B$}{$\threepcname$: 3PC Arithmetic to Binary Conversion}{fig:A2B3}
    
	%----
	\justify 
	
	\algoHeadacmart{Preprocessing:} 
	\begin{enumerate}[itemsep=0mm]
		%------
		\item Sample random values using $\piSRNG$:
		\begin{align*}
            &\Partysubset{0,1}: r_{0,1} 
		\end{align*}
        
            \item Communicate:
		\begin{align*}
            &\Party{0} \rightarrow \Party{2}: \msgset{0}{} = (- \lv{a}{}) \oplus r_{0,1} 
		\end{align*}
            
		\item Locally compute:
		\begin{align*}
            &\Partysubset{0,1}: \lv{b}{1} = 0 &
            &\Partysubset{0,2}: \lv{b}{2} = 0 &
            &\Partysubset{0,1}: \lv{c}{1} = r_{0,1} &
            &\Partysubset{0,2}: \lv{c}{2} = \msgset{0}{} 
		\end{align*}
  \end{enumerate} 
	\justify
	
    \algoHeadacmart{Online:} 
	\begin{enumerate}[itemsep=0mm]
		%-------
		\item Locally compute:
		\begin{align*}
            &\Party{1}: \mv{b}{2} = \mv{a}{2} + \lv{a}{1} &
            &\Party{2}: \mv{b}{1} = \mv{a}{1} + \lv{a}{2} \\
            &\Party{1}: \mv{c}{2} = \lv{c}{1} & 
            &\Party{2}: \mv{c}{1} = \lv{c}{2} 
		\end{align*}
            \end{enumerate} 
          \begin{center}
          \rule{\textwidth}{0.2pt} \\ % horizontal line
          End of protocol-specific primitive \\[-0.5em]
          \rule{\textwidth}{0.2pt} \\ % horizontal line
          \end{center}
          \centering Jointly compute:
             \begin{align*}
            &\Partyset: \sh{a}^B = \sh{b}^B + \sh{c}^B 
		\end{align*}
             
		%------   
\end{protocolbox}
%------------------

%----------------------
\paragraph{4PC}
%----------------------
Each party first obtains a share of $c = \sh{-\lv{a}{}}^B$. \Partysubset{0,1,3} sample $r_{0,1,3}$ and \Party{0} sends $\msgset{0,3}{} = (-\lv{a}{}) \oplus r_{0,1,3}$ to \Party{2} in the preprocessing phase. \Party{3} compares its view of $\msgset{0,3}{}$ with \Party{2}. The parties then define their shares of $b$ as shown in Figure \ref{fig:Bit2A4}.
\Partysubset{1,2} locally compute a share of $b = (a + \lv{a}{})^B$. \Party{2} sends $\msgset{1,2}{} = (a + \lv{a}{}) \oplus r_{1,2,3}$ to \Party{0}. 
\Party{0} and \Party{1} compare their views of $\msgset{1,2}{}$. The parties then define their shares as shown in Figure \ref{fig:Bit2A4}.
The parties proceed to compute $\sh{a}^B = \sh{b}^B + \sh{c}^B$ using a Boolean adder. 
To verify the correctness, observe that $\sh{-\lv{a}{}} = \mv{c}{} \oplus \lv{c}{} = \mw{c}{} \oplus \lw{c}{}$. Likewise, $\sh{a + \lv{a}{}} = \mv{b}{} \oplus \lv{b}{}  = \mw{b}{} \oplus \lw{b}{}$. Hence, the parties successfully created two Boolean shares $\sh{b}$ and $\sh{c}$ such that $\sh{a}^B = \sh{b}^B + \sh{c}^B$.

\begin{protocolbox}{$\piAtoB{_{,4PC}}(\vl{\sh{a}^A}) \rightarrow \sh{a}^B$}{$\fourpcname$: 4PC Arithmetic to Binary Conversion}{fig:A2B4}
    
	%----
	\justify 
	
	\algoHeadacmart{Preprocessing:} 
	\begin{enumerate}[itemsep=0mm]
		%------
		\item Sample random values using $\piSRNG$:
		\begin{align*}
            &\Partysubset{0,1,3}: r_{0,1,3} & &\Partysubsetverify{1,2,3}: r_{1,2,3} 
		\end{align*}
        		\item Locally compute:
		\begin{align*} 
  & \Party{0},\Partyverify{3}: \msgset{0,3}{} = -\lv{a}{} \oplus r_{0,1,3} 
            \end{align*}
            \item Communicate:
		\begin{align*}
            &\Party{0} \rightarrow \Party{2}: \msgset{0,3}{}
		\end{align*}
            
		\item Locally compute:
		\begin{align*} 
            &\Party{1}: \lv{b}{1} = 0 &
            &\Party{2}: \lv{b}{2} = 0 &
            &\Party{0},\Partyverify{3}: \lv{b}{} = 0 \\
            &\Party{1}: \lv{c}{1} = r_{0,1,3} &
            &\Party{2}: \lv{c}{2} = \msgset{0,3}{} &
            &\Party{0},\Partyverify{3}: \lv{c}{} = (-\lv{a}{}) \\
            &\Partyverify{3}: \lw{b}{} = r_{1,2,3}  & 
            &\Partyverify{3}: \lw{c}{} = 0  & 
             &\Partysubsetverify{1,2}: \msgset{1,2}{} = \mv{a}{} \oplus r_{1,2,3} 
            \end{align*}
  
  \end{enumerate} 
    \algoHeadacmart{Online:} 
	\begin{enumerate}[itemsep=0mm]
		%-------
             \item Communicate:
		\begin{align*}
            &\Partyverify{2} \rightarrow \Partyverify{0}: \msgset{1,2}{} 
		\end{align*}
            
          \item Locally compute:
		\begin{align*}   
            &\Partysubset{1,2}: \mv{b}{} = \mv{a}{} &\Party{0}: \mw{b}{} = \msgset{1,2}{} &
            &\Partysubset{1,2}: \mv{c}{} = 0 &
            &\Party{0}: \mw{c}{} = \lv{c}{}
            \end{align*}
             
        \item Compare views using $\piCompareView$:
        \begin{align*}
            &\Partysubsetverify{2,3}: \msgset{0,3}{} & 
            &\Partysubsetverify{0,1}: \msgset{1,2}{} 
		\end{align*}
  \end{enumerate} 
          \begin{center}
          \rule{\textwidth}{0.2pt} \\ % horizontal line
          End of protocol-specific primitive \\[-0.5em]
          \rule{\textwidth}{0.2pt} \\ % horizontal line
          \end{center}
          \centering Jointly compute:
             \begin{align*} 
            &\Partyset: \sh{a}^B = \sh{b}^B + \sh{c}^B  
		\end{align*}
        
		%------   
\end{protocolbox}
%------------------

%----------------------
\subsubsection{Bit to Arithmetic Conversion}
\label{sec:other-bit2A}
%----------------------
To promote a shared bit $\sh{a^{\bitb}} = \mv{a}{} \oplus \lv{a}{}$ in the Boolean domain to a shared bit $\sh{a^b}^A = \mv{c}{} + \lv{c}{}$ in the arithmetic domain, the parties construct an $XOR$-sharing of $\sh{b}^A = a \oplus \lv{a}{}$ and $\sh{c}^A = \lv{a}{}$. This way, $\sh{a}^A = \sh{b}^A \oplus \sh{c}^A$ Then, they perform a private $XOR$ of the resulting shares in the arithmetic domain. Note that $\mv{a}{} \oplus \lv{a}{} = \mv{a}{} + \lv{a}{} - 2 \mv{a}{} \lv{a}{}$. Figures \ref{fig:Bit2A3} and \ref{fig:Bit2A4} show the formal protocols for our 3PC and 4PC protocols respectively.

%----------------------
\paragraph{3PC}
%----------------------
Each party first obtains a share of $c = \sh{\lv{a}{}}^A$. \Partysubset{0,1} sample $r_{0,1}$ and \Party{0} sends $\msgset{0}{} = \lv{a}{} + r_{0,1}$ to \Party{2} in the preprocessing phase. The parties then define their shares of $b$ as shown in Figure \ref{fig:Bit2A3}.
All parties locally compute $\sh{b} = (a \oplus \lv{a}{})^A$. By computing an $XOR$ of $\sh{b}^A$ and $\sh{c}^A$ in the arithmetic domain, the parties obtain an arithmetic share of $a$.
To verify the correctness, observe that $\sh{\lv{a}{}} = \mv{c}{1} - \lv{c}{1} = \mv{c}{2} - \lv{c}{2}$. Likewise, $\sh{a \oplus \lv{a}{}} = \mv{b}{1} - \lv{b}{1}  = \mv{b}{2} - \lv{b}{2}$. Hence, the parties successfully created two arithmetic shares $\sh{b}^A$ and $\sh{c}^A$ such that $\sh{a}^A = \sh{b}^A \oplus \sh{c}^A$.

%----------------------
\paragraph{4PC}
%----------------------
Each party first obtains a share of $c = \sh{\lv{a}{}}^A$. \Partysubset{0,1,3} sample $r_{0,1,3}$ and \Party{0} sends $\msgset{0,3}{} = \lv{a}{} + r_{0,1,3}$ to \Party{2} in the preprocessing phase. \Party{3} compares its view of \msgset{0,3}{} with \Party{2}. The parties then define their shares of $c$ as shown in Figure \ref{fig:Bit2A4}. 
\Partysubset{1,2} locally compute a share of $b = [a \oplus \lv{a}{}]^A$. \Party{2} sends $\msgset{1,2}{} = a \oplus \lv{a}{} + r_{1,2,3}$ to $\Party{0}$. 
\Party{0} and \Party{2} compare their views of $\msgset{1,2}{}$. The parties then define their shares of $b$ and $c$ as shown in Figure \ref{fig:Bit2A4}.
By computing an $XOR$ of $\sh{b}^A$ and $\sh{c}^A$ in the arithmetic domain, the parties obtain an arithmetic share of $a$.
To verify the correctness, observe that $\sh{\lv{a}{}} = \mv{c}{} - \lv{c}{}$ and $\sh{\lv{a}{}} = \mw{c}{} - \lw{c}{}$. Likewise, $\sh{a \oplus \lv{a}{}} = \mv{b}{} - \lv{b}{}$ and $\sh{a \oplus \lv{a}{}} = \mw{b}{} - \lw{b}{}$. Hence, the parties successfully created two arithmetic shares $\sh{b}$ and $\sh{c}$ such that $\sh{a}^A = \sh{b}^A \oplus \sh{c}^A$.

\begin{protocolbox}{$\piBittoA{_{,3PC}}(\vl{\sh{a^b}}^B) \rightarrow \sh{a^b}^A$}{$\threepcname$: 3PC Bit to Arithmetic}{fig:Bit2A3}
    
	%----
	\justify 
	
	\algoHeadacmart{Preprocessing:} 
	\begin{enumerate}[itemsep=0mm]
		%------
		\item Sample random values using $\piSRNG$:
		\begin{align*}
            &\Partysubset{0,1}: r_{0,1} 
		\end{align*}
        
            \item Communicate:
		\begin{align*}
            &\Party{0} \rightarrow \Party{2}: \msgset{0}{} = \lv{a}{} + r_{0,1} 
		\end{align*}
            
		\item Locally compute:
		\begin{align*}
            &\Partysubset{0,1}: \lv{b}{1} = 0 &
            &\Partysubset{0,2}: \lv{b}{2} = 0 &
            &\Partysubset{0,1}: \lv{c}{1} = r_{0,1} &
            &\lv{c}{2} = - \msgset{0}{} 
		\end{align*}
  \end{enumerate} 
	\justify
	
    \algoHeadacmart{Online:} 
	\begin{enumerate}[itemsep=0mm]
		%-------
		\item Locally compute:
		\begin{align*}
            &\Party{1}: \mv{b}{2} = \mv{a}{2} \oplus \lv{a}{1} &
            &\Party{2}: \mv{b}{1} = \mv{a}{1} \oplus \lv{a}{2} \\
            &\Party{1}: \mv{c}{2} = - r_{0,1} & 
            &\Party{2}: \mv{c}{1} = \msgset{0}{} 
		\end{align*}
 \end{enumerate} 
          \begin{center}
          \rule{\textwidth}{0.2pt} \\ % horizontal line
          End of protocol-specific primitive \\[-0.5em]
          \rule{\textwidth}{0.2pt} \\ % horizontal line
          \end{center}
          \centering Jointly compute:
             \begin{align*} 
            &\Partyset: \sh{a}^A = \sh{b}^A \oplus \sh{c}^A  
		\end{align*}
             
		%------   
\end{protocolbox}
%------------------

\begin{protocolbox}{$\piBittoA{_{,4PC}}(\vl{\sh{a^b}}^B) \rightarrow \sh{a^b}^A$}{$\fourpcname$: 4PC Bit to Arithmetic}{fig:Bit2A4}
    
	%----
	\justify 
	
	\algoHeadacmart{Preprocessing:} 
	\begin{enumerate}[itemsep=0mm]
		%------
		\item Sample random values using $\piSRNG$:
		\begin{align*}
            &\Partysubset{0,1,3}: r_{0,1,3} & &\Partysubsetverify{1,2,3}: r_{1,2,3} 
		\end{align*}
         \item Locally Compute:
		\begin{align*}
            &\Party{0},\Partyverify{3}: \msgset{0,3}{} = \lv{a}{} + r_{0,1,3} 
		\end{align*}
            \item Communicate:
		\begin{align*}
            &\Party{0} \rightarrow \Party{2}: \msgset{0,3}{} 
		\end{align*}
            
		\item Locally compute:
		\begin{align*} 
            &\Party{1}: \lv{b}{1} = 0 &
            &\Party{2}: \lv{b}{2} = 0 &
            &\Party{0}: \lv{b}{} = 0 \\
            &\Party{1}: \lv{c}{1} = r_{0,1,3} &
            &\Party{2}: \lv{c}{2} = - \msgset{0,3}{} &
            &\Partysubset{0},\Partyverify{3}: \lv{c}{} = -\lv{a}{} \\
            &\Partyverify{3}: \lw{b}{} = r_{1,2,3} &
              &\Partyverify{3}: \lw{c}{} = 0 &
              &\Partysubsetverify{1,2}: \msgset{1,2}{} = \mv{a}{} + r_{1,2,3}
            \end{align*}
  
  \end{enumerate} 
	\justify
    \algoHeadacmart{Online:} 
	\begin{enumerate}[itemsep=0mm]
		%-------
             \item Communicate:
		\begin{align*}
            &\Partyverify{2} \rightarrow \Partyverify{0}: \msgset{1,2}{}
		\end{align*}
            
          \item Locally compute:
		\begin{align*}   
            &\Partysubset{1,2}: \mv{b}{} = \mv{a}{} &\Party{0}: \mw{b}{} = \msgset{1,2}{} &
            &\Partysubset{1,2}: \mv{c}{} = 0 &
            &\Party{0}: \mw{c}{} = \lv{a}{} 
            \end{align*}
             
        \item Compare views using $\piCompareView$:
        \begin{align*}
            &\Partysubsetverify{2,3}: \msgset{0,3}{} & 
            &\Partysubsetverify{0,1}: \msgset{1,2}{} 
		\end{align*}

 \end{enumerate} 
          \begin{center}
          \rule{\textwidth}{0.2pt} \\ % horizontal line
          End of protocol-specific primitive \\[-0.5em]
          \rule{\textwidth}{0.2pt} \\ % horizontal line
          \end{center}
          \centering Jointly compute:
             \begin{align*} 
            &\Partyset: \sh{a}^A = \sh{b}^A \oplus \sh{c}^A  
		\end{align*}
        
		%------
\end{protocolbox}
%------------------

%============================
\section{$\fourpcname$: Security Proof for 4PC}
\label{sec:security-4PC}
%============================
This section provides the security proof for our 4PC protocol $\fourpcname$ described in~\secref{4PC}. We consider a circuit $\circuit$ representing a function $\func{\cdot}$ to be evaluated. We establish security using the real-world/ideal-world simulation paradigm~\cite{Goldreich2004,EPRINT:Lindell16}. 

The security is evaluated by comparing an adversary's capabilities in the real-world execution of the protocol with those in an ideal-world execution involving a trusted third party. In the ideal world, parties send inputs to the trusted third party via secure channels, the trusted third party performs the computation, and returns the output to the parties. A protocol is considered secure if an adversary in the real world can only do what it could also do in the ideal world.

\begin{systembox}{$\FuncFour{\abort}$}{Ideal 4PC functionality for computing $\func{\cdot}$ with abort.}{fig:Fabort}
    \smallskip
    %----
    \justify 
    \noindent $\FuncFour{\abort}$ interacts with the parties in $\Partyset$ and the adversary $\Adv$. $\FuncFour{\abort}$  is parameterized by a 4-input function $\func{\cdot}$, represented by a publicly known circuit $\circuit$. Every honest party $\Party{i} \in \Partyset$ sends its input $x_i$ to the functionality. Corrupt parties may send arbitrary inputs as instructed by $\Adv$. While sending the inputs, $\Adv$ is also allowed to send a special $\abort$ command.
	%----
    \justify 
    \noindent \textbf{Input:} On message $(\Input, x_i)$ from $\Party{i}$, do the following: if $(\Input, \ast)$ already received from $\Party{i}$, then ignore the current message. Otherwise, record $x_i' = x_i$ internally. If $x_i$ is outside $\Party{i}$'s input domain, consider $x_i' = \abort$.
    %----
    \justify 
    \noindent \textbf{Output:} If there exists an $i \in \{0, 1, 2, 3\}$ such that $x_i' = \abort$, send $(\Output, \bot)$ to $\Partyset$. Otherwise, compute $y = f(x_0', x_1', x_2', x_3')$ and send $(\Output, y)$ to $\Adv$. Receive $(\select, \Partysubset{\Phi})$ from $\Adv$, where $\Partysubset{\Phi}$ denotes a subset of honest parties. Send $(\Output, \bot)$ to every honest party $\Party{i} \in \Partysubset{\Phi}$. Send $(\Output, y)$ to all the remaining parties.
\end{systembox}

Let $\Adv$ be a probabilistic polynomial time ($\PPT$) real-world adversary corrupting at most one party in $\Partyset$, $\Sim$ be the corresponding ideal world adversary, and $\Func{}$ be the ideal functionality. Let $\Ideal_{\Func{}, \Sim}(1^k, z)$ be the joint output of the honest parties and $\Sim$ from the ideal execution with security parameter $\kappa$ and auxiliary input $z$. Similarly, $\Real_{\Pi, \Sim}(1^{\kappa}, z)$ be the joint output of the honest parties and $\Adv$ from the real world execution. A protocol $\Pi$ securely realizes $\Func{}$ if for every $\PPT$ adversary $\Adv$, there is an ideal world adversary $\Sim$ corrupting the same parties such that $\Ideal_{\Func{}, \Sim}(1^k, z)$ and $\Real_{\Pi, \Sim}(1^{\kappa}, z)$ are computationally indistinguishable. Figure~\ref{fig:Fabort} shows the ideal functionality $\FuncFour{\abort}$ for computing a function $\func{\cdot}$ in the 4PC setting with abort security.

\begin{systembox}{$\FuncSRNG(\Partysubset{\Phi}, \ell') \rightarrow \randval{\Phi} \in \Z{\ell'}$}{Shared random value generator (SRNG) functionality}{fig:funcsrng}
	%-----
	$\FuncSRNG$ interacts with the parties in $\Partysubset{\Phi}$ and the adversary $\Adv$. 	
	\justify
	\algoHeadacmart{Procedure:} $\FuncSRNG$ samples a random value $\randval{\Phi} \in \Z{\ell'}$ and sends the message $(\Output, \randval{\Phi})$ to all the parties in $\Partysubset{\Phi}$.
\end{systembox}

We now provide the details for our simulation. For simplicity, we focus on the circuit evaluation process that includes input sharing, linear operations, multiplications and output reconstruction. 
Moreover, we provide the simulation for each of these stages separately and carrying out these simulations in the topological order of the circuit provides the simulation for the entire computation.
We provide proofs in the $\Funcsetup$, $\FuncSRNG$-hybrid model, where $\Funcsetup$ and $\FuncSRNG$~(\figref{funcsrng}) denote the ideal functionalities for shared-key setup and $\piSRNG$~(\figref{srng}) protocol respectively.

%--------------------------------
\subsubsection*{\textbf{Simulation Strategy}}
\label{sec:simulation-strategy}
%--------------------------------
The strategy for simulating the computation of the function $\func{\cdot}$ (represented by the circuit $\circuit$) is as follows: The simulation starts with the simulator emulating the shared-key setup ($\funcn{setup}$) functionality and distributing the corresponding keys to the adversary. Next is the input sharing phase, where the simulator $\Sim$ computes the input for the adversary $\Adv$ using the known keys and the messages received from $\Adv$, and sets the inputs of the honest parties to $0$ for the simulation. $\Sim$ then invokes the ideal functionality $\FuncFour{\abort}$ on behalf of $\Adv$ with the extracted input and receives the output $y$. Note that with the inputs of $\Adv$ now known, $\Sim$ can compute all the intermediate values for each building block. Finally, $\Sim$ simulates each of the building blocks in topological order. 
To keep track of honest parties that $\abort$ during the simulation due to incorrect messages by $\Adv$, $\Sim$ maintains a list of parties denoted as $\Partysubset{\Phi}$, initially set to $\emptyset$.

%--------------------------------
\subsubsection*{\textbf{Input Sharing}}
\label{sec:security-4PC-input}
%--------------------------------
To simulate the input sharing protocol $\piShare$~(cf.~\figref{piShare}), the ideal world adversary $\Sim_{\piShare}$ emulates $\Funcsetup$ and obtains all the PRF keys. It then gives the respective PRF keys to the adversary $\Adv$. We consider the case of corrupt $\Party{0}$ and $\Party{3}$ and the simulation for corrupt $\Party{1}$ and $\Party{2}$ follows similar to $\Party{0}$. To avoid repetition, we omit the details of $\Funcsetup$ in the simulation steps.

\begin{simulatorbox}{$\Sim_{\piShare}^{\Party{0}}$}{Simulator $\Sim_{\piShare}^{\Party{0}}$ for corrupt $\Party{0}$}{fig:sim_piShare_P0}
	%=========================
	\justify 
	\algoHeadacmart{Preprocessing:} 
	\begin{enumerate}[itemsep=0mm]
		%-------
        \item \emph{Case I: $\Party{I} = \Party{0}$.} $\Sim_{\piShare}^{\Party{0}}$ executes the steps of $\piSRNG$ using the respective PRF keys shared with $\Adv$ to obtain $\lv{x}{1}$, $\lv{x}{2}$, and $\lw{x}{}$. 
		%-------
        \item \emph{Case II: $\Party{I} \ne \Party{0}$.} $\Sim_{\piShare}^{\Party{0}}$ executes the steps of $\piSRNG$ using the respective PRF keys shared with $\Adv$ to obtain $\lv{x}{1}$, and $\lv{x}{2}$. It samples $\lw{x}{}$ randomly from $\Z{\ell}$. 
        %-------
    \end{enumerate}
    %=========================
	\justify 
	\algoHeadacmart{Online:} 
	\begin{enumerate}[itemsep=0mm]
        %-------
        \item \emph{Case I: $\Party{I} = \Party{0}$.}
        \begin{enumerate}[itemsep=0mm]
            %---------
            \item $\Sim_{\piShare}^{\Party{0}}$ receives $\mverify{x}{}$ from $\Party{0}$ on behalf of $\Party{1}$ and $\Party{2}$
            %---------
            \item If the received values are consistent, $\Sim_{\piShare}^{\Party{0}}$ computes input of $\Party{0}$~(= $\Adv$) as $x = \mverify{x}{} - \lv{x}{1} - \lv{x}{2} - \lw{x}{}$. Else, it sets $x = \abort$.
            %---------
            \item $\Sim_{\piShare}^{\Party{0}}$ honestly executes the steps of $\piCompareView$ using the messages it received from $\Party{0}$. If $\Sim_{\piShare}^{\Party{0}}$ receives inconsistent hash from $\Party{0}$ on behalf of $\Party{j} \in \Partysubset{1,2}$, it adds $\Party{j}$ to $\Partysubset{\Phi}$.
            %---------
        \end{enumerate}
        %%
		%-------
        \item \emph{Case II: $\Party{I} \ne \Party{0}$.}
        \begin{enumerate}[itemsep=0mm]
            %---------
            \item $\Sim_{\piShare}^{\Party{0}}$ sets $x = 0$ and honestly executes the steps of $\piShare$ on behalf of $\Party{1}$ and $\Party{2}$. If $\Sim_{\piShare}^{\Party{0}}$ receives inconsistent hash from $\Party{0}$ on behalf of $\Party{j} \in \Partysubset{1,2}$, it adds $\Party{j}$ to $\Partysubset{\Phi}$.
            %---------
        \end{enumerate}
        %%
        %-------
    \end{enumerate}
    %=========================
    \hrule\smallskip
    \small \centering \textit{$\Partysubset{I}$ denotes the input party with the secret $x \in \Z{\ell}$.}
\end{simulatorbox}

Shares unknown to $\Adv$ are randomly sampled in the simulation, whereas in the real protocol, they are sampled using the pseudorandom function (PRF). Therefore, the indistinguishability of the simulation is guaranteed by reducing it to the security of the PRF.

For the case when $\Party{I} \ne \Party{0}$ in~\figref{sim_piShare_P0}, $\Sim_{\piShare}^{\Party{0}}$ sets the input of the honest parties $\Partysubset{1,2,3}$ as $x = 0$. Given that $\Adv$ (who is $\Party{0}$) only receives $\mverify{x}{}$ in the online phase but does not possess $\lw{x}{}$, the shares of $x = 0$ are indistinguishable to $\Adv$ from the actual shares of the honest parties in the real protocol.

\smallskip
\begin{simulatorbox}{$\Sim_{\piShare}^{\Party{3}}$}{Simulator $\Sim_{\piShare}^{\Party{3}}$ for corrupt $\Party{3}$}{fig:sim_piShare_P3}
	%=========================
	\justify 
	\algoHeadacmart{Preprocessing:} 
	\begin{enumerate}[itemsep=0mm]
		%-------
        \item $\Sim_{\piShare}^{\Party{3}}$ executes the steps of $\piSRNG$ using the respective PRF keys shared with $\Adv$ to obtain $\lv{x}{1}$, $\lv{x}{2}$, and $\lw{x}{}$. 
        %-------
    \end{enumerate}
    %=========================
	\justify 
	\algoHeadacmart{Online:} 
	\begin{enumerate}[itemsep=0mm]
        %-------
        \item \emph{Case I: $\Party{I} = \Party{3}$.}
        \begin{enumerate}[itemsep=0mm]
            %---------
            \item $\Sim_{\piShare}^{\Party{3}}$ receives $\mverify{x}{}$ from $\Party{3}$ on behalf of $\Party{0}$, $\Party{1}$ and $\Party{2}$
            %---------
            \item If the received values are consistent, $\Sim_{\piShare}^{\Party{3}}$ computes input of $\Party{3}$~(= $\Adv$) as $x = \mverify{x}{} - \lv{x}{1} - \lv{x}{2} - \lw{x}{}$. Else, it sets $x = \abort$.
            %---------
        \end{enumerate}
        %%
		%-------
        \item \emph{Case II: $\Party{I} \ne \Party{3}$.} There is nothing to simulate as $\Party{3}$ does not receive any value during the online phase.
        %-------
    \end{enumerate}
    %=========================
    \hrule\smallskip
    \small \centering \textit{$\Partysubset{I}$ denotes the input party with the secret $x \in \Z{\ell}$.}
\end{simulatorbox}

%--------------------------------
\subsubsection*{\textbf{Linear Operations}}
\label{sec:security-4PC-linear}
%--------------------------------
Since linear operations are local, they do not require communications to be simulated. The simulator $\Sim$ carries out the local operations on behalf of the honest parties.

%--------------------------------
\subsubsection*{\textbf{Multiplication}}
\label{sec:security-4PC-mult}
%--------------------------------
To simulate protocol $\piMult$~(cf.~\figref{piMult4PC}), $\Sim_{\piMult}$ uses the PRF keys obtained during the emulation of $\Funcsetup$ and executes the $\piSRNG$ protocol. Once the random values are obtained, $\Sim_{\piMult}$ executes the remaining steps of $\piMult$ honestly. For this, $\Sim_{\piMult}$ uses the inputs extracted during the simulation of $\piShare$ and thereby learns all the intermediate values of the circuit. While simulating the steps of $\piCompareView$, $\Sim_{\piMult}$ keeps track of the honest parties who receive incorrect messages from $\Adv$ and adds them to $\Partysubset{\Phi}$. 

\smallskip
\begin{simulatorbox}{$\Sim_{\piMult}^{\Party{0}}$}{Simulator $\Sim_{\piMult}^{\Party{0}}$ for corrupt $\Party{0}$}{fig:sim_piMult_P0}
	%=========================
	\justify 
	\algoHeadacmart{Preprocessing:} 
	\begin{enumerate}[itemsep=0mm]
		%-------
        \item $\Sim_{\piMult}^{\Party{0}}$ executes the steps of $\piSRNG$ using the respective PRF keys shared with $\Adv$ to obtain $\lv{c}{1}$, $\lv{c}{2}$, and $r_{0,1,3}$. It samples $\lw{c}{}$ and $r_{1,2,3}$ randomly from $\Z{\ell}$. 
        %-------
        \item $\Sim_{\piMult}^{\Party{0}}$ computes the values $\lv{c}{}$, $\msgset{0,3}{}$ and $\msgset{3}{}$ honestly.
        %-------
        \item $\Sim_{\piMult}^{\Party{0}}$ receives $\msgset{0,3}{}$ from $\Party{0}$ on behalf of $\Party{2}$. If the received value is inconsistent with its locally computed version, $\Sim_{\piMult}^{\Party{0}}$ sets the input of party $\Party{0}$ as $x = \abort$.
        %-------
        \item $\Sim_{\piMult}^{\Party{0}}$ sends $\msgset{3}{}$ to $\Party{0}$ on behalf of $\Party{3}$. 
        %-------
    \end{enumerate}
    %=========================
	\justify 
	\algoHeadacmart{Online:} 
	\begin{enumerate}[itemsep=0mm]
        %-------
        \item $\Sim_{\piMult}^{\Party{0}}$ executes the steps  honestly on behalf of $\Party{1}$ and $\Party{2}$.
        %-------
        \item $\Sim_{\piMult}^{\Party{0}}$ sends $\msgset{1,2}{}$ to $\Party{0}$ on behalf of $\Party{2}$.
		%-------
        \item $\Sim_{\piMult}^{\Party{0}}$ interacts with $\Party{0}$ honestly for $\piCompareView$ instances corresponding to $\msgset{1,2}{}$, and $\mverify{c}{}$. If it receives an inconsistent value from $\Party{0}$ in any of the two instances, $\Sim_{\piMult}^{\Party{0}}$ adds the respective honest parties to $\Partysubset{\Phi}$.
        %-------
    \end{enumerate}
    %=========================
\end{simulatorbox}

We now provide the simulation for corrupt $\Party{0}$ and $\Party{2}$. The case for $\Party{1}$ is similar to $\Party{2}$, and for $\Party{3}$, the role is minimal. Hence, we avoid providing simulation details for those cases. We begin with the case of corrupt $\Party{0}$.

The simulation discussed above captures two scenarios. If $\Adv$ sends an incorrect message to any of the honest parties during the protocol, it is equivalent to $\Adv$ inputting $\abort$ to the ideal functionality $\FuncFour{\abort}$. In this case, $\Sim_{\piMult}^{\Party{0}}$ simulates this behavior by setting $\Adv$'s input to $\abort$. 

In the other scenario, where $\Adv$ provides inconsistent values to honest parties during $\piCompareView$, this is equivalent to $\Adv$ choosing those honest parties to receive $\abort$ in the ideal world. $\Sim_{\piMult}^{\Party{0}}$ simulates this by identifying such instances and adding those honest parties to the list $\Partysubset{\Phi}$ that it maintains. After simulating all the individual protocols, $\Sim_{\piMult}^{\Party{0}}$ will input this list to $\FuncFour{\abort}$ on behalf of $\Adv$.

\begin{simulatorbox}{$\Sim_{\piMult}^{\Party{2}}$}{Simulator $\Sim_{\piMult}^{\Party{2}}$ for corrupt $\Party{2}$}{fig:sim_piMult_P2}
	%=========================
	\justify 
	\algoHeadacmart{Preprocessing:} 
	\begin{enumerate}[itemsep=0mm]
		%-------
        \item $\Sim_{\piMult}^{\Party{2}}$ executes the steps of $\piSRNG$ using the respective PRF keys shared with $\Adv$ to obtain $\lv{c}{2}$, $\lw{c}{}$, and $r_{1,2,3}$. It samples $\lv{c}{1}$ and $r_{0,1,3}$ randomly from $\Z{\ell}$. 
        %-------
        \item $\Sim_{\piMult}^{\Party{2}}$ sends $\msgset{0,3}{}$ to $\Party{2}$ on behalf of $\Party{0}$. 
        %-------
    \end{enumerate}
    %=========================
	\justify 
	\algoHeadacmart{Online:} 
	\begin{enumerate}[itemsep=0mm]
        %-------
        \item $\Sim_{\piMult}^{\Party{2}}$ executes the steps  honestly on behalf of $\Party{0}$ and $\Party{1}$.
        %-------
        \item $\Sim_{\piMult}^{\Party{2}}$ sends $\msgset{1}{}$ to $\Party{2}$ on behalf of $\Party{1}$.
        %-------
        \item $\Sim_{\piMult}^{\Party{2}}$ receives $\msgset{2}{}$ and $\msgset{1,2}{}$ from $\Party{2}$ on behalf of $\Party{1}$ and $\Party{0}$ respectively. If any of the received values are inconsistent with their locally computed version, $\Sim_{\piMult}^{\Party{2}}$ sets the input of party $\Party{2}$ as $x = \abort$.
		%-------
        \item $\Sim_{\piMult}^{\Party{2}}$ interacts with $\Party{2}$ honestly for $\piCompareView$ instances corresponding to $\msgset{0,3}{}$, and $\mverify{c}{}$. If it receives an inconsistent value from $\Party{2}$ in any of the two instances, $\Sim_{\piMult}^{\Party{2}}$ adds the respective honest parties to $\Partysubset{\Phi}$.
        %-------
    \end{enumerate}
    %=========================
\end{simulatorbox}

\medskip
%-------------------
\paragraph{Heterogeneous Multiplication}
%-------------------
The simulation steps for the 4PC multiplication protocol in the heterogeneous setting, $\piMultH$ (see \figref{piMult4PCH}), are very similar to those of $\piMult$ discussed in this section. The primary difference from a simulation perspective is one communication message $\msgset{3}{}$ in $\piMult$, which is replaced with a message from $\Party{2}$ to $\Party{0}$ in $\piMultH$. Therefore, we omit the formal details for $\piMultH$. 

\medskip
%--------------------------------
\subsubsection*{\textbf{Output Reconstruction}}
\label{sec:security-4PC-output}
%--------------------------------
To simulate the output reconstruction protocol $\piRec$~(cf.~\figref{piRec}), $\Sim_{\piRec}$ utilizes the information about $\Adv$'s inputs that the ideal world adversary extracted during the simulation of the input sharing protocol. Additionally, it uses the list $\Partysubset{\Phi}$, which it maintained to track the honest parties who were aborted during the simulation. We consider the case of corrupt $\Party{0}$ and $\Party{3}$ and the simulation for other parties are similar.

\begin{simulatorbox}{$\Sim_{\piRec}^{\Party{0}}$}{Simulator $\Sim_{\piRec}^{\Party{0}}$ for corrupt $\Party{0}$}{fig:sim_piRec_P0}
	%=========================
	\justify 
	\algoHeadacmart{Postprocessing:} 
	\begin{enumerate}[itemsep=0mm]
        %-------
        \item $\Sim_{\piRec}^{\Party{0}}$ invokes $\FuncFour{\abort}$ on $(\Input, x)$ on behalf of $\Adv$ to obtain the function output $y$. Here, $x$ denotes the extracted inputs of $\Party{0}$ during simulation of input sharing.
        %-------
        \item \emph{Case I: $\Party{O} = \Party{0}$.}
        \begin{enumerate}[itemsep=0mm]
            %---------
            \item If $y \ne \bot$, $\Sim_{\piRec}^{\Party{0}}$ computes and sends $\mv{y}{} = y + \lv{y}{1} + \lv{y}{2}$ to $\Party{0}$ on behalf of $\Party{2}$. Else, terminate.
            %---------
            \item $\Sim_{\piRec}^{\Party{0}}$ honestly executes the steps of $\piCompareView$ for $\mv{y}{}$ and $\lv{y}{}$ on behalf of $\Party{1}$ and $\Party{3}$ respectively. If $\Sim_{\piRec}^{\Party{0}}$ receives inconsistent hash from $\Party{0}$ on behalf of $\Party{j} \in \Partysubset{1,3}$, it adds $\Party{j}$ to $\Partysubset{\Phi}$.
            %---------
        \end{enumerate}
        %%
		%-------
        \item \emph{Case II: $\Party{O} \ne \Party{0}$.} 
        \begin{enumerate}[itemsep=0mm]
            %---------
            \item $\Sim_{\piRec}^{\Party{0}}$ receives $\lv{y}{}$ from $\Party{0}$ on behalf of $\Party{O}$. If the received value is not consistent with the value it holds (obtained as part of simulating $\Funcsetup$), it adds $\Party{O}$ and $\Party{3}$ to $\Partysubset{\Phi}$.
            %---------
        \end{enumerate}
        %%
        %-------
        \item $\Sim_{\piRec}^{\Party{0}}$ sends $(\select, \Partysubset{\Phi})$ to $\FuncFour{\abort}$ on behalf of $\Adv$ and terminates.
        %---------
    \end{enumerate}
    %=========================
    \hrule\smallskip
    \small \centering \textit{$\Partysubset{O}$ denotes the output party.}
\end{simulatorbox}

\begin{simulatorbox}{$\Sim_{\piRec}^{\Party{3}}$}{Simulator $\Sim_{\piRec}^{\Party{3}}$ for corrupt $\Party{3}$}{fig:sim_piRec_P3}
	%=========================
	\justify 
	\algoHeadacmart{Postprocessing:} 
	\begin{enumerate}[itemsep=0mm]
        %-------
        \item $\Sim_{\piRec}^{\Party{3}}$ invokes $\FuncFour{\abort}$ on $(\Input, x)$ on behalf of $\Adv$ to obtain the function output $y$. Here, $x$ denotes the extracted inputs of $\Party{3}$ during simulation of input sharing.
        %-------
        \item \emph{Case I: $\Party{O} = \Party{3}$.}
        \begin{enumerate}[itemsep=0mm]
            %---------
            \item If $y \ne \bot$, $\Sim_{\piRec}^{\Party{3}}$ computes and sends $\mv{y}{} = y + \lv{y}{1} + \lv{y}{2}$ to $\Party{3}$ on behalf of $\Party{2}$. Else, terminate.
            %---------
            \item $\Sim_{\piRec}^{\Party{3}}$ sends $\lv{y}{}$ to $\Party{3}$ on behalf of $\Party{0}$.
            %---------
            \item $\Sim_{\piRec}^{\Party{3}}$ honestly executes the steps of $\piCompareView$ for $\mv{y}{}$ on behalf of $\Party{1}$. If $\Sim_{\piRec}^{\Party{3}}$ receives inconsistent hash from $\Party{3}$ on behalf of $\Party{1}$, it adds $\Party{1}$ to $\Partysubset{\Phi}$.
            %---------
        \end{enumerate}
        %%
		%-------
        \item \emph{Case II: $\Party{O} \ne \Party{3}$.} 
        \begin{enumerate}[itemsep=0mm]
            %---------
            \item $\Sim_{\piRec}^{\Party{3}}$ honestly executes the steps of $\piCompareView$ for $\lv{y}{}$ (obtained as part of simulating $\Funcsetup$) on behalf of $\Party{O}$. If $\Sim_{\piRec}^{\Party{3}}$ receives inconsistent hash from $\Party{3}$ on behalf of $\Party{O}$, it adds $\Party{O}$ to $\Partysubset{\Phi}$.
            %---------
        \end{enumerate}
        %%
        %-------
        \item $\Sim_{\piRec}^{\Party{3}}$ sends $(\select, \Partysubset{\Phi})$ to $\FuncFour{\abort}$ on behalf of $\Adv$ and terminates.
        %---------
    \end{enumerate}
    %=========================
    \hrule\smallskip
    \small \centering \textit{$\Partysubset{O}$ denotes the output party.}
\end{simulatorbox}

\newpage
%============================
\section{Computational Complexity}
\label{sec:others}
%============================
\tabref{comp_full} extends \tabref{comp_score} in \secref{Introduction} by showing the specific number of operations each party needs to perform locally in our protocols and related ones. While our 3PC protocol performs similarly to related work, we note that we shift a large amount of that computation to $\Party{0}$ which is only active in the preprocessing phase. The online phase is generally considered a bottleneck for MPC deployments as it is latency-critical. Hence shifting computation and communication to the preprocessing phase may become more attractive for deployment.
Our 4PC protocol reduces the computational complexity significantly with similar considerations: The majority of the computation is assigned to $\Partysubset{0,3}$ who only carry out non-latency critical operations.

\begin{table}[htb!]
\centering
\small
\captionsetup{font=small}
\caption{Operations and communication for multiplications (Extended)}
\label{tab:comp_full}
\vspace{-3mm}
\begin{threeparttable}
\begin{tabular}{c|l|l|ll|ll}
\hline
\multirow{2}{*}{Setting} & \multirow{2}{2cm}{Protocol} & \multirow{2}{*}{Party} & \multicolumn{2}{c|}{Operation} & \multicolumn{2}{c}{Com}\tnote{a} \\
& & & Add & Mult\tnote{b} & Pre.\tnote{c} & On. \tnote{c} \\ \hline
\multirow{12}{*}{3PC} & \multirow{4}{2cm}{Replicated \cite{CCS:AFLNO16}} & $P_0$ & 4 & 2 (+1) & 0 & 1 \\
& & $P_1$ & 4 & 2 (+1) & 0 & 1 \\
& & $P_2$ & 4 & 2 (+1) & 0 & 1 \\
& & Total & 12 & 6 (+3) & 0 & 3 \\ \cline{2-7}
& \multirow{4}{2cm}{ASTRA \cite{CCSW:CCPS19}} & $P_0$ & 2 & 1 & 1 & 0 \\
& & $P_1$ & 4 & 2 & 0 & 1 \\
& & $P_2$ & 5 & 2\tnote{d} & 0 & 1 \\
& & Total & \textbf{11} & 5 & 1 & \textbf{2} \\ \cline{2-7}
& \multirow{4}{2cm}{$\threepcname$~(\textbf{This work})} & $P_0$ & 4 & 2 & 1 & 0 \\
& & $P_1$ & 4 & 2 & 0 & 1 \\
& & $P_2$ & 3 & 1 & 0 & 1 \\
& & Total & \textbf{11} & \textbf{5} & 1 & \textbf{2} \\ \hline
\multirow{15}{*}{4PC} & \multirow{5}{2.2cm}{Fantastic Four \cite{USENIX:Dalskov0K21}} & $P_0$ & 15 & 9 & 0 & 0-3 \\
& & $P_1$ & 15 & 9 & 0 & 0-3 \\
& & $P_2$ & 15 & 9 & 0 & 0-3 \\
& & $P_3$ & 15 & 9 & 0 & 0-3 \\
& & Total & 60 & 36 & 0 & 6 \\ \cline{2-7}
& \multirow{5}{2.2cm}{Tetrad \cite{NDSS:KotiPRS22}} & $P_0$ & 14 & 5 & 1 & 0 \\
& & $P_1$ & 12 & 8 & 0 & 1 \\
& & $P_2$ & 12 & 8 & 0 & 2 \\
& & $P_3$ & 14 & 9 & 1 & 0 \\
& & Total & 52 & 30 & 2 & \textbf{3} \\ \cline{2-7}
& \multirow{5}{2.2cm}{$\fourpcname$~(\textbf{This work})} & $P_0$ & 7 & 3 & 1 & 0 \\
& & $P_1$ & 5 & 3 & 0 & 1 \\
& & $P_2$ & 6 & 3 & 0 & 2-3\tnote{e} \\
& & $P_3$ & 7 & 3 & 1-0\tnote{e} & 0 \\
& & Total & \textbf{25} & \textbf{12} & 2-1\tnote{e} & \textbf{3-4}\tnote{e} \\ \hline
\end{tabular}
\begin{tablenotes}
\item[a] Number of ring elements sent by the respective party.
\item[b] Replicated 3PC additionally requires a division operation (+1) per party in the arithmetic domain.
 \item[c] \textit{Pre.} refers to Preprocessing, \textit{On.} refers to Online Phase.
\item[d] Reduced from 3 to 2 using a straightforward optimization (c.f. \secref{3PC}).
\item[e] Right entry refers to heterogeneous variant. Additional online communication is part of consant-round verification. 
\end{tablenotes}
\end{threeparttable}
\end{table}

\section{Evaluation of Additional Protocols}
\label{sec:app-eval}

\begin{figure}[htb!]
    \centering
      \begin{subfigure}[b]{0.45\textwidth}
        \centering
        \includegraphics[width=\linewidth, trim = 0cm 4.5cm 0cm 4.5cm, clip]{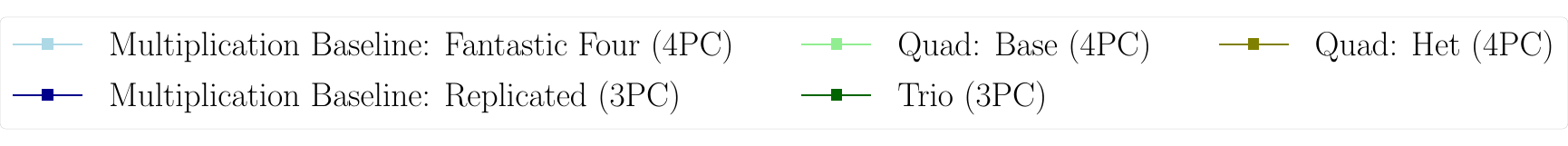}
    \end{subfigure}
    \begin{subfigure}[b]{0.22\textwidth}
        \centering
        \includegraphics[width=\linewidth, trim = 0cm 0cm 7.9cm 0cm, clip]{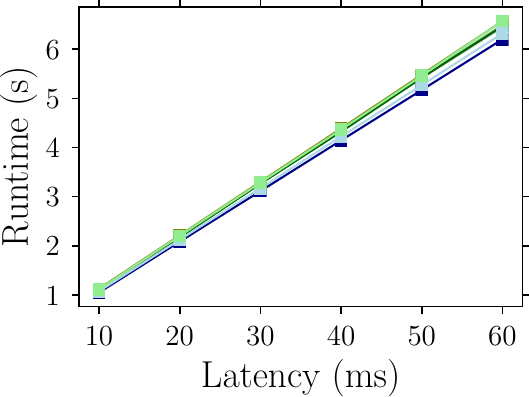}
        \vspace{-5mm}
        \captionsetup{font=footnotesize}
        \caption{Multiplication: Restricting latency between all Links}
        \label{fig:mult_lat}
    \end{subfigure}
    \hspace{0.01\textwidth}
    \begin{subfigure}[b]{0.22\textwidth}
        \centering
        \includegraphics[width=\linewidth, trim = 0cm 0cm 7.9cm 0cm, clip]{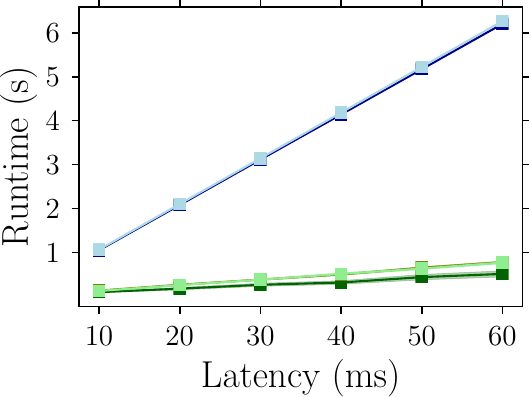}
        \vspace{-5mm}
        \captionsetup{font=footnotesize}
        \caption{Multiplication: Restricting latency between $5/6$ of all Links}
        \label{fig:mult_latp}
    \end{subfigure}
    \vspace{0.01\textwidth}
    \begin{subfigure}[b]{0.22\textwidth}
        \centering
        \includegraphics[width=\linewidth, trim = 0cm 0cm 9.2cm 0cm, clip]{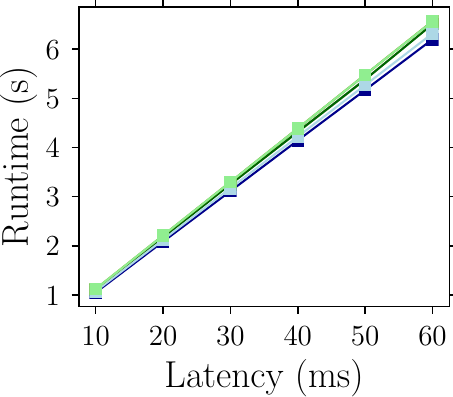}
        \vspace{-5mm}
        \captionsetup{font=footnotesize}
        \caption{Mult + Trunc: Restricting the latency between all Links}
        \label{fig:fixed_mult_lat}
    \end{subfigure}
        \hspace{0.01\textwidth}
    \begin{subfigure}[b]{0.22\textwidth}
        \centering
        \includegraphics[width=\linewidth, trim = 0cm 0cm 9.2cm 0cm, clip]{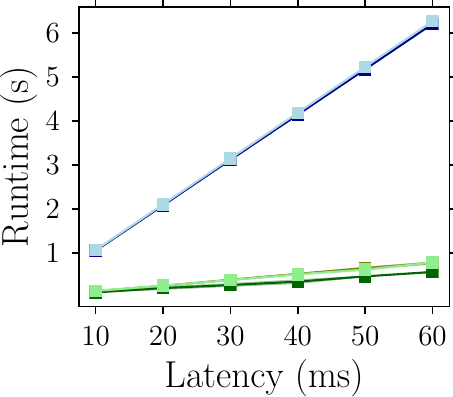}
        \vspace{-5mm}
        \captionsetup{font=footnotesize}
        \caption{Mult + Trunc: Restricting latency between $5/6$ of all Links}
        \label{fig:fixed_mult_latp}
    \end{subfigure}
    \vspace{0.01\textwidth}
        \begin{subfigure}[b]{0.22\textwidth}
        \centering
        \includegraphics[width=\linewidth, trim = 0cm 0cm 9.2cm 0cm, clip]{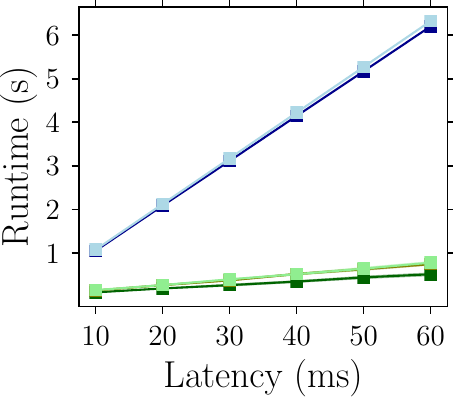}
        \vspace{-5mm}
        \captionsetup{font=footnotesize}
        \caption{A2B (protocol-specific): Restricting latency between all Links}
        \label{fig:a2bit_lat}
    \end{subfigure}
    \hspace{0.01\textwidth}
    \begin{subfigure}[b]{0.22\textwidth}
        \centering
        \includegraphics[width=\linewidth, trim = 0cm 0cm 9.2cm 0cm, clip]{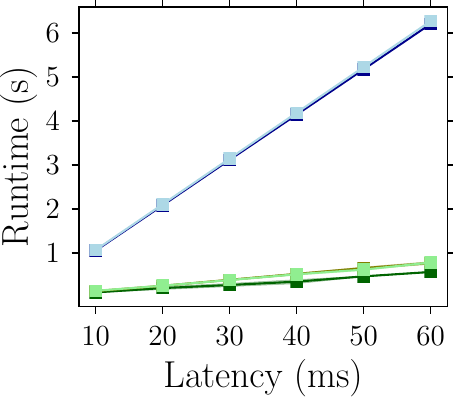}
        \vspace{-5mm}
        \captionsetup{font=footnotesize}
        \caption{Bit2A (protocol-specific): Restricting latency between all Links}
        \label{fig:bit2a_lat}
    \end{subfigure}
    %\hspace{0.01\textwidth}
    \vspace{-2mm}
    \captionsetup{font=small}
    \caption{Runtimes of 50 sequential operations under various latency restrictions}
    \label{fig:latencies}
\end{figure}

To demonstrate the utility of our additional protocols under latency-restricted settings for mixed circuits and fixed point arithmetic, we first compare 50 sequential evaluations of our multiplication and multiplication + truncation protocols against a baseline of the Fantastic Four and Replicated multiplication protocols which are one-round protocols as well. Figures \ref{fig:mult_lat} and \ref{fig:fixed_mult_lat} show that if we restrict the latency between all links in our setup, our protocols perform almost identically to the baseline. When we however simply allow one link between the parties to be unaffected by the link throttling, figures \ref{fig:mult_latp} and \ref{fig:fixed_mult_latp} show, that our protocols are only marginally affected by the restriction compared to the baseline. 
f
Our A2B and Bit2A protocol-specific primitives do not require any latency-critical communication between the parties and are thus classified by us as 0-round protocols. Figures \ref{fig:a2bit_lat} and \ref{fig:bit2a_lat} show that even if we restrict the latency of all links between parties, these are marginally affected compared to the one-round baseline. Note that these are only the protocol-specific computations required before performing an arithmetic XOR or Boolean addition (c.f. Table \ref{tab:additional_protocols}) to complete the share conversion as described in \secref{other-A2B} and \secref{other-bit2A}. The high-level protocol-agnostic circuits can be evaluated with \piMult which is already assessed by Figure \ref{fig:mult_latp}. The results show that any high-level circuit that depends on our additional protocols can be expected to have similar runtime advantages in real-world network settings as shown in \secref{bench-bottlenecks}.
\vspace{-1mm}

%------------------------------
%------------------------------

%--------------------------

\end{document}